\documentclass[preprint]{aastex63}
\usepackage{xspace}
\usepackage{graphicx}
\usepackage{natbib}
\usepackage{url}
\usepackage{float}
\raggedright
\newcommand{\gbonraoblurb}{The Green Bank Observatory and National
Radio Astronomy Observatory are facilities of the National Science
Foundation operated under cooperative agreements by Associated
Universities, Inc.}
\newcommand{\degree}{\ensuremath{\,^\circ}}

\newcommand{\gyr}{\ensuremath{\,{\rm Gyr}}}

\newcommand{\mhz}{\ensuremath{\,{\rm MHz}}}
\newcommand{\ghz}{\ensuremath{\,{\rm GHz}}}

\newcommand{\K}{\ensuremath{\,{\rm K}}}
\newcommand{\mk}{\ensuremath{\,{\rm mK}}}
\newcommand{\m}{\ensuremath{\,{\rm m}}}

\newcommand{\cm}{\ensuremath{\,{\rm cm}}}

\newcommand{\percc}{\ensuremath{\,{\rm cm^{-3}}}}

\newcommand{\pc}{\ensuremath{\,{\rm pc}}}
\newcommand{\kms}{\ensuremath{\,{\rm km\, s^{-1}}}}

\newcommand{\hr}{\,hr}

\newcommand{\microK}{\ensuremath{\rm \,\mu K}}

\newcommand{\msun}{\ensuremath{{\rm \,M_\odot}}\xspace}     
\newcommand{\zsun}{\ensuremath{{\rm \,Z_\odot}}\xspace}     
\newcommand{\te}{\ensuremath{T_{\rm e}}}

\newcommand{\tsys}{\ensuremath{T_{\rm sys}}}

\newcommand{\hrrl}[1]{H#1}

\newcommand{\halpha}   {\hrrl{91}\ensuremath{\alpha}}
\newcommand{\hbeta}    {\hrrl{114}\ensuremath{\beta}}
\newcommand{\hbbeta}   {\hrrl{115}\ensuremath{\beta}}
\newcommand{\hgamma}   {\hrrl{130}\ensuremath{\gamma}}
\newcommand{\hggamma}  {\hrrl{131}\ensuremath{\gamma}}
\newcommand{\hdelta}   {\hrrl{144}\ensuremath{\delta}}
\newcommand{\hepsilon} {\hrrl{154}\ensuremath{\epsilon}}
\newcommand{\heepsilon}{\hrrl{152}\ensuremath{\epsilon}}
\newcommand{\hzeta}    {\hrrl{164}\ensuremath{\zeta}}
\newcommand{\heta}     {\hrrl{171}\ensuremath{\eta}}
\newcommand{\htheta}   {\hrrl{180}\ensuremath{\theta}}
\newcommand{\hiota}    {\hrrl{187}\ensuremath{\iota}}
\newcommand{\hkappa}   {\hrrl{193}\ensuremath{\kappa}}
\newcommand{\hlambda}  {\hrrl{198}\ensuremath{\lambda}}
\newcommand{\hmu}      {\hrrl{203}\ensuremath{\mu}}
\newcommand{\hnu}      {\hrrl{208}\ensuremath{\nu}}
\newcommand{\hxi}      {\hrrl{213}\ensuremath{\xi}}
\newcommand{\homicron} {\hrrl{221}\ensuremath{o}}
\newcommand{\hpi}      {\hrrl{222}\ensuremath{\pi}}
\newcommand{\hrho}     {\hrrl{227}\ensuremath{\rho}}
\newcommand{\hsigma}   {\hrrl{232}\ensuremath{\sigma}}
\newcommand{\htau}     {\hrrl{236}\ensuremath{\tau}}
\newcommand{\hupsilon} {\hrrl{238}\ensuremath{\upsilon}}
\newcommand{\hphi}     {\hrrl{239}\ensuremath{\phi}}
\newcommand{\hchi}     {\hrrl{247}\ensuremath{\chi}}
\newcommand{\hpsi}     {\hrrl{249}\ensuremath{\psi}}
\newcommand{\homega}   {\hrrl{249}\ensuremath{\omega}}
\newcommand{\expo}[1]{\ensuremath{10^{#1}}}
\newcommand{\nexpo}[2]{\ensuremath{#1 \times 10^{#2}}}

\newcommand{\hii}{H\,{\sc ii}}

\newcommand{\yp}[1]{\ensuremath{y_{#1}^{+}}}

\newcommand{\deut}{\ensuremath{^{2}{\rm H}}}
\newcommand{\he}[1]{\ensuremath{^{#1}{\rm He}}}
\newcommand{\hep}[1]{\ensuremath{^{#1}{\rm He}^{+}}}

\newcommand{\li}[1]{\ensuremath{{}^#1{\rm Li}}}

\newcommand{\her}[1]{\ensuremath{{}^#1{\rm He}/{\rm H}}}
\newcommand{\hepr}[1]{\ensuremath{{}^#1{\rm He}^{+}/\,{\rm H^+}}}
\newcommand{\heppr}[1]{\ensuremath{{}^#1{\rm He}^{++}/\,{\rm H^+}}}
\newcommand{\carbon}[1]{\ensuremath{^{#1}{\rm C}}}
\newcommand{\cratio}{\ensuremath{{}^{12}{\rm C}/\,^{13}{\rm C}}}
\newcommand{\threec}[1]{3C\thinspace #1}
\newcommand{\ic}[1]{IC\thinspace #1}
\newcommand{\ngc}[1]{NGC\thinspace #1}

\newcommand{\gsim}{\ensuremath{\gtrsim}}
\newcommand{\lsim}{\ensuremath{\lesssim}}
\newcommand{\urltilda}{\kern -.15em\lower .7ex\hbox{\~{}}\kern .04em}

\shorttitle{Planetary Nebulae \hep3\ Emission}
\shortauthors{Bania \& Balser}
\begin{document}

\setlength{\parindent}{15pt} 

\title{Green Bank Telescope Observations of ${\bf ^3He^{\bf +}}$: Planetary Nebulae}

\correspondingauthor{T.~M. Bania}
\email{bania@bu.edu, dbalser@nrao.edu}

\author[0000-0003-4866-460X]{T. M. Bania}
\affiliation{Institute for Astrophysical Research, Astronomy
  Department, Boston University,\\
  725 Commonwealth Ave., Boston, MA 02215, USA.}

\author[0000-0002-2465-7803]{Dana S. Balser}
\affiliation{National Radio Astronomy Observatory, 520 Edgemont
  Rd., Charlottesville, VA 22903, USA.}

\begin{abstract}
  We use the Green Bank Telescope to search for \hep3\ emission from a
  sample of four Galactic planetary nebulae: \ngc{3242}, \ngc{6543},
  \ngc{6826}, and \ngc{7009}.  During the era of primordial
  nucleosynthesis the light elements \deut, \he3, \he4, and \li7\ were
  produced in significant amounts and these abundances have since been
  modified primarily by stars.  Observations of \hep3\ in \hii\ regions
  located throughout the Milky Way disk reveal very little variation in
  the \her3\ abundance ratio---the ``\he3\ Plateau''---indicating that
  the net effect of \he3\ production in stars is negligible.  This is in
  contrast to much higher \her3\ abundance ratios reported for some
  planetary nebulae. This discrepancy is known as the ``\he3\ Problem''.
  We use radio recombination lines observed simultaneously with the
  \hep3\ transition to make a robust assessment of the spectral
  sensitivity that these observations achieve.  We detect spectral lines
  at $\sim$\,1 -- 2 \mk\ intensities, but at these levels instrumental
  effects compromise our ability to measure accurate spectral line
  parameters.  We do not confirm reports of previous detections of
  \hep3\ in \ngc{3242} nor do we detect \hep3\ emission from any of our
  sources.
  This result calls into question all reported detections of
  \hep3\ emission from any planetary nebula.  The \her3\ abundance upper
  limit we derive here for \ngc{3242} is inconsistent with standard
  stellar production of \he3\ and thus requires that some type of extra
  mixing process operates in low-mass stars.
\end{abstract}


\section{The \he3\ Problem}\label{sec:intro}

The \he3\ abundance in Milky Way sources provides important constraints
for many fields of astrophysics including cosmology, stellar evolution,
and Galactic chemical evolution.  The abundance of \he3\ is derived from
measurements of the hyperfine transition of \hep3\ which has a rest
wavelength of 3.46\cm\ (8.665\ghz). The present \he3\ abundance results
from a combination of Big Bang Nucleosynthesis (BBN) and stellar
nucleosynthesis \citep[see, e.g.,][]{1994WR}. Abundances derived for the
light elements \deut, \he3, \he4, and \li7\ can be compared with BBN
models for their primordial production to give an estimate for the
primordial baryon to photon ratio, $\eta$. Observations made by the
{\em Wilkinson Microwave Anisotropy Probe (WMAP)} also yield a value for
$\eta$.  The concordance of these values for $\eta$ is a triumph of
observational cosmology. Together, these analyses derive a
\he3\ primordial abundance of (\her3)${}_{\rm p}$ =
\nexpo{(1.00\pm0.07)}{-5} by number \citep{2002Nature, 2003RTMC,
  2004bbns}. Subsequent to the BBN era this primordial \he3\ abundance
will be modified by nuclear processing in many generations of stars.

We are studying the evolution of the \he3\ abundance over cosmic time by
using observations of \hep3\ emission to derive the \he3\ abundance in
Milky Way \hii\ regions and planetary nebulae (PNe). \hii\ regions are
zero-age objects compared to the age of the Galaxy. Their
\he3\ abundances chronicle the results of billions of years of Galactic
chemical evolution (GCE). PNe abundances arise from material that has
been ejected from low-mass (M \lsim\ 2 \msun) and intermediate-mass (M
$\sim$ 2--5 \msun) stars.
Standard stellar evolution models\footnote[3]{I.e., models that only
  consider convection as a mixing mechanism in stellar interiors.}
predict the production of significant amounts of \he3\ in low-mass stars
with peak abundances of $^{3}{\rm He/H} \sim
\nexpo{\rm few}{-3}$ by number \citep{1967aIben, 1967bIben, 1972Rood}.
As the star ascends the red giant (RGB) and asymptotic giant (AGB)
branches, the convective zone subsumes the \he3\ enriched material which
is expected to be expelled into the interstellar medium (ISM) via
ejected planetary nebula shells \citep[see, e.g.,][]{1970PNeEject}.

\citet[hereafter RST]{RST1976} first identified the interstellar
\he3\ abundance as being a significant constraint on stellar evolution
models. Using yields from standard stellar models, RST found a
\he3\ enrichment of the primordial abundance due to stellar processing. 
The \her3\ abundance in the ISM is thus an important GCE diagnostic.
RST argued that standard stellar nucleosynthesis in low-mass stars
predicts that: (1) the protosolar \her3\ abundance should be less than
that found in the present ISM; (2) the \her3\ abundance should grow 
with source metallicity; and (3) there should be a \her3\ abundance
gradient across the Galactic disk with the highest abundances occurring
in the highly-processed inner Galaxy. None of these predictions is
confirmed by observations.

\hii\ regions sample the result of the chemical evolution of the Milky
Way since its formation.  After 40 years of effort \citep[see our papers
  starting and ending with][]{1979RWS, BB2018} our observations still
yield \he3\ abundances for \hii\ regions that are inconsistent with
RST's expectations.  Furthermore, the \hii\ region \her3\ abundances
together with those derived for protosolar material \citep{1993Geiss}
and the local solar neighborhood \citep{1996LISM} all indicate a value
for \her3\ $\sim$\,\nexpo{2}{-5} by number.  \hii\ region abundances
thus show no evidence for stellar \he3\ enrichment during the last
4.5\gyr.  Furthermore, there is no large \he3\ abundance gradient across
the Milky Way disk and, finally, there is no trend of \he3\ abundance
with source metallicity (``The \he3\ Plateau'').

Nonetheless, there are three PNe with published \hep3\ detections:
\ngc{3242} \citep{1992Nature, 1997PNe, 1999n3242}, J\thinspace 320
\citep{2006J320}, and \ic{418} \citep{2016IC418}. The \her3\ abundance
ratios derived for these detections range from \nexpo{2}{-4} to
\nexpo{6}{-3} by number, consistent with the yields predicted by
standard stellar evolution models
\citep[e.g.,][]{1972Rood}. Furthermore, these abundances are an order of
magnitude higher than the abundances found in \hii\ regions \citep[see,
  e.g.][]{BB2018}. Thus although {\em some} PNe have apparently produced
\he3\ according to standard stellar nucleosynthetic expectations, there
is no evidence for substantial \he3\ enrichment in the Milky Way ISM.
This conundrum has long been dubbed ``The \he3\ Problem''
\citep{1995Galli}.

\citet[RBW hereafter]{1984RBW} suggested that the \he3\ problem could be
related to striking chemical abundance anomalies in red giant
stars. They posited that some extra-mixing process during the RGB stage
in stellar interiors might reduce the \he3\ abundance, and this might
also explain the depletion of \li7\ in main-sequence stars and the low
\cratio\ abundance ratios in low-mass RGB stars \citep[also
  see][]{1995Charbonnel}.

Resolving ``The \he3\ Problem'' requires that the vast majority of
low-mass stars fail to enrich the ISM with \he3\ produced during their
nucleosynthetic lifetimes.  GCE models can account for ``The
\he3\ Plateau'' only if \gsim\ 90 \% of solar analog stars are
non-producers of \he3 \citep{1997Galli, 1998Tosi, 2000Palla,
  2002Chiappini}.  Exactly what produces this extra-mixing has been
actively investigated for many years. \citet{BB2018} give a
comprehensive review of the various physical mechanisms proposed and the
current status of this research. The most significant recent advance was
made by \citet{2007aCharZahn} who argue that the thermohaline
instability, a double-diffusive instability, is an important mechanism.
The best stellar evolutionary models for low and intermediate-mass stars
now include both the thermohaline instability and rotation-induced
mixing \citep{2010thermohaline, 2011Lagarde}.  \citet{2012Lagarde} used
their model \he3\ yields together with GCE models to predict a modest
enrichment of \he3\ with time and \her3\ abundance ratios about a factor
of two higher in the central regions of the Milky Way relative to the
outer regions.

The scatter of the \he3\ Plateau abundances determined by
\citet{2002Nature}, however, is large and spans the range of abundances
predicted by \citet{2012Lagarde}. \citet{BB2018} improved the derived
\her3\ abundances for Galactic \hii\ regions by studying a sample of 5
morphologically simple nebulae located over a wide range of
Galactocentric radii: $4.4\,{\rm kpc} < R_{\rm gal} < 16.2\,{\rm
  kpc}$. Their goal was to derive accurate \her3\ abundance ratios for a
small sample of sources to uncover any trend in the \her3\ abundance
with $R_{\rm gal}$, and to compare their results with the predictions of
\citet{2012Lagarde}.

\citet{BB2018} determined a shallow \her3\ radial gradient of $-0.116
\pm\ 0.022\, \times \,$\expo{-5}$\,$kpc$^{-1}$, consistent with the
overall trend predicted by \citet{2012Lagarde}.  Their \her3\ abundance
ratios, however, are typically slightly less than the models that
include thermohaline mixing.  Furthermore, \cite{BB2018} could not
obtain \her3\ abundances with sufficient accuracy to determine whether
or not strong magnetic fields in some stars could inhibit the
thermohaline instability as predicted by \citet{2007bCharZahn}.

A critical constraint on all these stellar evolution models, however,
is the \her3\ abundance derived for a sample of just three PNe.
Observations of \hep3\ emission are very challenging.  Accurate
measurement of these weak, broad spectral lines requires significant
integration time and a stable, well-behaved spectrometer
\citep{1994Balser}. This is particularly true when observing
\hep3\ emission from PNe since the lines are weaker and broader than for
\hii\ regions.

Our goal here is to confirm our previous \hep3\ detection for \ngc{3242}
and to derive accurate \her3\ abundance ratios for it and a small sample
of other PNe.  The Green Bank Observatory's
(GBO)\footnote[4]{\gbonraoblurb} Green Bank Telescope (GBT) X-band
(3\cm\ wavelength) spectrometer is in principle an order of magnitude
more sensitive than any previous \cm-wavelength radio frequency
telescope/spectrometer system.  This increased sensitivity results from
the GBT's unique 100\m\ clear aperture, unblocked optics that enable
unprecedented dynamic range measurements together with its X-band
receiver, and auto-correlation spectrometer (ACS).

\clearpage
\begin{figure}
\centering
\includegraphics[angle=0,scale=0.4]{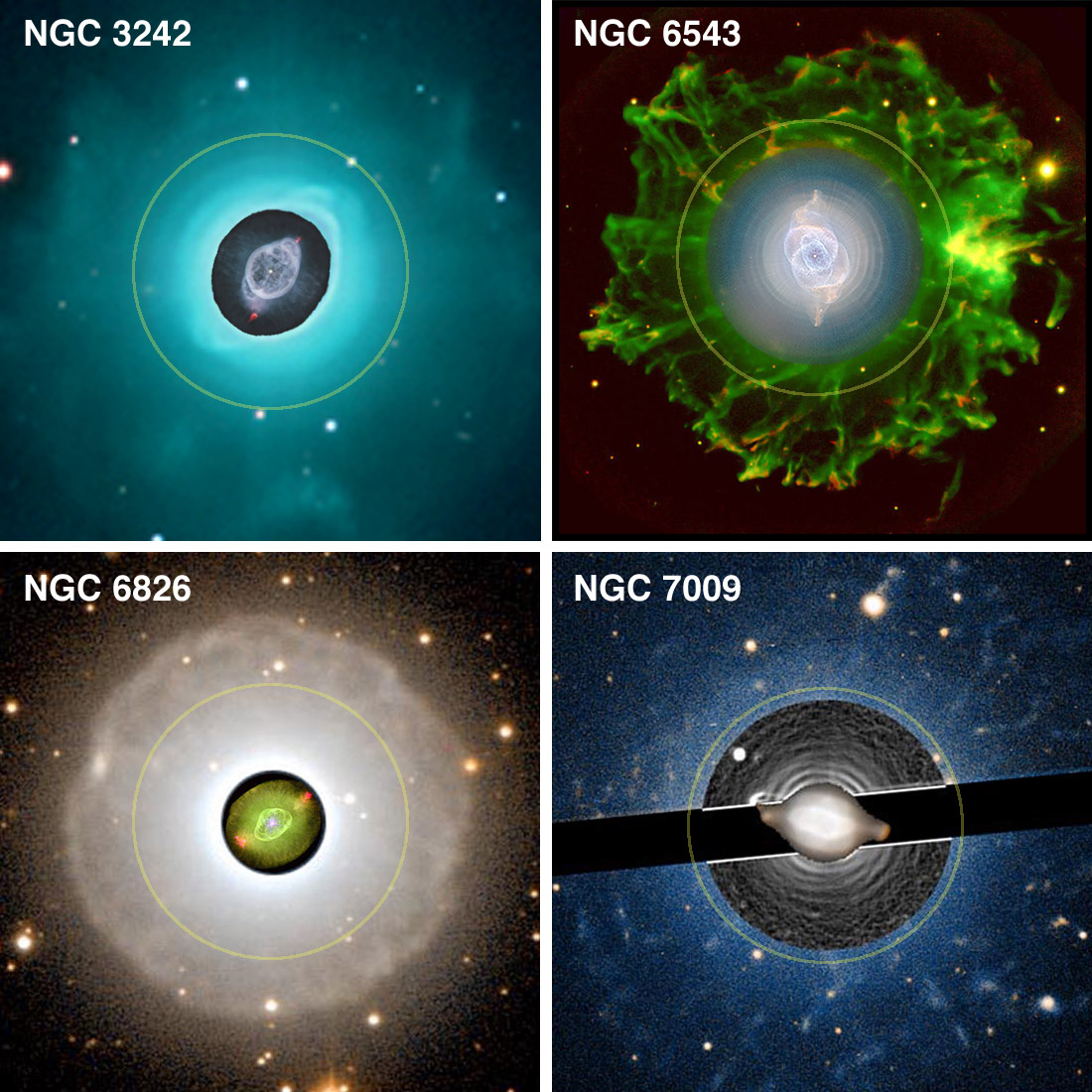}
\caption{  
  Public domain optical images of the planetary nebulae studied
  here. North is up in each $\sim$\,180\arcsec\ square image.  The
  yellow circles are centered on the exciting stars and show the GBT
  beam size.  Images of the PN cores have been overlayed atop more
  sensitive wide field images of the PN halos. The halos shown here are
  photometrically uncalibrated ground-based images, and the optical
  filters vary from one object to another.  Furthermore, in order to
  make the halos visible, their images have been intensity flattened so
  what is seen here is the logarithm of the surface
  brightness. \citet{2016PNrings} describe the general properties of
  faint optical rings in PNe.
}
\label{fig:PNeMosaic}
\end{figure}

\section{The Planetary Nebula Sample}\label{sec:sample}

We are observing a sample of PNe that is purposefully biased to maximize
the likelihood of detecting \hep3\ emission \citep[see,
  e.g.,][]{1999He3abundance}.  Our PN targets are sources that both show
evidence for {\em no} extra-mixing and also have physical properties
that, for a given \her3\ abundance, would produce the largest
\hep3\ intensities.  We have chosen PNe targets based on their estimated
\hep3\ intensity using an assumed constant \her3\ abundance together
with the following criteria:
(1) Sources with the strongest expected \hep3\ intensities.  This
  requires measurements of PN brightness, angular size, and
  distance. Ordinarily the radio continuum was used to estimate the
  brightness.
(2) Sources that are located $\gsim$~500\pc\ away from Galactic plane.
  This increases the likelihood that a given PN is drawn from the
  old/thick disk or halo population. On average, such PNe would have low
  progenitor masses, longer main-sequence lifetimes, and, thus, higher
  \her3\ abundances.
(3) Sources without any indication in the nebular gas of nuclear
  processing on the RGB or AGB.  High He, N, or \carbon{13} abundances
  would suggest that the gas had undergone CNO processing and,
  presumably, \he3\ depletion.
(4) Sources where most of the He is singly ionized.

Here, we used the GBT to search for \hep3\ emission from 
four Galactic PNe: \ngc{3242}, \ngc{6543}, \ngc{6826}, and \ngc{7009}.
Table~\ref{tab:sample} summarizes properties of our PN sample.  Listed
are the source name, equatorial coordinates, optical heliocentric source
velocity, $V_{\rm opt}$, heliocentric distance, $D_{\rm sun}$, and the
effective temperature of the central exciting star, $T_{\rm eff}$. The
$D_{\rm sun}$ values are {\it Gaia} DR2 parallax distances from
\citet{2019PNeDist}. The effective temperatures stem from
\citet{2018SBJ}.

\begin{deluxetable}{cccccc}[h]
\tablecolumns{5} \tablewidth{0pt}
\tablecaption{Galactic Planetary Nebula Sample \label{tab:sample}}
\tablehead{
      \colhead{} & \colhead{R.A. (J2000)} & \colhead{Decl. (J2000)} &
      \colhead{{$V_{\rm opt}$}\tablenotemark{a}} &
      \colhead{$D_{\rm sun}$\tablenotemark{b}}   &
      \colhead{$T_{\rm eff}$\tablenotemark{c}}   \\
      \colhead{Source} & \colhead{(hh:mm:ss.s)} & \colhead{(dd:mm:ss)} &
      \colhead{(\kms)} & \colhead{(kpc)} & \colhead{(K)} 
}
\startdata
\ngc{3242} & 10:24:46.2 & $-$18:38:34 & \phn$+$4.6  &
     1.466~${^{+0.219}_{-0.168}}$ & 90,000 \\
\ngc{6543} & 17:58:33.4 & $+$66:37:59 & $-$66.1     &
     1.625~${^{+0.212}_{-0.167}}$ & 68,000 \\
\ngc{6826} & 19:44:48.3 & $+$50:31:30 & \phn$-$6.2  &
     1.575~${^{+0.128}_{-0.103}}$ & 46,000 \\
\ngc{7009} & 21:04:10.8 & $-$11:21:57 & $-$46.6     &
     1.154~${^{+0.177}_{-0.136}}$ & 82,000 \\
\enddata
\tablenotetext{{\rm a}}{Heliocentric radial velocity using the optical
  definition of velocity and the barycentric reference frame \citep{2006Vopt}.}
\tablenotetext{{\rm b}}{{\it Gaia} DR2 parallax from
  \citet{2019PNeDist}.}
\tablenotetext{{\rm c}}{Effective temperature of central exciting star
  from Table~8 in \citet{2018SBJ}.}
\end{deluxetable}

We have previously reported detections of \hep3\ emission from
\ngc{3242} using independent observations made with the Max Planck
Institut f\"ur Radioastronomie (MPIfR) 100\m\ and the National Radio
Astronomy Observatory (NRAO) 140 Foot telescopes.  Moreover, the
composite \hep3\ 100\m\ spectrum of \ngc{6543}+\ngc{7009} suggests
\hep3\ emission at the $\sim$~1\mk\ intensity level. Too, observations
made with the Arecibo 305\m\ telescope give a composite PNe spectrum
including \ngc{6543} that showed a hint of \hep3\ emission at this same
level \citep{2007bozo}.

Our sample PNe are shown in Figure~\ref{fig:PNeMosaic} where images of
the PN cores have been overlayed atop more sensitive wide field images
of the PN halos produced by thousands of years of AGB winds injecting
gas into the surrounding nebula at 10--15\kms.  Notable and oddly, each
of the cores in these four PNe contain pairs of small, [N II] emitting
regions --- ``Fast Low-Ionization Emission Regions'' (FLIERs) --- seen
as red knots on opposite sides of the central star.  FLIERs are volumes
of gas moving at supersonic outflow speeds that are significantly higher
than the nebular flows in which they are embedded, and their ionizations
are much lower. In general FLIERs are rare, so the ubiquitous presence
of FLIERS in all four of our targets is somewhat surprising.

All of our PNe are surrounded by extended low density halos that
are expanding due to episodic mass loss from their evolved stellar
cores.  An AGB star steadily blows off its outer layers. After this,
although the mass loss rate from the faster post-AGB outflows is higher
than that carried by the AGB winds, the duration of this mass ejection
is quite brief compared to the AGB ejection phase. Because of this, at
the end of its evolution nearly all of the mass of a PN is ejected by
the slow AGB winds and this material ultimately ends up residing in its
halo.

Our GBT observations measure the combined emission from the entire
core~$+$~halo complex of our sample PNe.  The angular resolution (the
half power beam width [HPBW]) of the GBT at 8665\ghz\ is plotted in
Figure~\ref{fig:PNeMosaic} as a yellow circle in these
$\sim$\,180\arcsec\ square images.  Since these PN core radii are
typically $\sim$~10\arcsec\ in size, approximately 80\% of the Gaussian
GBT beam area is sampling only halo gas.


\section{GBT Observations and Data Analysis}\label{sec:obs}

We used the GBT to search for \hep3\ emission from our PN sample during
a series of observing sessions held between 2004 and 2011. These
measurements targeted both PNe and \hii\ regions. Here we focus on our
PNe sample; \cite{BB2018} analyzed the \hii\ regions and discussed 
the implications of their derived \he3\ abundances.
Observations were made with the X-band (8-10\ghz) receiver and either
the Digital Continuum Receiver (DCR) or Autocorrelation Spectrometer
(ACS) backend.  Before 2011 we targeted all four PNe in our sample and
used a slightly different tuning configuration of the ACS than that
employed during the 2011 epoch.  In 2011 we only observed \ngc{3242} and
\ngc{6543}.  This results in significantly larger total integration
times for the \hep3\ and radio recombination line (RRL) spectra for
these two nebulae compared with those measured for \ngc{6826} and
\ngc{7009}. Full details for the 2011 observations, tuning
configuration, and data analysis strategy are described by
\cite{BB2018}.

\begin{deluxetable}{cccccc}[h]
\tablecolumns{5} \tablewidth{0pt}
\tablecaption{Continuum Emission Properties at 8665\mhz \label{tab:continuum}}
\tablehead{
      \colhead{} &
      \colhead{T$_\alpha$\tablenotemark{a}} &
      \colhead{$\Theta_\alpha$\tablenotemark{b}} &
      \colhead{T$_\delta$\tablenotemark{a}} &
      \colhead{$\Theta_\delta$\tablenotemark{b}} \\
      \colhead{Source} & \colhead{(K)} & \colhead{(arcsec)} &
      \colhead{(K)} & \colhead{(arcsec)} 
}
\startdata
\ngc{3242} & 
    1.25~$\pm$~0.01 & 85.8~$\pm$~0.9 &
    1.24~$\pm$~0.01 & 91.6~$\pm$~0.8  \\
\ngc{6543} & 
    1.41~$\pm$~0.01 & 85.5~$\pm$~0.3 &
    1.43~$\pm$~0.01 & 84.2~$\pm$~0.5  \\
\ngc{6826} &
    0.65~$\pm$~0.01 & 87.7~$\pm$~0.8 &
    0.66~$\pm$~0.01 & 88.7~$\pm$~0.8  \\
\ngc{7009} &
    1.08~$\pm$~0.01 & 84.3~$\pm$~0.9 &
    1.08~$\pm$~0.01 & 83.5~$\pm$~0.9  \\
\enddata
\tablenotetext{{\rm a}}{R.A. and Decl. intensities in antenna temperature units.}
  \tablenotetext{{\rm b}}{R.A. and Decl. full width at half maximum (FWHM)
  angular sizes.}
\end{deluxetable}

\vspace{-2cm}

\subsection{Observing Protocol and Calibration}

During each observing session the pointing and focus were updated every
$\sim$\,2\hr\ using DCR observations of the continuum emission from a
calibration source located within $\sim$\,15\degree\ of the target
position.  To calibrate the antenna temperature intensity scale, a noise
signal with an intensity of $\sim$\,5--10\% of the total system
temperature, \tsys, was injected into the signal path.  For the GBT
X-band system these noise diodes have intensities $T_{\rm cal} \sim
2$\K.  To check the flux density calibration, we observed the continuum
emission from the flux density calibrator \threec{286}.  We adopt the
\citet{2000Peng} flux density for \threec{286} and assume a telescope
  gain at X-band of $2\,{\rm K}\,{\rm Jy}^{-1}$ \citep{2001Gigho}.
  Based on these measurements we deem the intensity scale to be accurate
  to within $\sim$\,5--10\%.

\subsection{Radio Continuum Emission}\label{sec:continuum}

The continuum emission from PNe stems from thermal free-free
bremsstrahlung radiation from the nebular plasmas. Since hydrogen is
overwhelmingly the most abundant element in these plasmas, analyzing the
continuum emission allows one to determine the PNe hydrogen abundances. 
Hydrogen abundances are essential for the derivation of the \her3\ ratio.

We measured the continuum emission properties of our PN sources with the
GBT using the DCR tuned to a rest frequency of 8665\mhz.  The continuum
emission intensity was sampled over a bandwidth of 320\mhz\ as the GBT
was slewed in a cross pattern --- first in R.A. and then in Decl. ---
centered at each PN position. Table~\ref{tab:continuum} summarizes
Gaussian fits to these measurements. Listed are the peak intensities and
angular sizes in each direction together with the $\pm~1\sigma$ errors
in the fits.


\begin{figure}[h]
\centering
\includegraphics[angle=-90,scale=0.3]{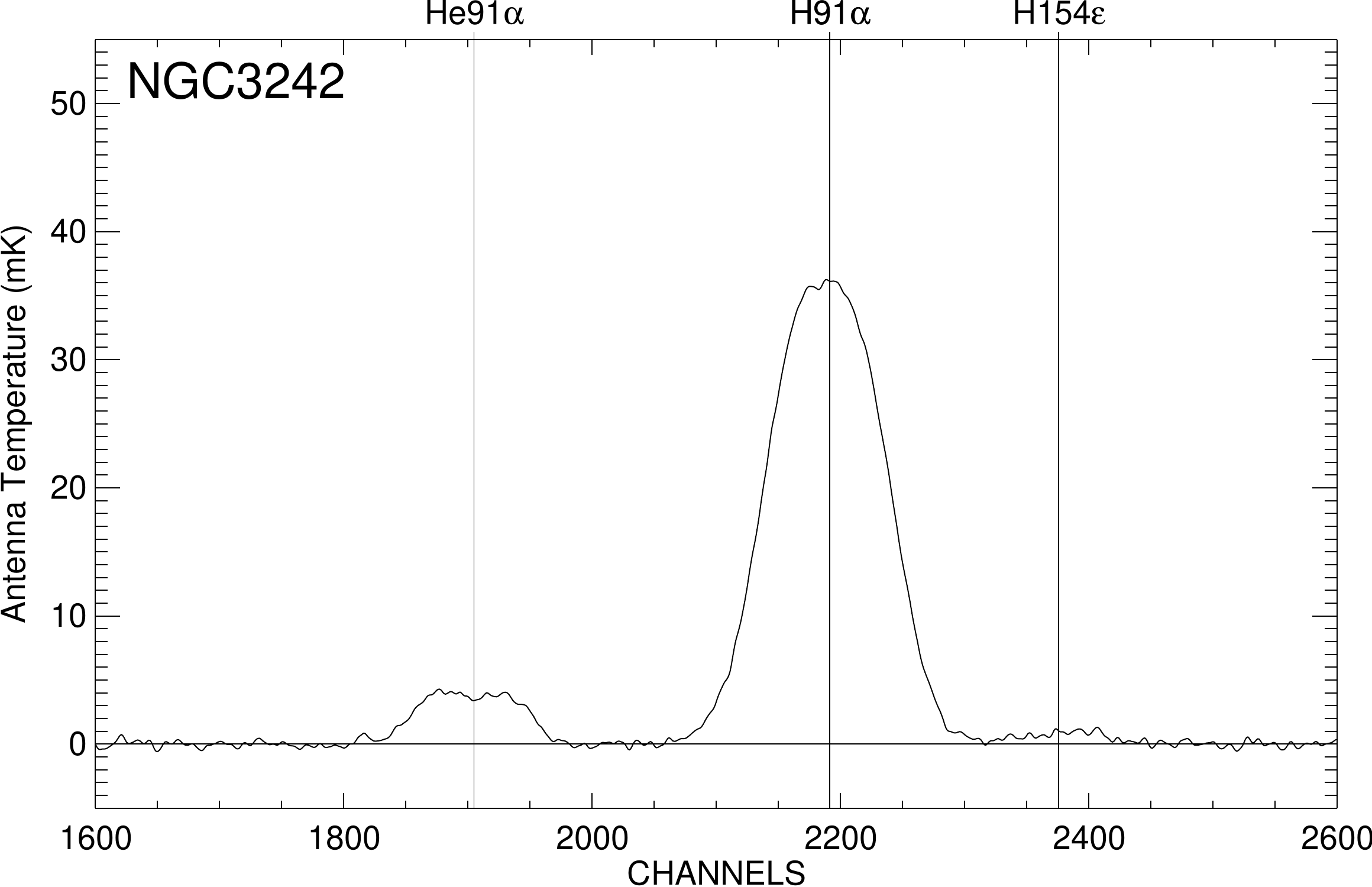}
\includegraphics[angle=-90,scale=0.3]{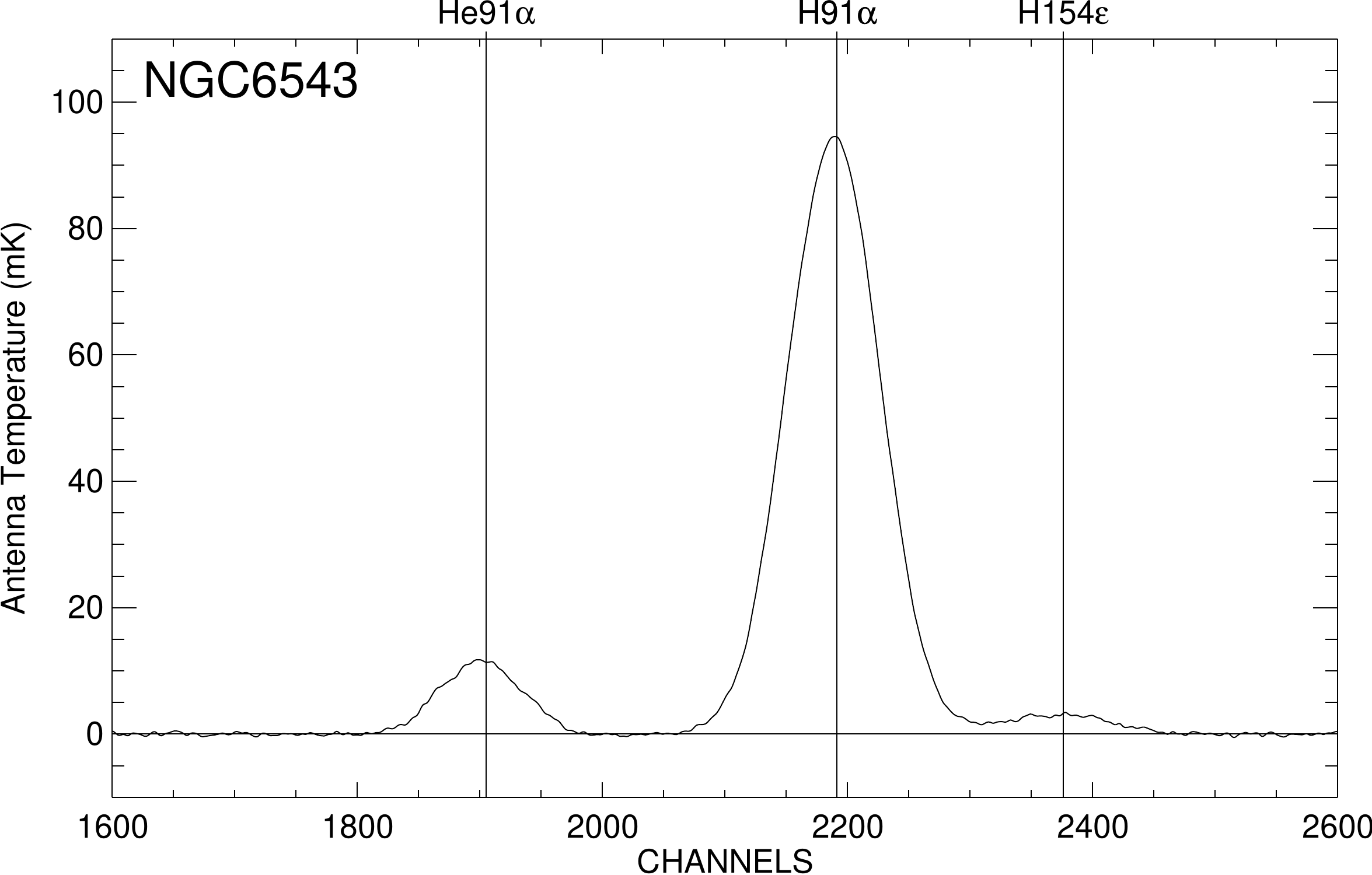}\\
\includegraphics[angle=-90,scale=0.3]{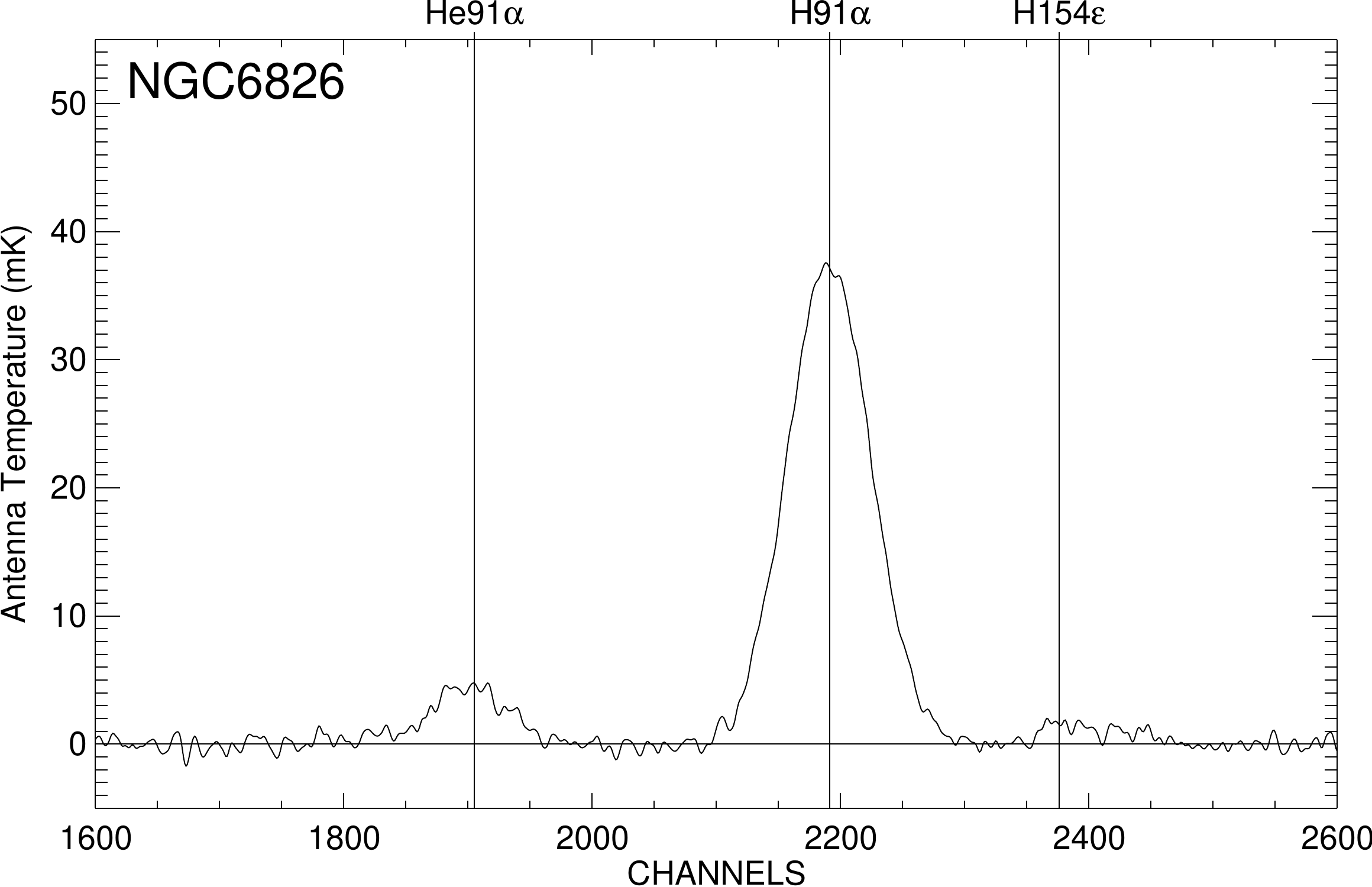}
\includegraphics[angle=-90,scale=0.3]{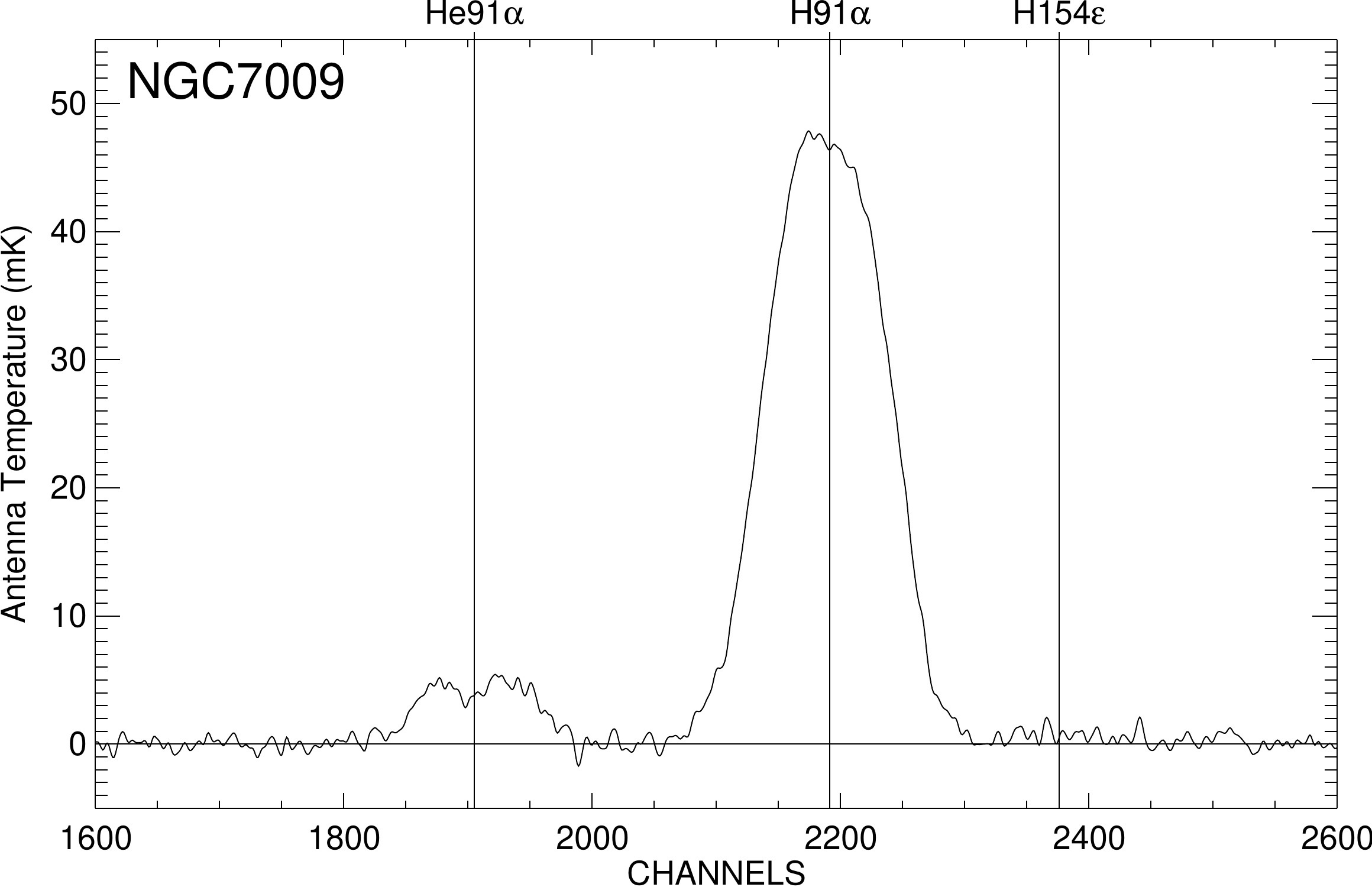}
\caption{
  Observed H91$\alpha$ spectra for the planetary nebula sample.  Shown
  clockwise starting from top left they are: \ngc{3242}, \ngc{6543},
  \ngc{7009}, and \ngc{6826}. Vertical lines flag, from left to right,
  the recombination line transitions: He91$\alpha$, H91$\alpha$, and
  H154$\epsilon$. Each spectrum spans a total bandwidth of 12.5\mhz.
  The spectra have been smoothed to a velocity resolution 5\kms\ and a
  5-th order model for the instrumental baseline has been subtracted.
}
\label{fig:halpha}
\end{figure}

\subsection{Radio Recombination Line Emission}\label{sec:lines}

To acquire our spectral line observations we employ total power position
switching by observing an Off-source position for 6 minutes and then the
target On-source position for 6 minutes, for a total time of 12 minutes. The
Off position is offset 6 minutes in R.A. relative to the On position so
that the telescope tracks the same sky path during each observation.

Each total power Off/On pair produces 16 independent spectra that are
observed simultaneously.  For all our observations the GBT ACS
spectrometer is configured with eight spectral windows (SPWs) at two
orthogonal, circular polarizations for a total of 16 SPWs. Each SPW has
a bandwidth of 50\mhz\ spanning 4096 spectral channels.  This gives a
spectral resolution per channel of 12.2 kHz, or a velocity resolution of
0.42\kms\ at 8665.65\mhz. The eight SPWs were tuned to different center
frequencies in order to observe both \hep3\ and also a host of different
RRL transitions that are used to assess spectrometer performance. For
all our observations 6 of the 8 SPW tunings were identical. The 2011
epoch observations replaced two RRL transition SPWs with \hep3\ tunings
displaced in frequency (see below).

There is a plethora of RRL transitions that lie within the frequency
ranges of our observed SPWs. These are compiled in
Table~\ref{tab:transitions} for transitions with principle quantum
numbers N $<$ 250 that have quantum number changes $\Delta$N $<$ 25. We
list these RRLs for completeness; transitions with $\Delta$N $>$ 7 are
too weak to be detected with the sensitivity achieved by the
observations reported here.

The data were analyzed using the single-dish software package
TMBIDL\footnote[5]{Version 8.1 see
  \url{https://github.com/tvwenger/tmbidl}.} \citep{TMBIDL}.  Before
averaging, each Off/On pair spectrum was visually inspected and
discarded if significant spectral baseline structure or radio frequency
interference (RFI) was present. Narrow band RFI that did not contaminate
the spectral lines was excised and the RFI-cleaned spectrum was included
in the average. Spectra were averaged in different ways to assess any
problems with the spectral baselines. For example, we made the following
tests: (1) inspected the average spectrum for each observing epoch to
search for any anomalies; (2) divided the entire data set into several
subsets to assess whether the noise was integrating down as expected;
(3) compared the two orthogonal polarizations, which should have nearly
identical spectra; and (4) compared the different SPWs for cases where
we had simultaneous, duplicate \hep3\ and RRL transition tunings
(see Section~\ref{sec:baseline}).

The GBT observations reported here are very deep integrations.  The
total integration time for \ngc{3242} and \ngc{6543} spectra is
$\sim$300\hr; for \ngc{6826} and \ngc{7009} it is $\sim$50\hr. This
gives typical root mean square (RMS) noise after smoothing to
5\kms\ resolution of $\sim$0.2\mk\ and $\sim$0.5\mk,
respectively. Integration times for different SPW tunings vary depending
on the ACS configuration for a particular observing epoch and on how
much data were excised during the averaging process.

To improve spectral sensitivity each spectrum was smoothed to a velocity
resolution of 5\kms.  Any sky continuum emission and baseline frequency
structure was then removed by fitting a fifth-order baseline model that
was then subtracted. We explored using different order baseline fits and
determined that 5-th order was the best compromise between under and
over fitting.  A clear indication of over-fitting, for example, would be
to achieve an impossible spectral RMS for the integration time.  The
RRL properties were measured by fitting a Gaussian function to the
spectra using a Levenberg-Markwardt \citep{2009Markwardt} least-squares
method to derive the peak intensity and the full-width at half-maximum
(FWHM) line width. As we shall show below, these measurements of RRL
properties make astrophysical sense and so indicate that we have not
over-fitted the removal of the instrumental baselines.

\begin{figure}[h]
\centering
\includegraphics[angle=-90,scale=0.3]{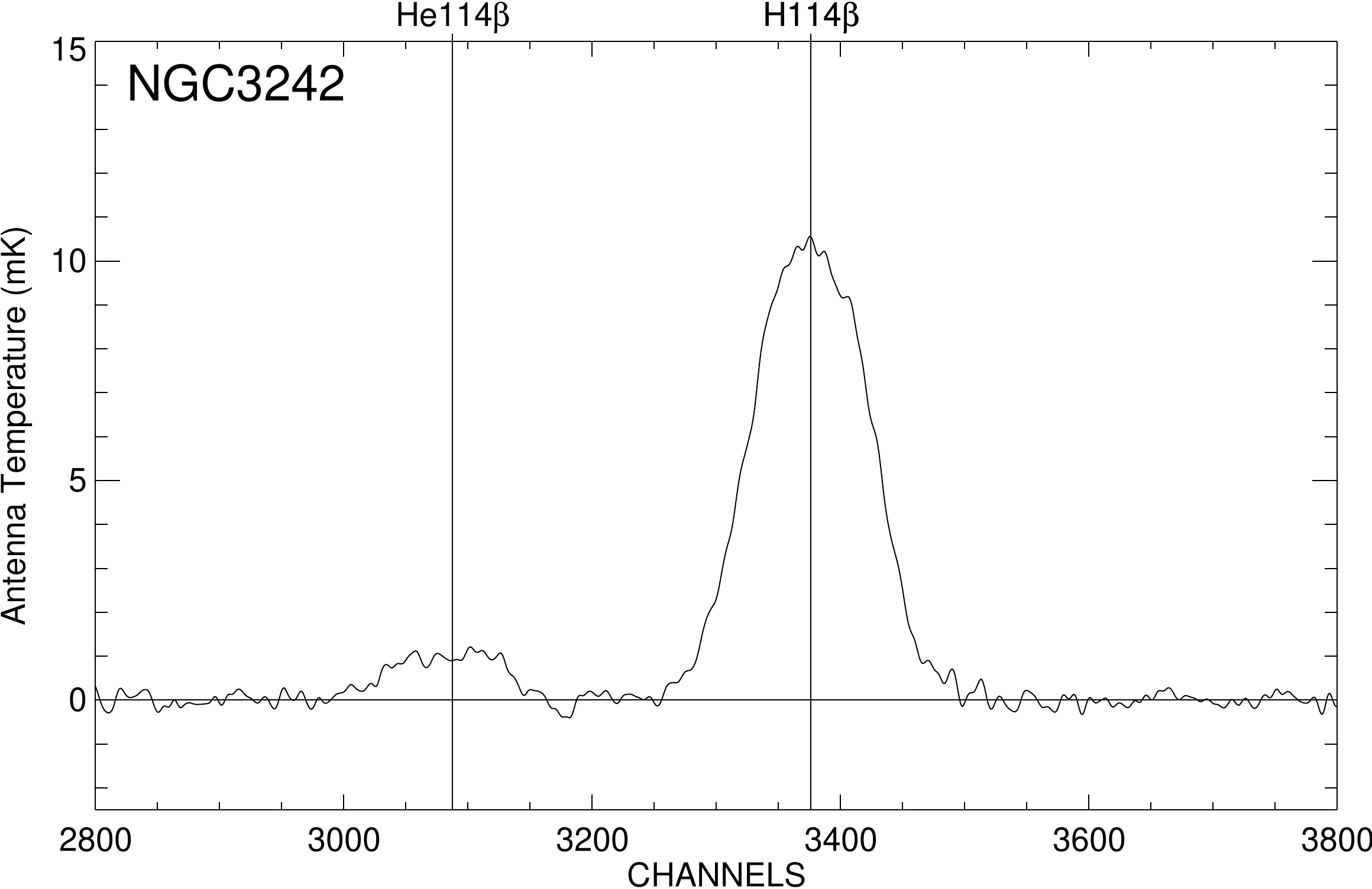}
\includegraphics[angle=-90,scale=0.3]{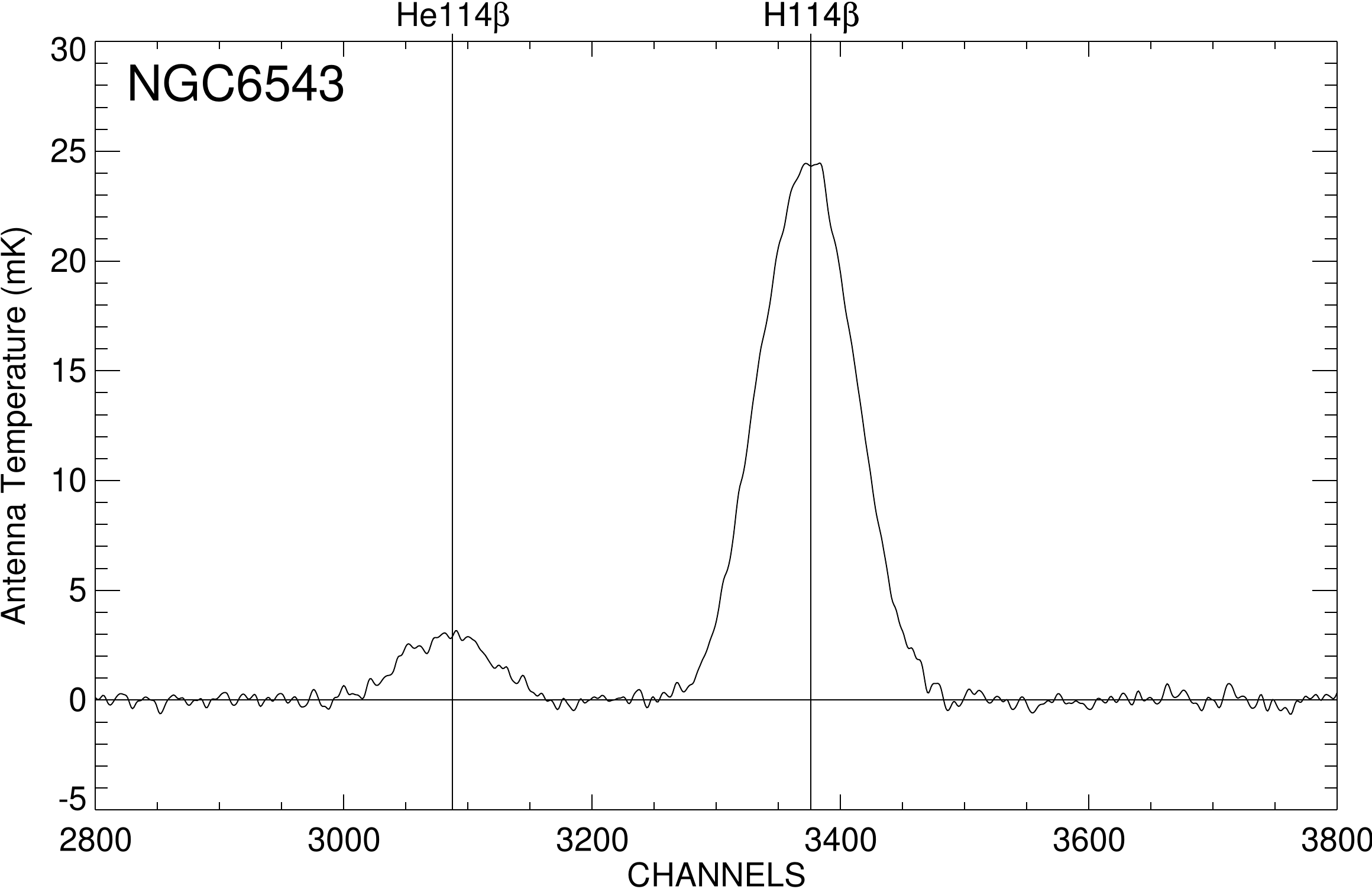}\\
\includegraphics[angle=-90,scale=0.3]{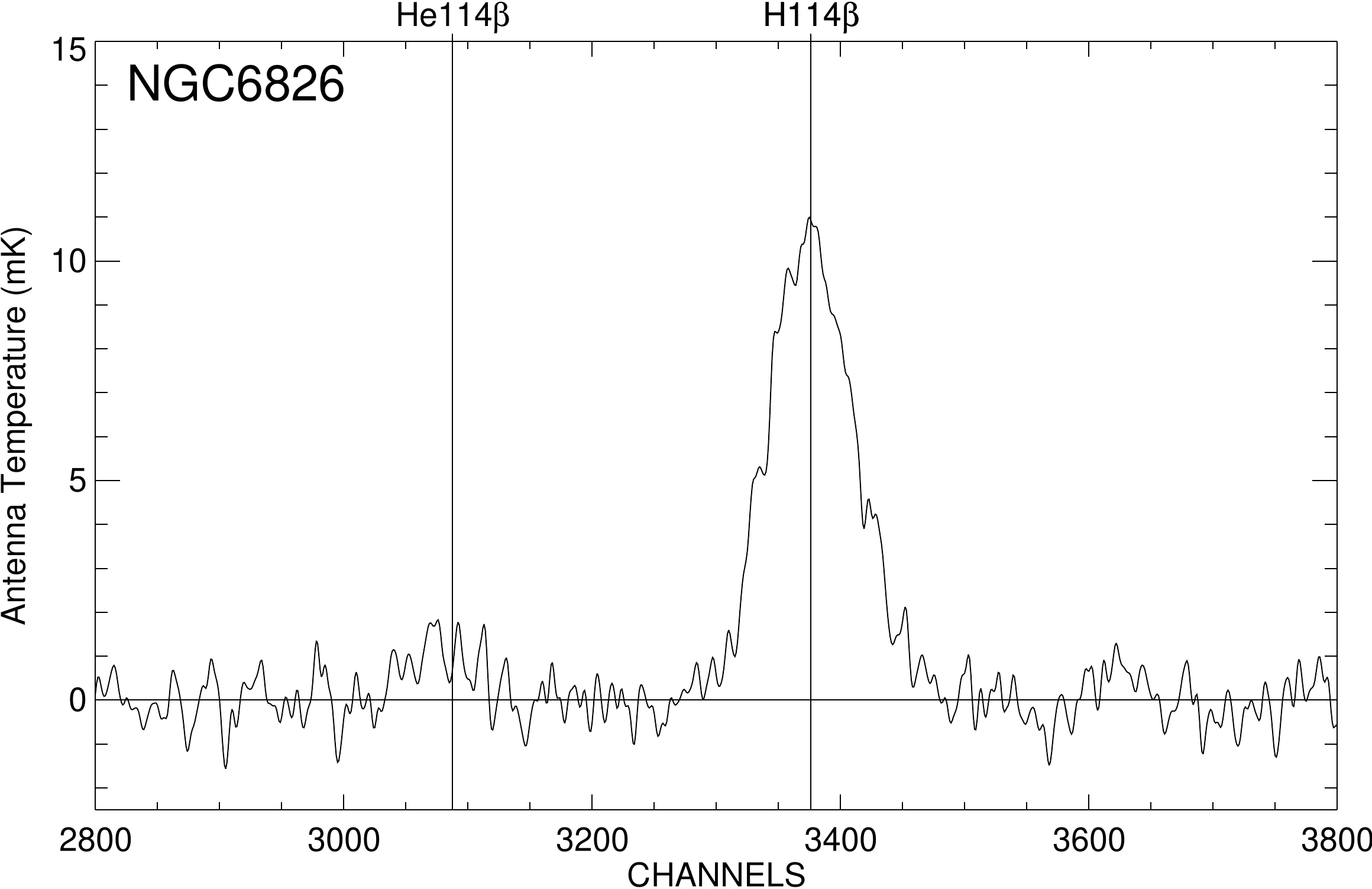}
\includegraphics[angle=-90,scale=0.3]{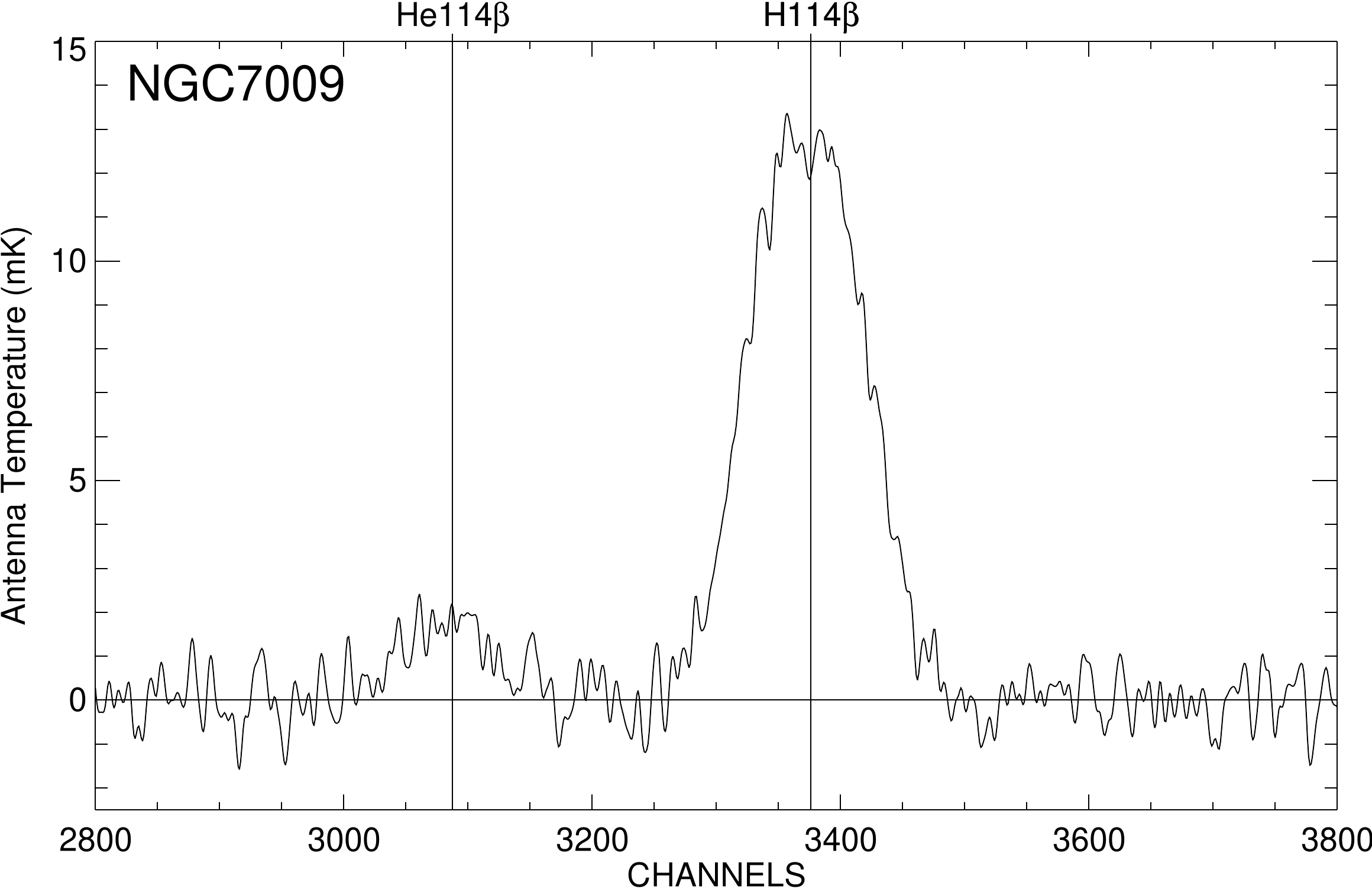}\\
\includegraphics[angle=-90,scale=0.3]{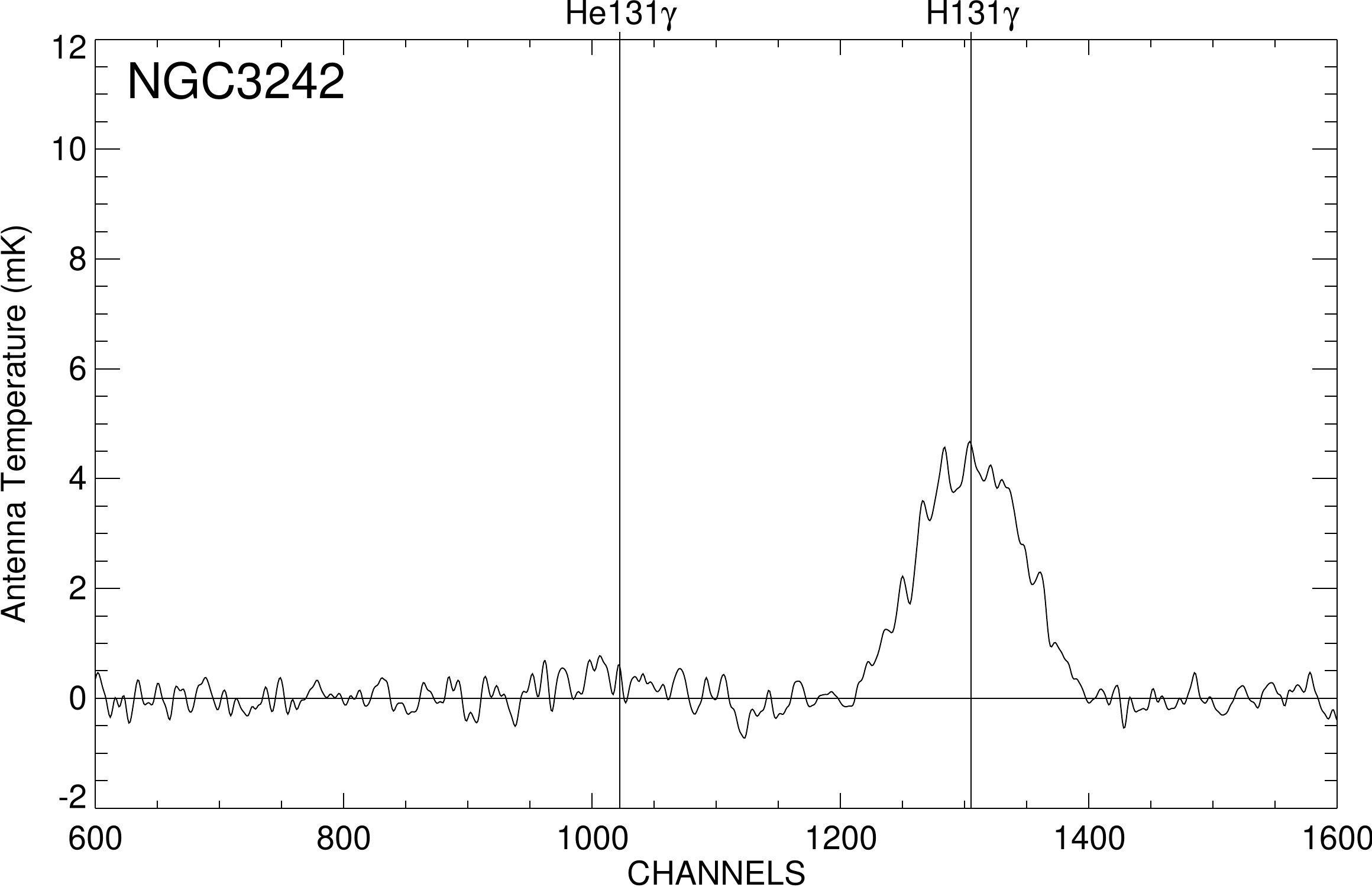}
\includegraphics[angle=-90,scale=0.3]{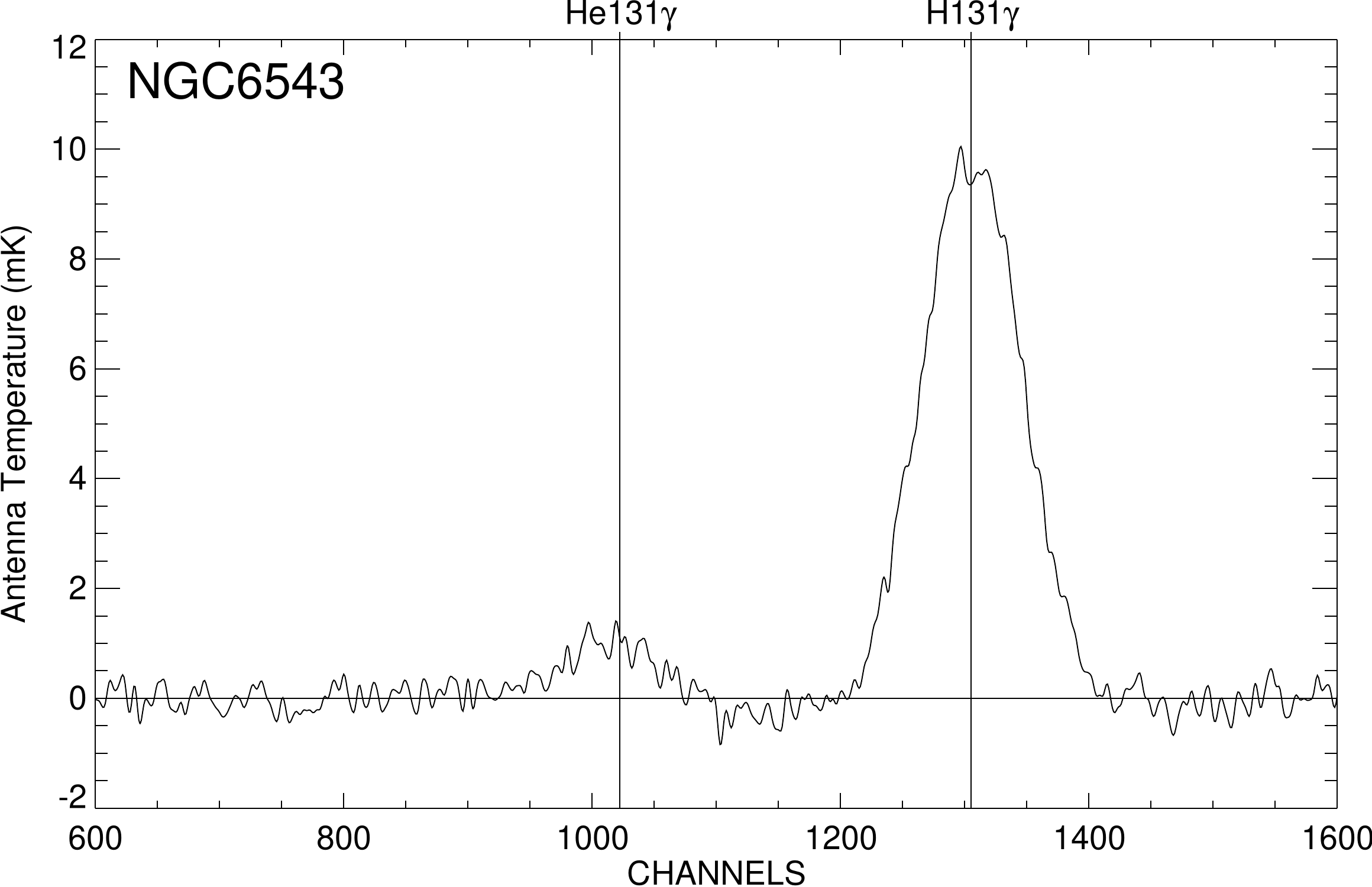}\\
\includegraphics[angle=-90,scale=0.3]{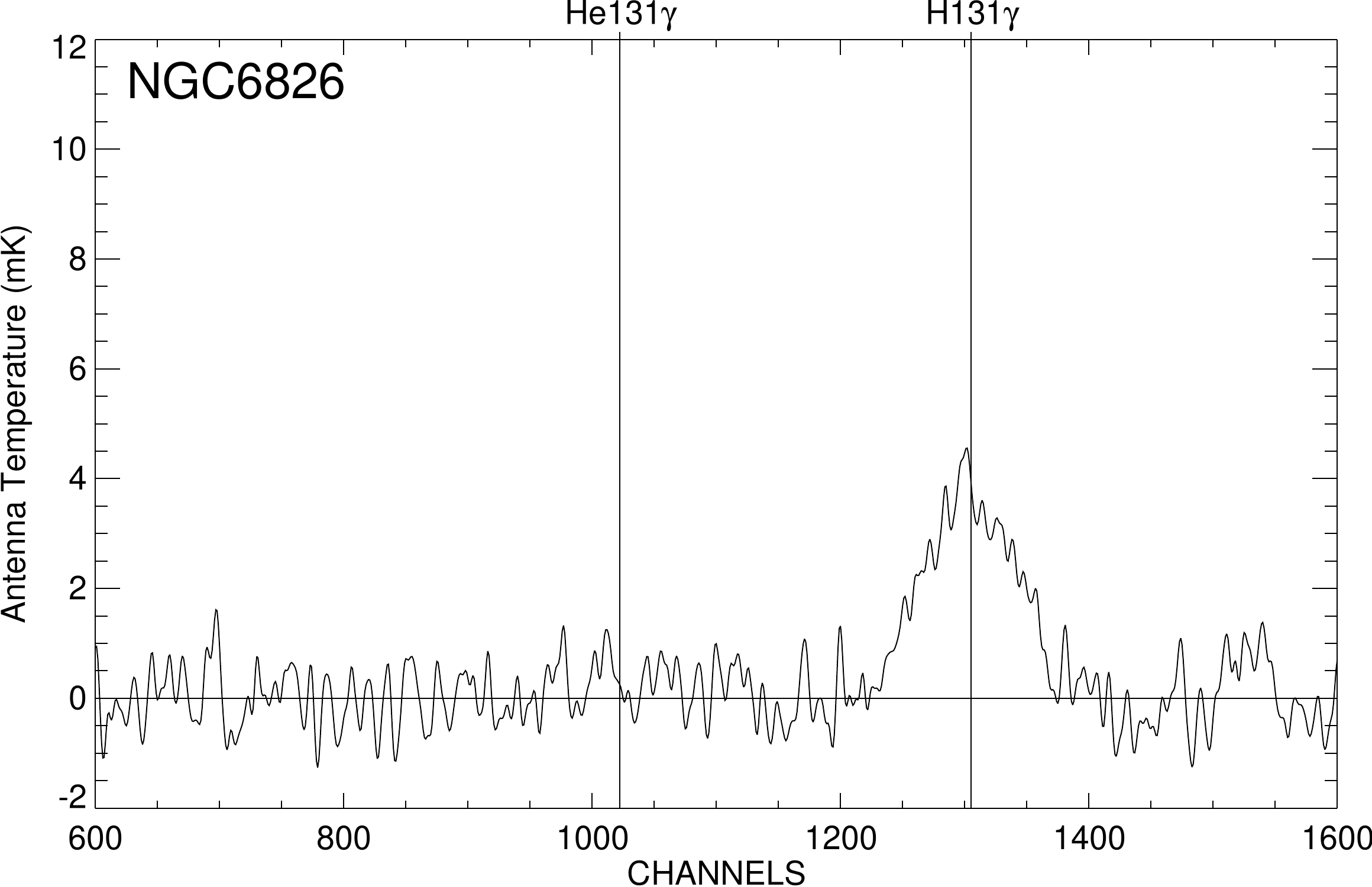}
\includegraphics[angle=-90,scale=0.3]{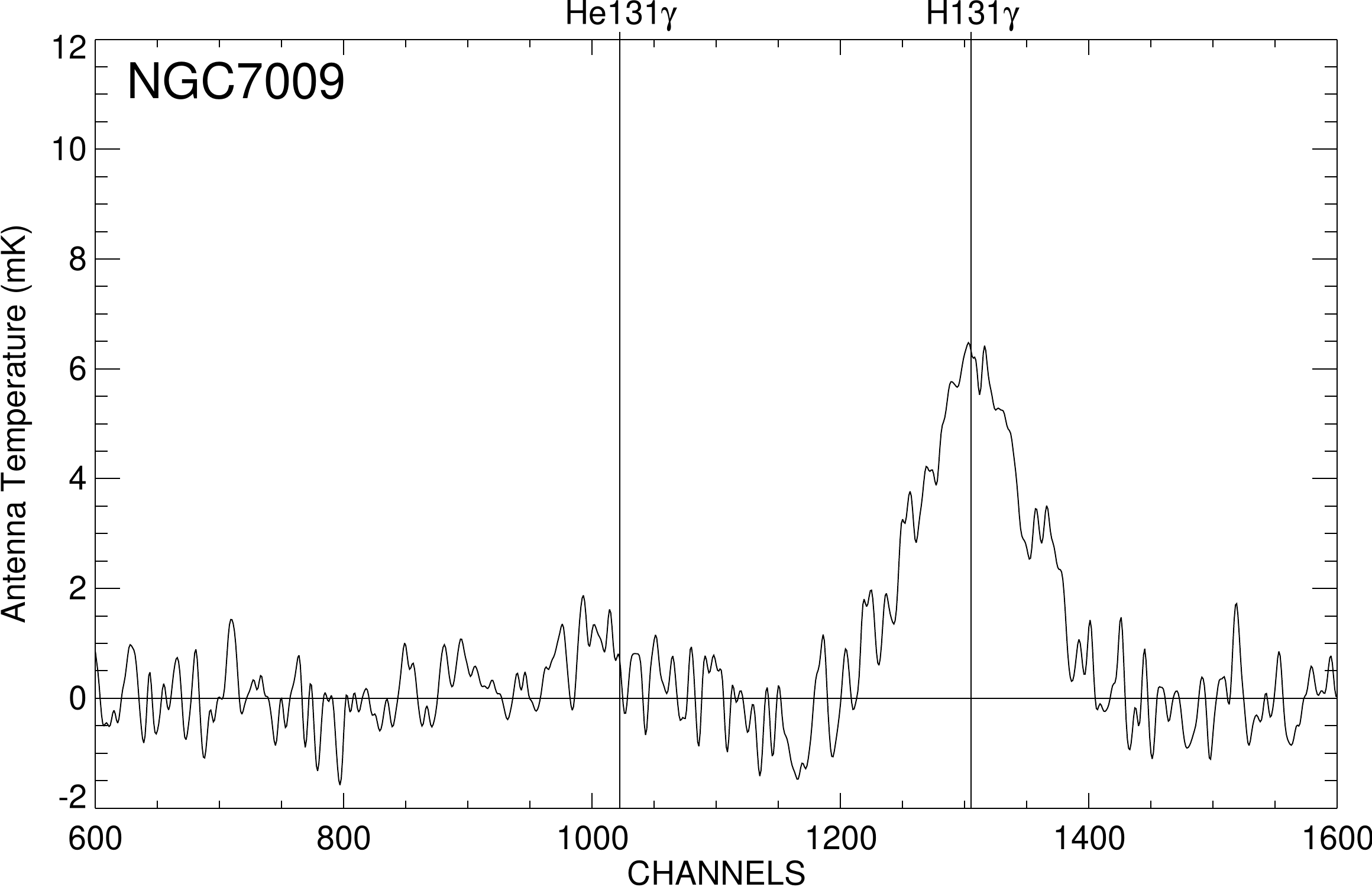}
\caption{
  Observed H114$\beta$ and H131$\gamma$ spectra and for the planetary
  nebula sample.  Each spectrum spans a total bandwidth of 12.5\mhz.
  The spectra have been smoothed to a velocity resolution 5\kms\ and a
  5-th order model for the instrumental baseline has been subtracted.
  For the 114$\beta$ spectra vertical lines flag, from left to right,
  the recombination line transitions: He114$\beta$ and H114$\beta$. For
  the 131$\gamma$ spectra vertical lines flag, from left to right, the
  recombination line transitions: He131$\gamma$ and H131$\gamma$.
}
\label{fig:beta_gamma}
\end{figure}

The H91$\alpha$, H114$\beta$, and H131$\gamma$ spectra analyzed in this
way are shown for our PN sample in Figure~\ref{fig:halpha} and
Figure~\ref{fig:beta_gamma}, respectively. Each
spectrum is a 12.5\mhz\ wide subset of its native 50\mhz\ bandwidth SPW.
The radio recombination line parameters measured for our PN sources are
given in Appendix~\ref{appen:A} for all RRL transitions detected at the
$\gsim$~3$\sigma$ level. Compiled there for each nebula is the RRL
transition, the change in principle quantum number, $\Delta$N, the line
intensity, $T_{\rm L}$ in \mk, and full width at half maximum (FWHM)
line width, $\Delta$V in \kms, together with the RMS noise in
\mk\ and total integration time, $t_{\rm intg}$ in hr, of the spectral
band containing the transition.  Also listed are the 1$\sigma$ errors of
the Gaussian fits to the intensity, $\sigma T_{\rm L}$, and line width,
$\sigma \Delta V$.

\begin{deluxetable}{cc}
\tablecolumns{2} \tablewidth{0pt}
\tablecaption{RRL Transitions\tablenotemark{a}\label{tab:transitions}}
\tablehead{
  \colhead{$\Delta$N} & \colhead{H$^+$ He$^+$ C$^+$} 
}\vspace{-10pt}
\startdata
1 &  91$\alpha$ 92$\alpha$ \\
2 & 114$\beta$ 115$\beta$  \\
3 & 130$\gamma$ 131$\gamma$ 132$\gamma$ \\
4 & 144$\delta$ $145\delta$ \\
5 & 152$\epsilon$ 154$\epsilon$ 155$\epsilon$ 156$\epsilon$ \\
6 & 164$\zeta$ 165$\zeta$ \\
7 & 171$\eta$  \\
8 & 179$\theta$ 180$\theta$ 181$\theta$ \\
9 & 186$\iota$ 187$\iota$ 188$\iota$\\
10 & 190$\kappa$ 193$\kappa$ \\
11 & 198$\lambda$ 199$\lambda$ \\
12 & 201$\mu$ 203$\mu$ 206$\mu$ \\
13 & 206$\nu$ 208$\nu$ 210$\nu$ *238$\nu$ \\
14 & 211$\xi$ 213$\xi$ 215$\xi$ \\
15 & 221$o$ \\
16 & 220$\pi$ 222$\pi$ 224$\pi$ \\
17 & 222$\rho$ 224$\rho$ 227$\rho$ 228$\rho$ 230$\rho$ \\
18 & 228$\sigma$ 231$\sigma$ 232$\sigma$ 234$\sigma$ \\
19 & 232$\tau$ *234$\tau$ 235$\tau$ 236$\tau$ 238$\tau$ \\
20 & 238$\upsilon$ \\
21 & 239$\phi$ 242$\phi$ \\
22 & 245$\chi$ 247$\chi$ 249$\chi$ \\
23 & 249$\psi$ \\
24 & 249$\omega$ \\
\enddata
\tablenotetext{a}{Transitions with N~$>$~250 or $\Delta$N~$>$~25 not listed.
  Not all transitions for all ionic species were detected.}
\tablecomments{All these recombination line transitions lie within the
  spectral windows observed. Transitions with $\Delta$N~$>$~7 are too
  weak to be detected with the sensitivity achieved here.}
\end{deluxetable}

\clearpage

\clearpage

\subsection{\hep3\ Emission}\label{sec:3Heobs}

\begin{figure}[h!]
\centering
\includegraphics[angle=-90,scale=0.35]{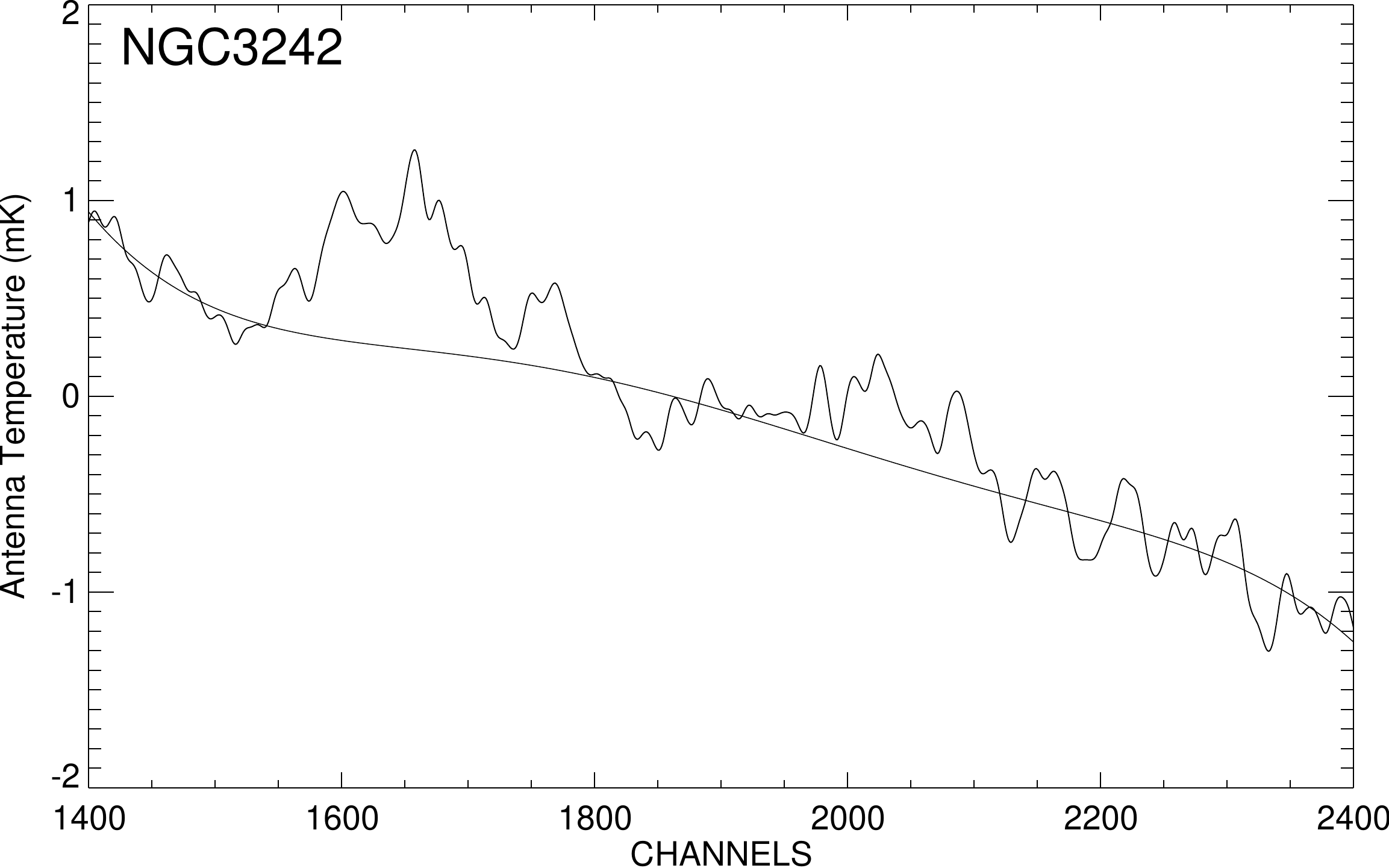}
\includegraphics[angle=-90,scale=0.35]{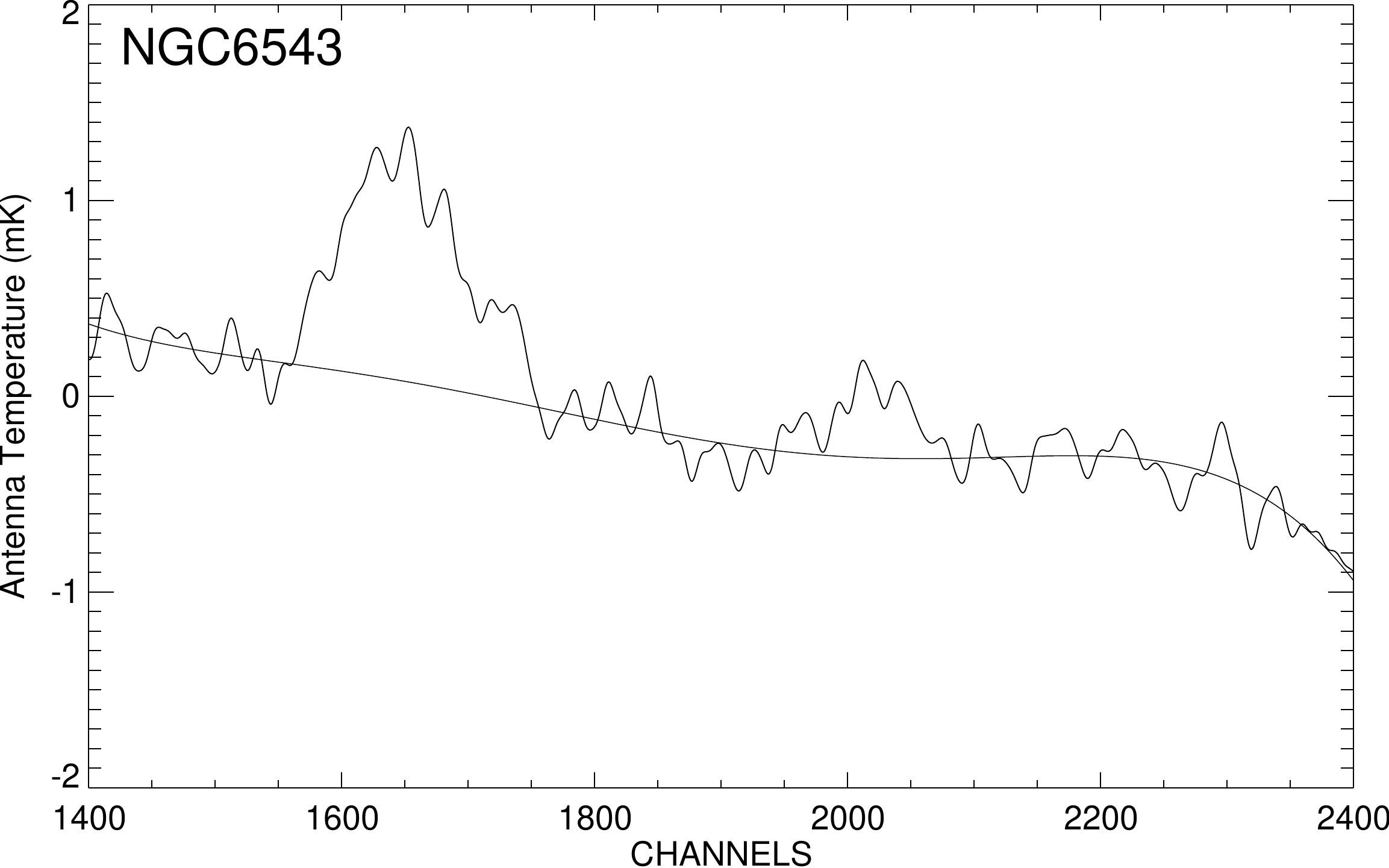}\\
\includegraphics[angle=-90,scale=0.35]{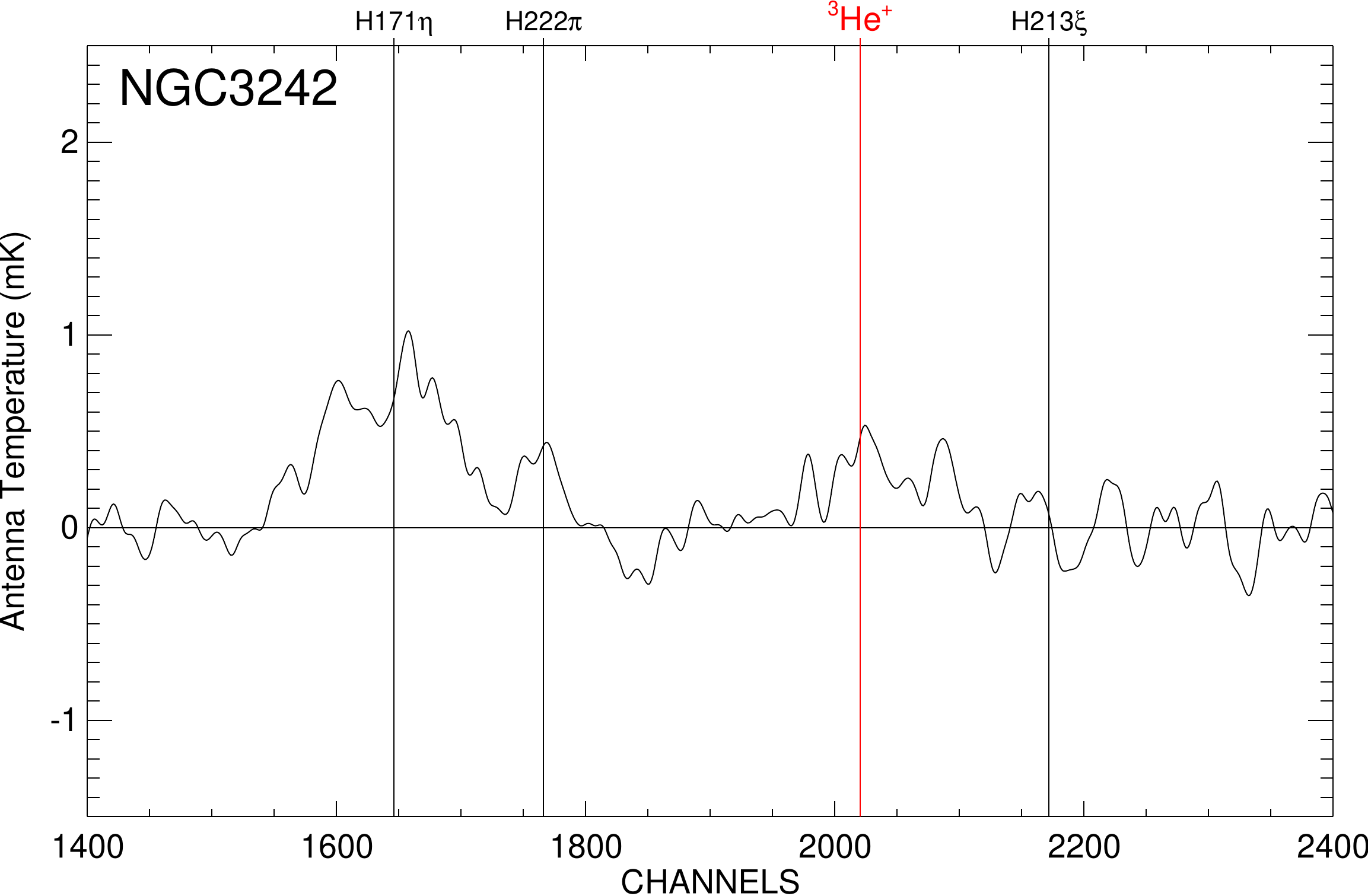}
\includegraphics[angle=-90,scale=0.35]{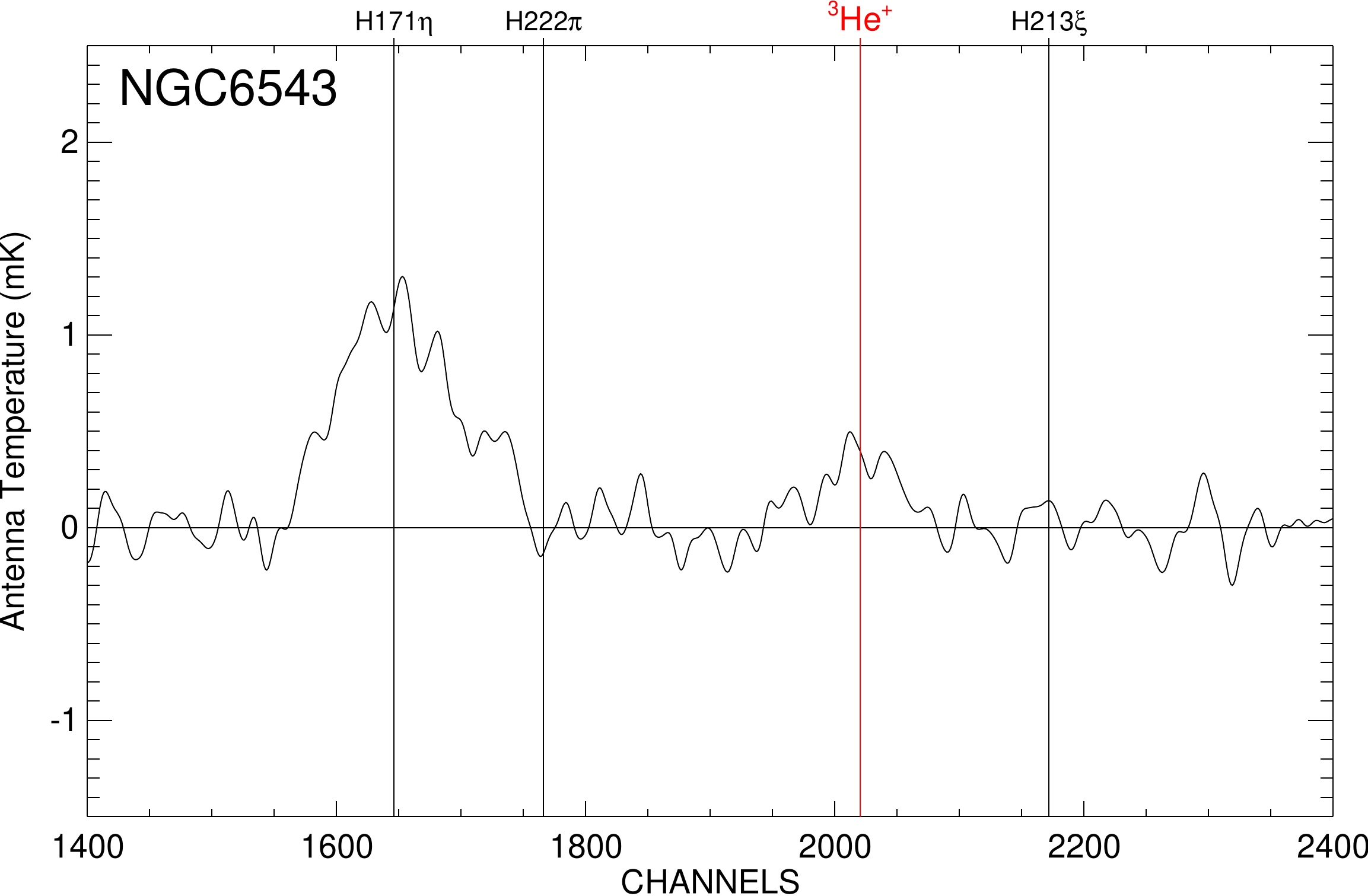}\\
\includegraphics[angle=-90,scale=0.35]{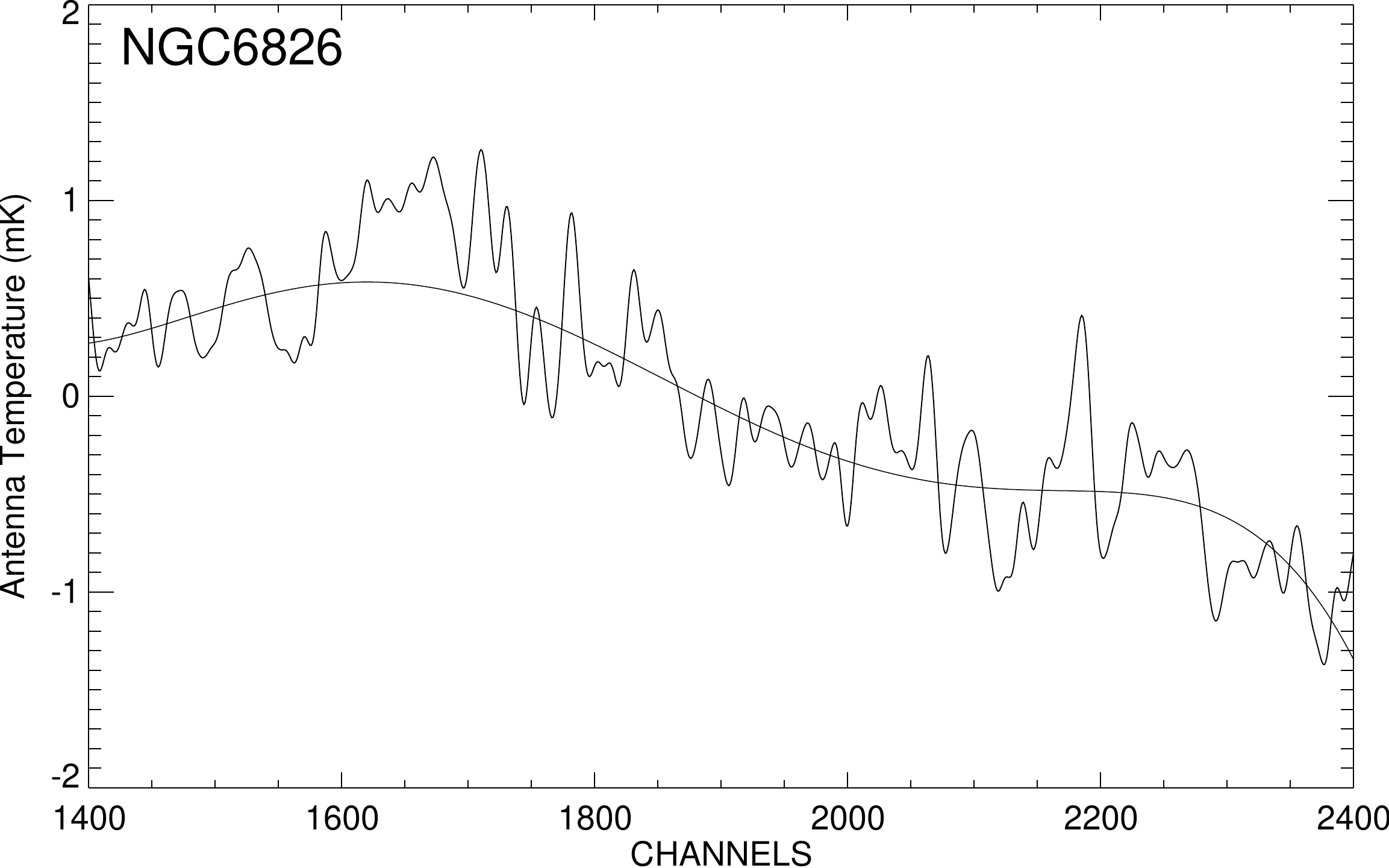}
\includegraphics[angle=-90,scale=0.35]{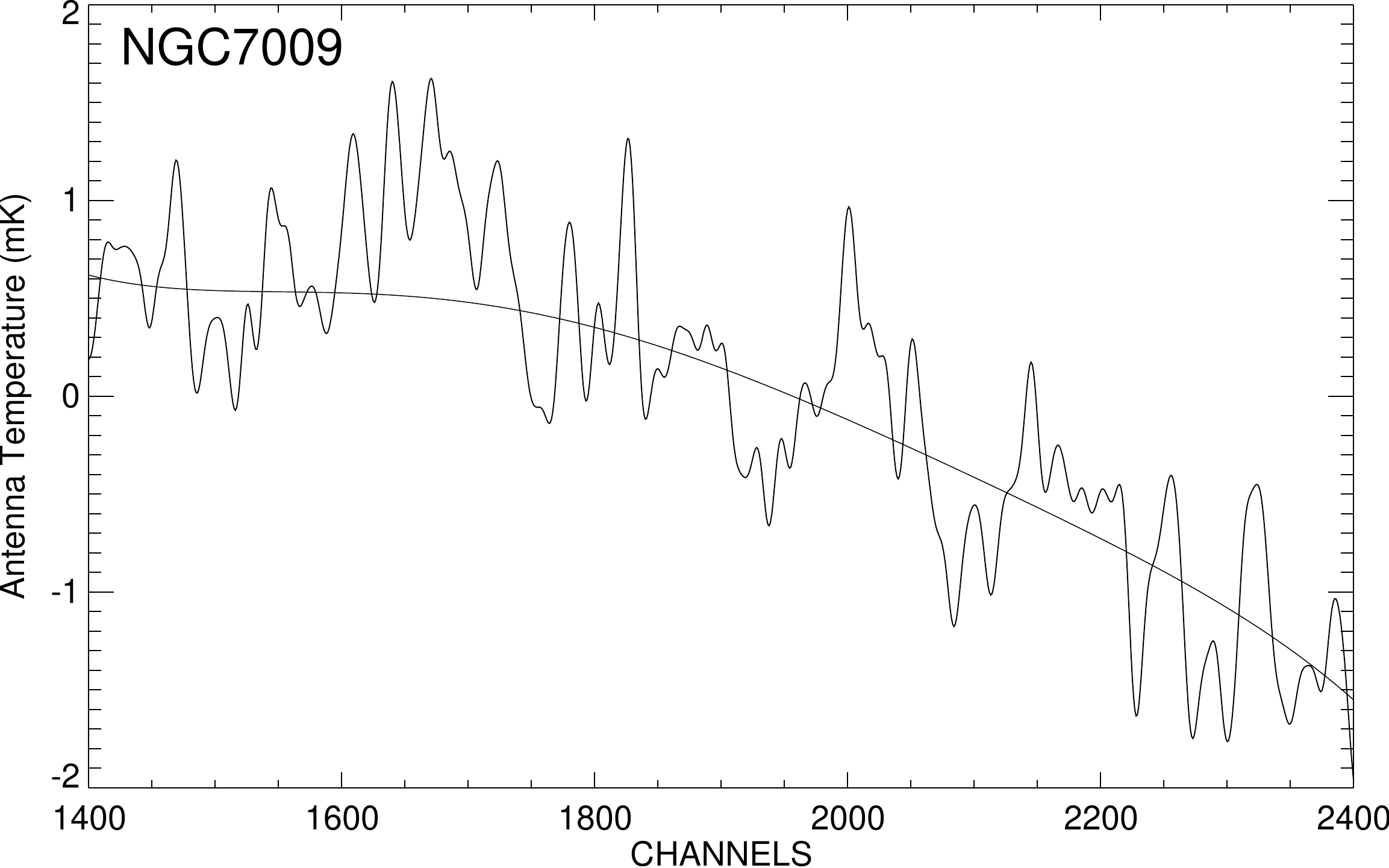}\\
\includegraphics[angle=-90,scale=0.35]{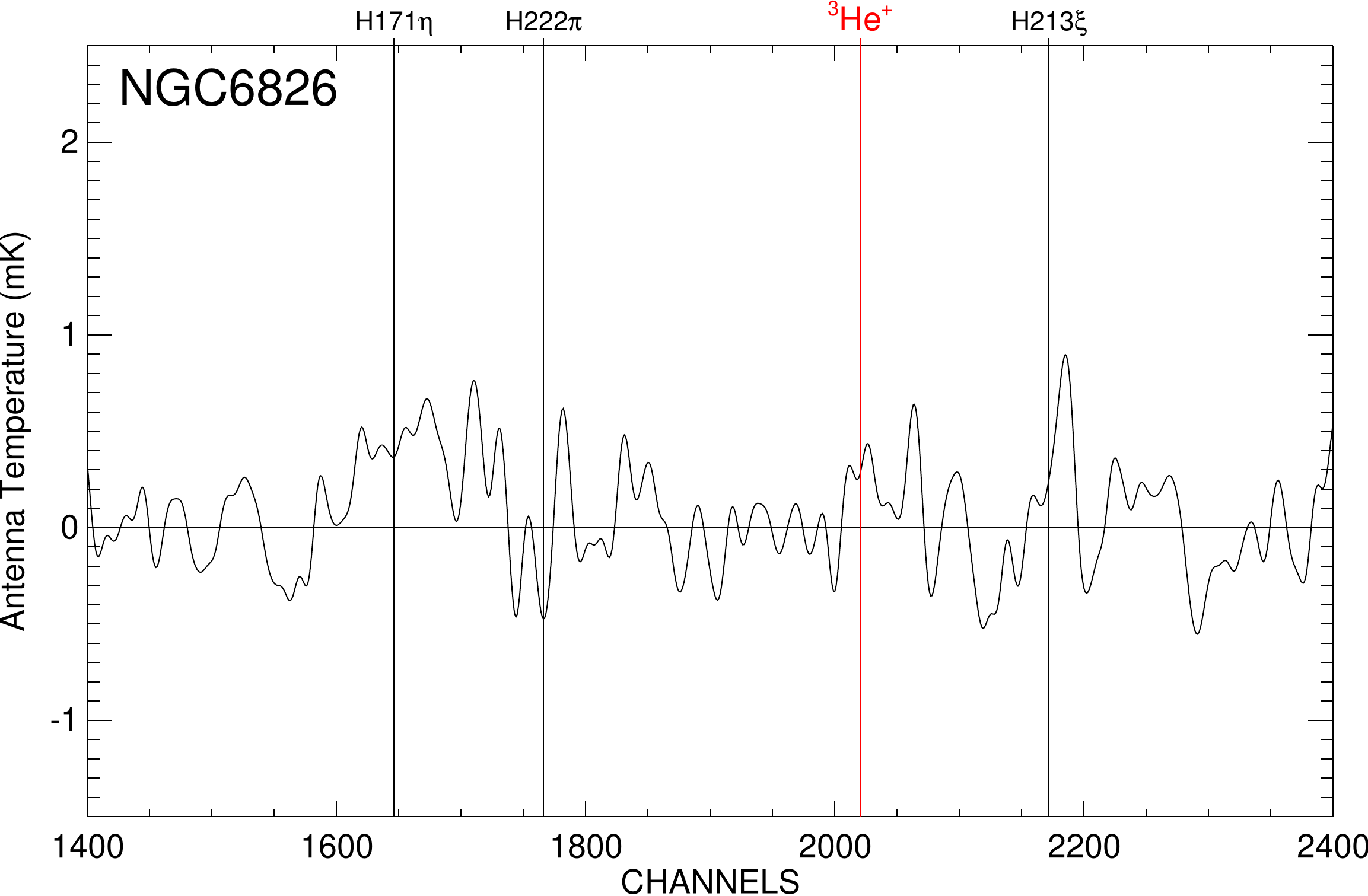}
\includegraphics[angle=-90,scale=0.35]{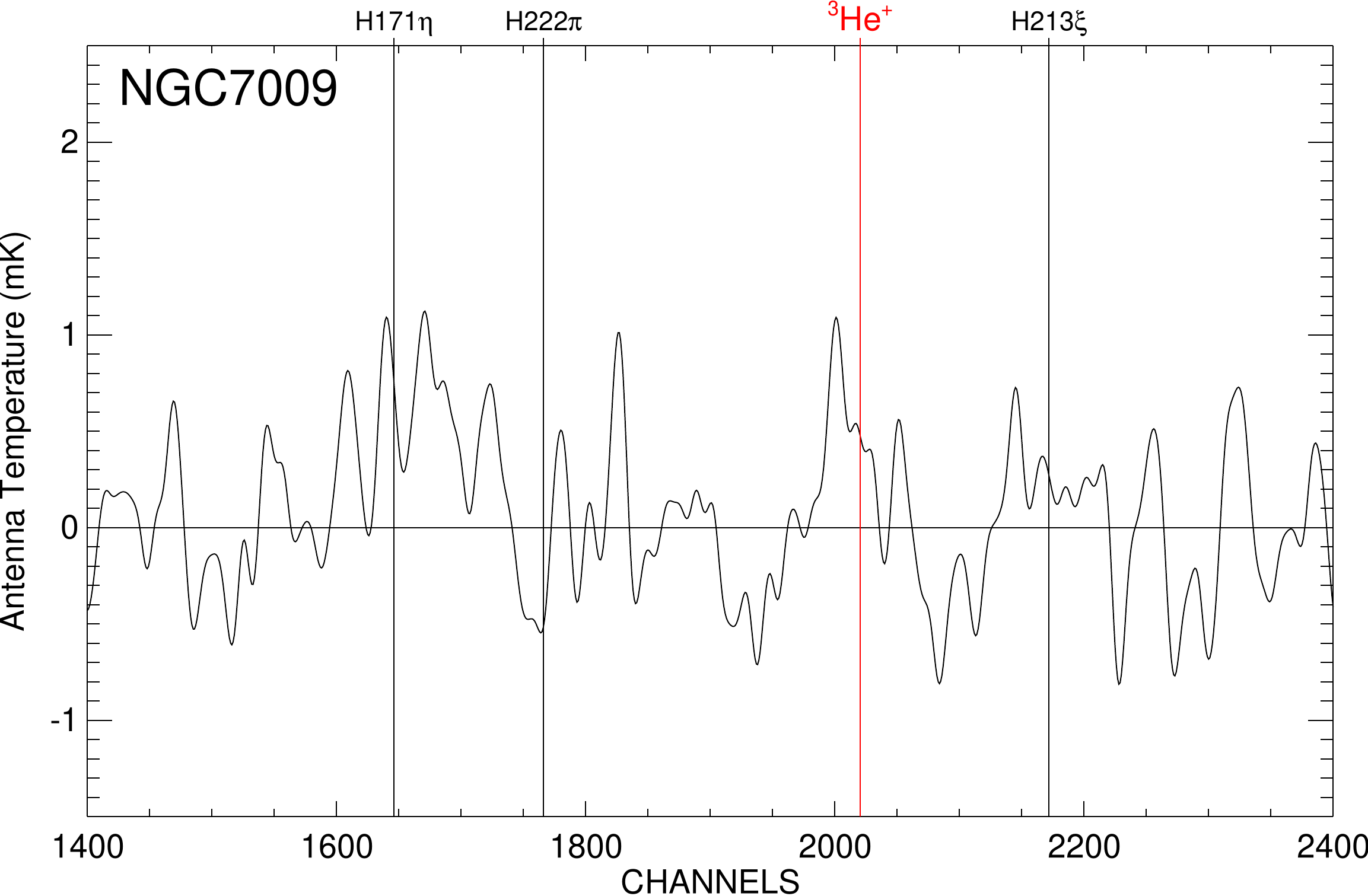}\\
\caption{\hep3\ spectra for the planetary nebula sample.
  {\em Top Panels:} The average spectrum smoothed to a 5\kms\ velocity
  resolution with the 5-th order model fit to the instrumental baseline
  superimposed. 
  {\em Bottom Panels:} The spectrum after the baseline model is subtracted.
  Vertical lines flag, from left to right, the \heta, \hpi, \hep3,
  and \hxi\ transitions. 
}
\label{fig:he3_all}
\end{figure}

Detecting \hep3\ emission from PNe turns out to be very
challenging. These emission lines are weaker and broader than those
produced by \hii\ regions. The \hep3\ spectra for our PN sample are
shown in Figure~\ref{fig:he3_all}.  As for the RRL analysis, these
spectra were smoothed to a 5\kms\ velocity resolution and a 5-th order
polynomial instrumental baseline model was subtracted from the data.
Each PN in Figure~\ref{fig:he3_all} has two spectra shown.  The top
panel displays the baseline model atop the smoothed spectrum and the
bottom panel plots the spectrum after this baseline model is subtracted.

We judge that we have not robustly detected \hep3\ emission from any of
our target PNe. The spectra show that any \hep3\ signal must have an
intensity less than 1\mk. Upper limits for \hep3\ emission from our
sample PNe are listed in Table~\ref{tab:he3}.  For \ngc{3242} and
\ngc{6543} we can fit Gaussian functions to their spectra to provide the
upper limits. These fits give 2.6\,$\sigma$ and 4.3\,$\sigma$
\hep3\ emission features for \ngc{3242} and \ngc{6543}, respectively.
For \ngc{6826} and \ngc{7009} no realistic fits could be made so their
upper limits are 3 times their RMS spectral noise.

Our RRL analysis in Section~\ref{sec:baseline} below indicates that, due
to instrumentally caused frequency structure in the baselines, spectral
line parameters at the $\lsim$~1\mk\ level cannot be measured accurately
with the sensitivity achieved by these GBT observations.  We explored
using different baseline model fitting regions and polynomial orders in
analyzing these \hep3\ spectra.  We can make the \hep3\ feature
stronger, but only at the expense of creating nonphysical \heta\ line
parameters.  The spectra in Figure~\ref{fig:he3_all} all use the same
fitting regions to model the instrumental baselines. These baseline
models, however, do not entirely remove instrumental baseline frequency
structure. That the noise in the \ngc{3242} $\sim$~300\hr\ spectrum is
larger than that in the $\sim$~250\hr\ \ngc{6543} spectrum makes it
clear that such structure still remains in these spectra. This is why we
are only quoting upper limits for \hep3\ emission from these
observations.

\begin{deluxetable}{lrccccr}
\tablecaption{Upper Limits for \hep3\ Emission from Planetary Nebulae\label{tab:he3}}
\tablewidth{0pt} \tablecolumns{7}
\tablehead{
  \colhead{} &
  \colhead{$T_{\rm L}$} &
  \colhead{$\sigma\,T_{\rm L}$} & 
  \colhead{$\Delta{V}$} &
  \colhead{$\sigma\,\Delta{V}$} & 
  \colhead{RMS} &
  \colhead{$t_{\rm intg}$} \\ 
  \colhead{Source} &
  \colhead{(mK)} &
  \colhead{(mK)} & 
  \colhead{(\kms)} &
  \colhead{(\kms)} &
  \colhead{(mK)} &
  \colhead{(hr)}
} 
\startdata 
\ngc{3242}\tablenotemark{a} & $<$0.36    & 0.02  & 48.0  & 2.94  & 0.145 & 301.7 \\
\ngc{6543}\tablenotemark{a} & $<$0.39    & 0.02  & 29.1  & 1.63  & 0.112 & 256.1 \\
\ngc{6826}\tablenotemark{b} & $<$0.77    & \dots & \dots & \dots & 0.257 &  51.4 \\
\ngc{7009}\tablenotemark{b} & $<$1.04    & \dots & \dots & \dots & 0.348 &  44.0 \\
\enddata
\tablenotetext{a}{Upper limit derived from Gaussian fit to
  the \hep3\ spectrum.}
\tablenotetext{b}{Upper limit is 3 times the RMS noise listed 
  here. This noise figure is calculated for the spectral regions used
  to model the instrumental baseline.}
\end{deluxetable}

\clearpage

\clearpage

\begin{figure}[b]
\centering
\includegraphics[angle=-90,scale=0.65]{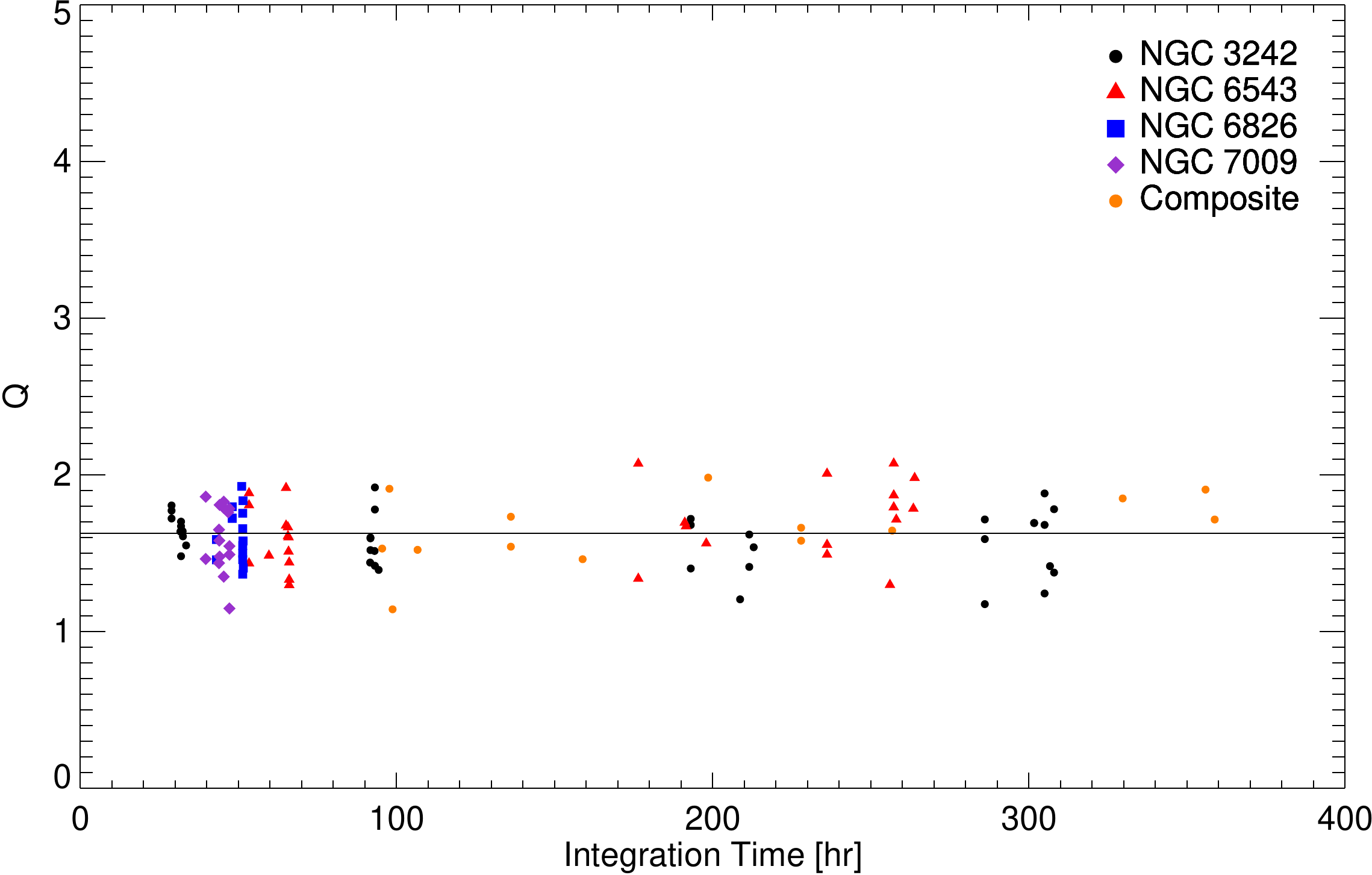} 
\caption{ The GBT/ACS spectrometer system performance for this
  experiment. Shown is the sensitivity metric Q [see text] plotted as a
  function of total integration time for various subsets of the RRL data
  for our PNe sample sources.  ``Composite'' spectra stem from averages
  of two or more PN.  These Q values do not change with integration time
  as they should. The horizontal solid line marks the data average:
  Q\,=\,1.62.  }
\label{fig:qradiom}
\end{figure}

\section{Instrumental Frequency Structure in GBT Spectral Baselines}\label{sec:baseline}

Instrumental effects limit the spectral sensitivity of most single-dish
radio telescopes. These effects cause frequency structure in spectra
that can mimic real emission lines.  For traditionally designed single
dish radio telescopes this frequency structure in spectral baselines is
primarily caused by reflections from the super structure.  The GBT was
specially designed with a clear aperture to significantly reduce
reflections from the secondary structure and therefore improve the image
fidelity and spectral purity. Nevertheless, baseline structure still,
unfortunately, exists and is primarily located within the electronics
\citep{Baselines}\footnote[6]{See
  \url{http://library.nrao.edu/public/memos/edir/EDIR_312.pdf}}.
The GBT X-band (3\cm\ wavelength) receiver is a heterodyne system
wherein radio waves from the sky are mixed with a local oscillator to
convert the signal to an intermediate frequency (IF). Unfortunately, the
GBT IF system transports analog signals over a $\sim$ 1 mile long path
length.  There are a host of electronic components in this path and each
can produce reflections and, therefore, standing waves. The spectral
baselines are significantly better for the GBT than for traditionally
designed, on-axis telescopes but they are still the limiting factor in
measuring accurate line parameters for weak, broad spectral lines.

For observing epochs starting in 2011 and later, we used the strategy
described by \citet{BB2018} to mitigate instrumental frequency structure
in the GBT IF system.  This technique simultaneously tunes four SPWs to
the \hep3\ transition with center frequencies offset by 0, 3, 6, and
9\mhz\ during data acquisition. The idea is to move the
\hep3\ transition around in IF frequency space and that when the four
SPWs are subsequently aligned and averaged, the result smooths out
instrumental frequency structure in the IF. \citet{BB2018} evaluated the
efficacy of this technique.  These baseline structure tests are germane
here because measurements of our planetary nebula and \hii\ region
targets were interleaved during all observing epochs.

An important test was to observe two extra-galactic calibration sources,
the quasars \threec{286} and \threec{84}, which have 9\ghz\ continuum
antenna temperatures of 10\K\ and 50\K, respectively.  The advantage of
using these extra-galactic sources is that they do not have any
measurable spectral lines at these frequencies. The observing procedures
and ACS configuration were the same as the observations for the
\hii\ region and PN targets.

These bright calibrators were chosen to amplify any instrumental
spectral artifacts.  The instrumental baseline structure for these
quasar observations is therefore expected to be much worse than for our
PNe target sources which have continuum antenna temperatures
$\lsim$\,1.5\K\ (see Table~\ref{tab:continuum}). \citet{BB2018} show
that the spectral baselines produced by this frequency offset observing
technique are clearly flatter.  Moreover, the amplitude of the remaining
baseline features roughly scales with continuum intensity most likely
due to the power mismatch between the On and Off-source observations.
Averaging over the four frequency offset \hep3\ SPWs does not reduce the
random thermal noise because the signals are correlated, but the
instrumental systematic noise is reduced.  The improvement is, however,
not dramatic. Careful scrutiny of early epoch data acquired without
using this technique gave us no reason not to include these data in the
final averages.

A major goal here is to assess the instrumental baseline frequency
structure that remains in our PNe data. These systematic effects, and
not random Gaussian RMS noise, limit the spectral sensitivity that we
can achieve with the GBT/ACS spectrometer system. We thus distinguish
between thermal radiometer noise, ``RN'', and systematic instrumental
baseline noise, ``BN''.  A radio spectrometer's RMS thermal noise should
obey the radiometer equation; the RMS of this RN should diminish as the
square root of the integration time: RMS $\propto {\rm t_{\rm
    intg}^{-1/2}}$. Larger RMS values for the same integration time are
manifestations of instrumental baseline frequency structure --- the BN.
As we have done for all the \hep3\ experiment telescopes, we use the
plethora of RRL measurements to assess the performance of the GBT/ACS
spectrometer system.

For example, the RRL line intensities should not vary with time, the
\hepr4\ abundance ratio should not be a function of principle quantum
number, and the lines should only be seen in emission.  RRL theory
predicts LTE line intensities and these can be used to assess
spectrometer performance. For any given nebula the entire ensemble of
RRLs should give an astrophysically self-consistent set of measured line
parameters.  Any deviations or inconsistencies are caused by
instrumental effects which, experience shows, inevitably appear at some
point as one attains ever increasing sensitivity levels. At the
sensitivity required by the \he3\ experiment, characterizing the
spectrometer's performance is an on-going, evolving process.

A spectrometer that is behaving properly only has RN and so will obey
the radiometer equation.  Since the ACS spectral bandwidth and sampling
mode are identical for all our observations, the radiometer equation
demands that the RMS noise in a spectrum should only depend on the
system temperature, \tsys, and the integration time, $t_{\rm intg}$:
${\rm RMS} \propto \tsys / \sqrt{t_{\rm intg}}$.  We therefore adopt the
sensitivity metric, Q, used by \citet{1994Balser} to evaluate the
performance of radio spectrometers where ${\rm Q} = (\sqrt{t_{\rm intg}}
\times {\rm RMS}) / \tsys$. Here, the RMS is that measured for each
spectrum.

A perfect radiometer would have a Q of unity, but the ACS is an
autocorrelator and its sensitivity is less than this ideal
performance. For our observations the ACS uses 9-level sampling and
uniform channel weighting. Thus, if the GBT/ACS spectrometer only has
RN, it should have $Q_{\rm ACS}$\,=\,1.63 at 8665\ghz\ for a spectral
resolution of 5\kms. Furthermore, the spectral RMS should integrate down
with increased observing time, $t_{\rm intg}$, and the Q value ought to
remain constant at $Q_{\rm ACS}$.
If there is, however, systematic instrumental baseline structure, BN,
then the spectrometer's Q value will be larger than $Q_{\rm ACS}$.  BN
will also manifest itself in the dispersion of Q values for a given
integration time.  Finally, although RN will integrate down with
increased observing time, BN will always be present so the spectral Q
will always remain above $Q_{\rm ACS}$ and the dispersion of Q values
will persist no matter how large $t_{\rm intg}$ gets.

The Q factors attained by our GBT observations are plotted in
Figure~\ref{fig:qradiom} as a function of $t_{\rm intg}$.  Spectra for
all our PNe targets and SPW tunings are contained in this plot.
Included are not only all the spectra whose RRL properties are listed in
Appendix~\ref{appen:A}, but also temporal subsets of the measurements
leading to these spectra together with composite spectra that stem from
averages of two or more PNe for the same RRL transitions. These latter
data are needed to provide a more complete sampling of $t_{\rm intg}$.

For the spectral integration times achieved by our \hep3\ observations,
Figure~\ref{fig:qradiom} shows that the GBT/ACS spectrometer's
performance --- characterized by its spectral Q factors --- is
independent of integration time.  A linear fit to these data yields an
essentially flat relationship between Q factor and integration time; the
slope across Figure~\ref{fig:qradiom} is 0.00020\,$\pm$\,0.00019. The
average Q factor is $<$Q$>$\,=\,1.62\,$\pm$\,0.20 ($\pm$12.3\%) which is
nearly identical to the linear fit's intercept of 1.59\,$\pm$\,0.03.
Thus Figure~\ref{fig:qradiom} demonstrates that the GBT/ACS
spectrometer's random thermal RMS noise --- its RN --- is integrating
down as expected from the radiometer equation.

This average Q factor is, remarkably, essentially identical to the
$Q_{\rm ACS}$\,=\,1.63 value expected for the ACS from pure radiometer
noise.  Nonetheless the systematic instrumental BN frequency structure
is manifested by the $\sim$\,12\% dispersion in the Q factor present for
any $t_{\rm intg}$ value.  The spectra used to craft
Figure~\ref{fig:qradiom} were all smoothed to a common velocity
resolution of 5\kms. These spectra, however, come from SPWs that sample
different places in IF space.  Each SPW is tuned to a different center
frequency. Because of this a 5\kms\ velocity resolution results in
slightly different frequency resolutions for the different SPWs and that
will produce some variation in Q for a fixed \tsys\ and $t_{\rm
  intg}$. Our ACS SPW tuning frequencies span the range from
8280\mhz\ to 8918\mhz. This produces an expected $\sim$\,4\% variation
in Q from different SPW tunings assuming all else is the same.  The
measured variation of the average Q for our experiment is $\sim$\,3
times larger than this.  This is due to BN: SPWs that stem from
different places in IF space show different instrumental frequency
structure.

\begin{figure}[h]
\centering
\includegraphics[angle=-90,scale=0.40]{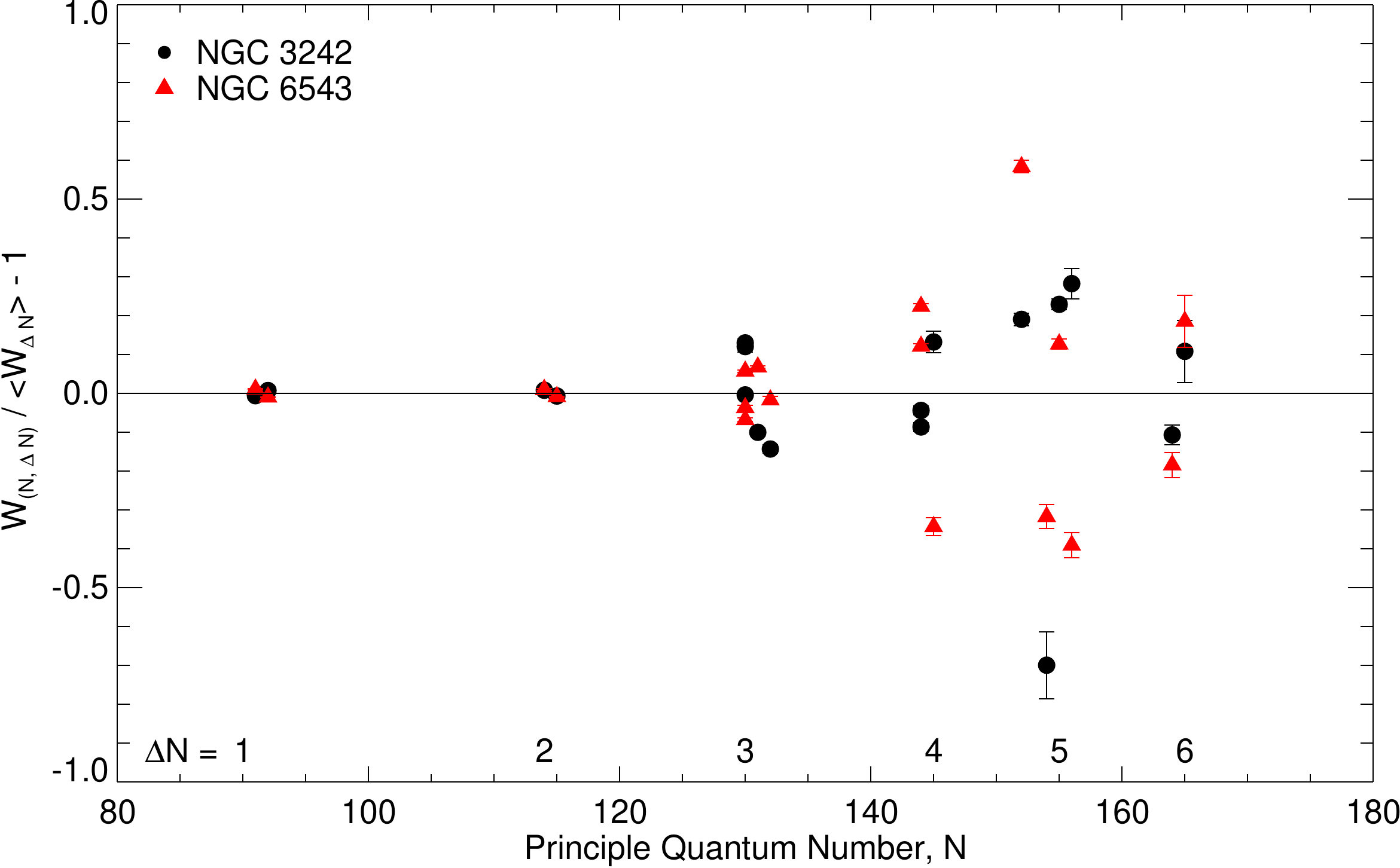}
\includegraphics[angle=-90,scale=0.40]{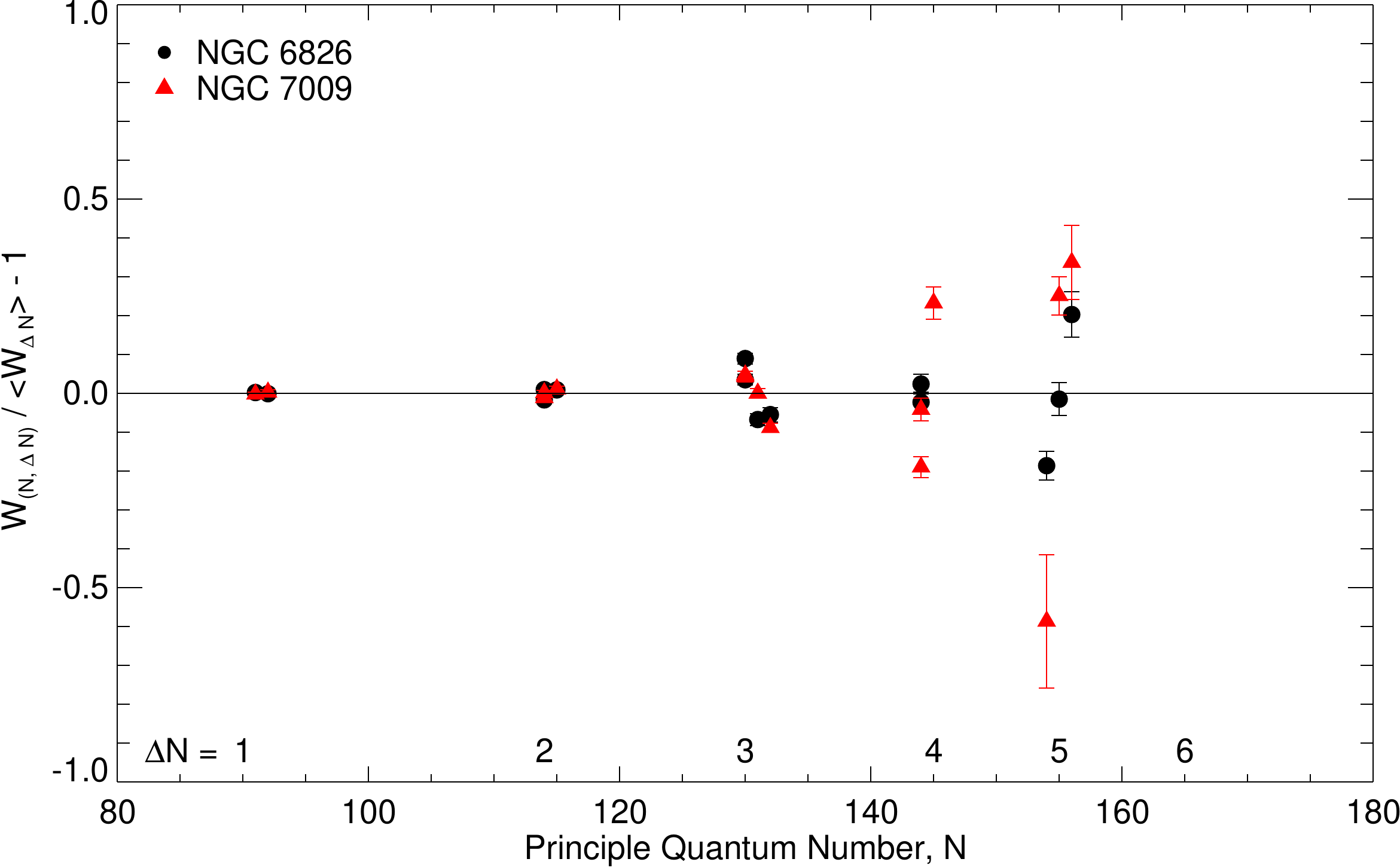}\\
\caption{
  Comparison of measured RRL parameters for transitions with either the
  same (N,\,$\Delta$N) or the same $\Delta$N with different, but
  adjacent N values. The ratio of the integrated intensity of a RRL
  transition, W(N,\,$\Delta$N) [\mk\kms], and $<{\rm W_{\rm \Delta
      N}}>$, the average integrated intensity of all observed
  transitions with the same $\Delta$N, is plotted fractionally,
  (W(N,\,$\Delta$N)/$<{\rm W_{\rm \Delta N}}>$) $-$ 1, as a function of
  the transition principal quantum number, N. The $\Delta$N of the
  transitions is also shown above the x-axis. At these high-N levels
  transitions with nearby N values emit, to first order, spectral lines
  with the same intensities.
}
\label{fig:bntest}
\end{figure}

Comparing measurements that ought to be identical provides another test
of radiometer performance. Here we do this in two ways by using
measurements of: (1) identical RRL transitions, and (2) RRL transitions
with the same $\Delta$N but with different, but adjacent principal
quantum numbers, N.  The first type of measurements are RRL transitions
with the same (N,\,$\Delta$N) observed simultaneously but in different
SPWs. These spectra sample different regions of IF and ACS baseband
space and are thus subject to different BN. These transitions are
labeled H114a/H114b ($\beta$ lines; $\Delta$N=2), H130a/H130b ($\gamma$
lines; $\Delta$N=3), and H144b/H144g ($\delta$ lines; $\Delta$N=4) in
Appendix~\ref{appen:A}. The second class of test RRLs exploits the fact
that at high N-levels the energy differences for $\Delta$N transitions
occurring from nearby N levels are nearly identical provided that the N
range is not too great. Thus the RRL intensities for adjacent N
transitions with the same $\Delta$N, e.g. \hbeta/\hbbeta, ought to be
close to unity.

This comparison is made in Figure~\ref{fig:bntest} using the RRL
measurements listed in Appendix~\ref{appen:A}. It plots the fractional
ratio, (W(N,\,$\Delta$N)/$<{\rm W_{\rm \Delta N}}>$) $-$ 1, as a
function of the transition principal quantum number, N, where
W(N,\,$\Delta$N) is the integrated intensity of a RRL transition in
\mk\kms\ and $<{\rm W_{\rm \Delta N}}>$ is the average integrated
intensity of all observed transitions with the same $\Delta$N. We use
the integrated intensity, W = $ {\rm T_{\rm pk}} \times \Delta {\rm
  V_{\rm FWHM}}$, to better characterize the RRL measurement.  Above the
Figure~\ref{fig:bntest} x-axis we also plot the $\Delta$N values of the
RRL transitions. As is clear from this figure our RRL plots with either
N or $\Delta$N x-axes are equivalent since N increases monotonically
from $\Delta$N=1, which for our data has N=91, to $\Delta$N=7 where
N=171 (see Table~\ref{tab:transitions}). When interpreting the
information presented in Figure~\ref{fig:bntest}, it is important to
keep in mind that the top panel PNe sources, \ngc{3242} and \ngc{6543},
have integration times between $\sim$\,250 and $\sim$\,300 hrs whereas
the bottom panel targets, \ngc{6826} and \ngc{7009}, have much smaller,
$\sim$\,50 hr, integration times.

The Figure~\ref{fig:bntest} RRL comparisons show that for the $\alpha$
($\Delta$N=1) and $\beta$ ($\Delta$N=2) transitions our integrated
intensity measurements agree to within $\lsim$1\% for all our PNe
targets.  These RRLs have the largest intensities; the weaker $\beta$
line intensities span $\sim$10\mk\ to $\sim$25\mk\ for our sample PNe.
For $\gamma$ ($\Delta$N=3) transitions, the comparison shows poorer
agreement and higher order $\Delta$N transitions evince a large
dispersion in the measured integrated intensities.  For $\Delta$N$>$3
transitions, the W ratio discrepancies exceed our $\sim$10\% intensity
scale calibration uncertainty, reaching differences as large as
$\sim$30\%.  The magnitude of these discrepancies are a manifestation of
BN at the $\sim$\,2\mk\ level.

\begin{figure}[h]
\centering
\includegraphics[angle=-90,scale=0.35]{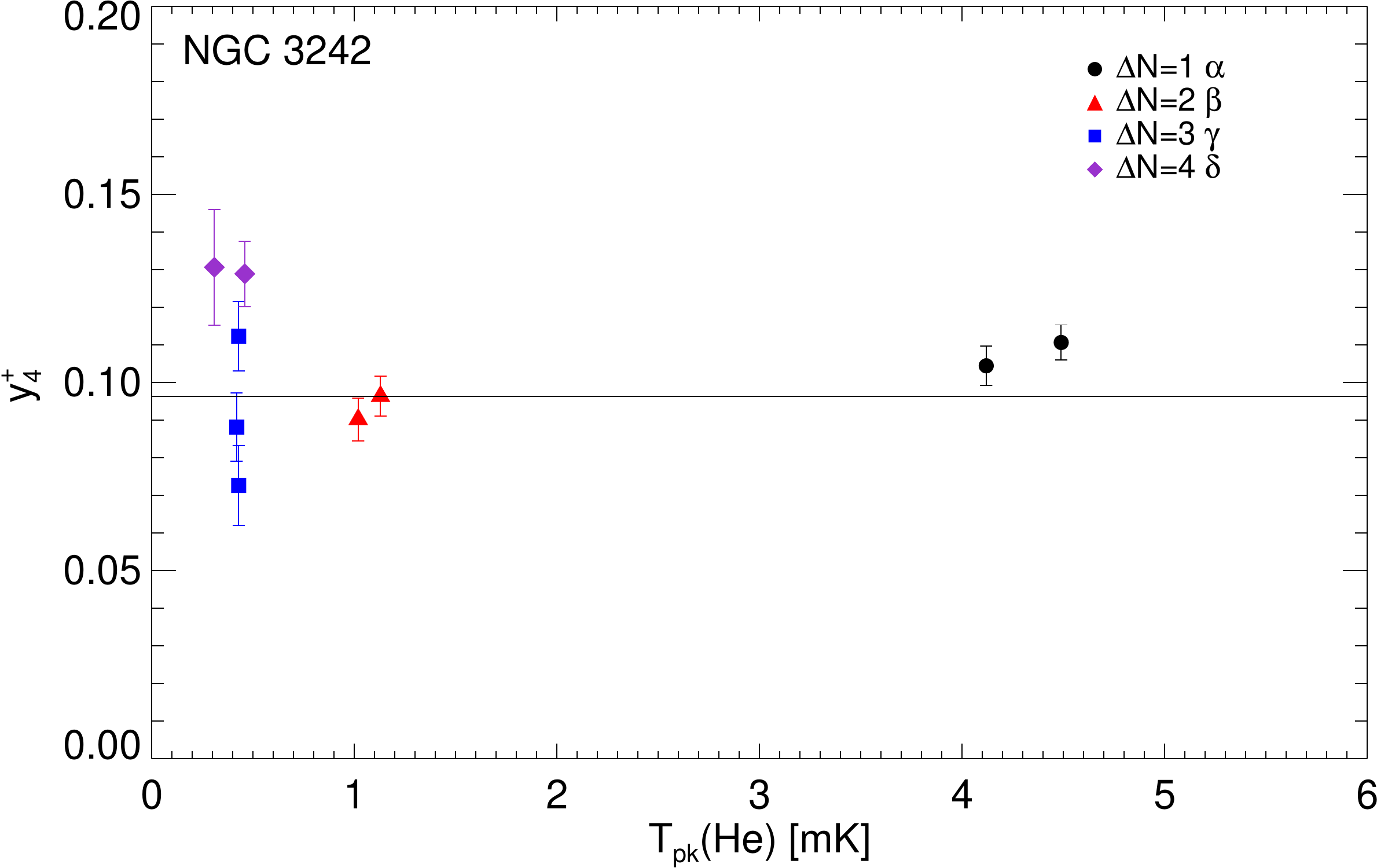}
\includegraphics[angle=-90,scale=0.35]{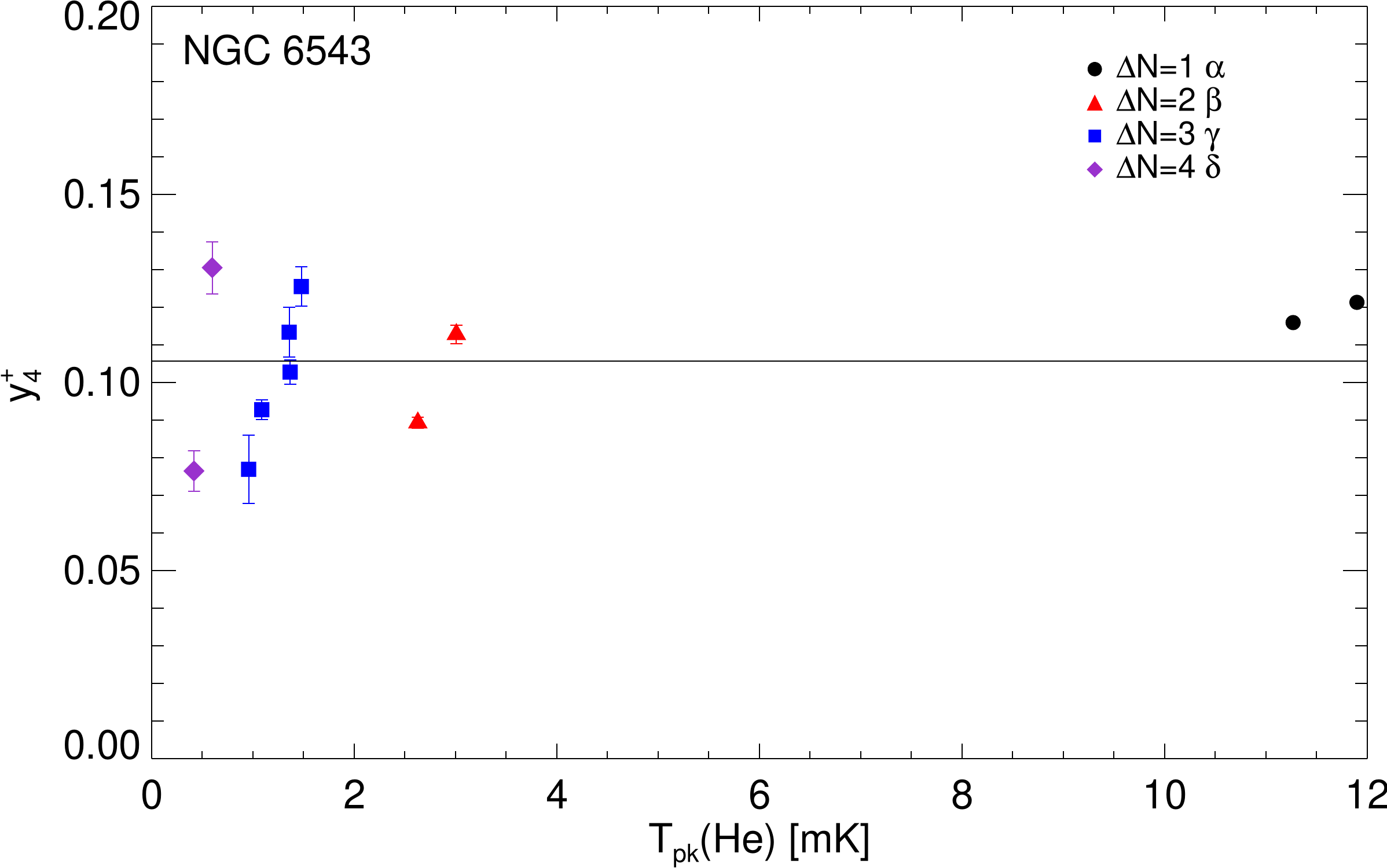}\\
\includegraphics[angle=-90,scale=0.35]{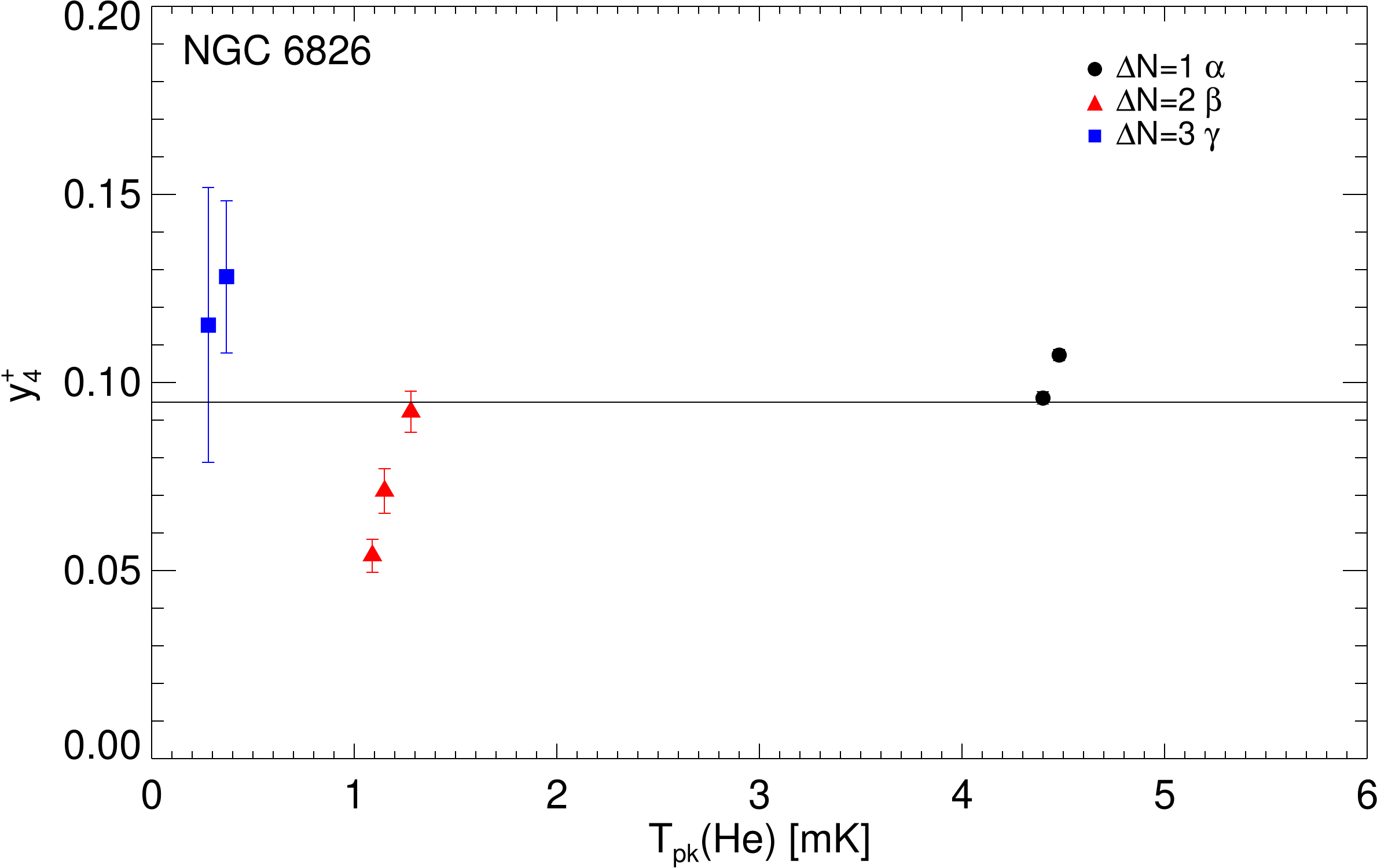}
\includegraphics[angle=-90,scale=0.35]{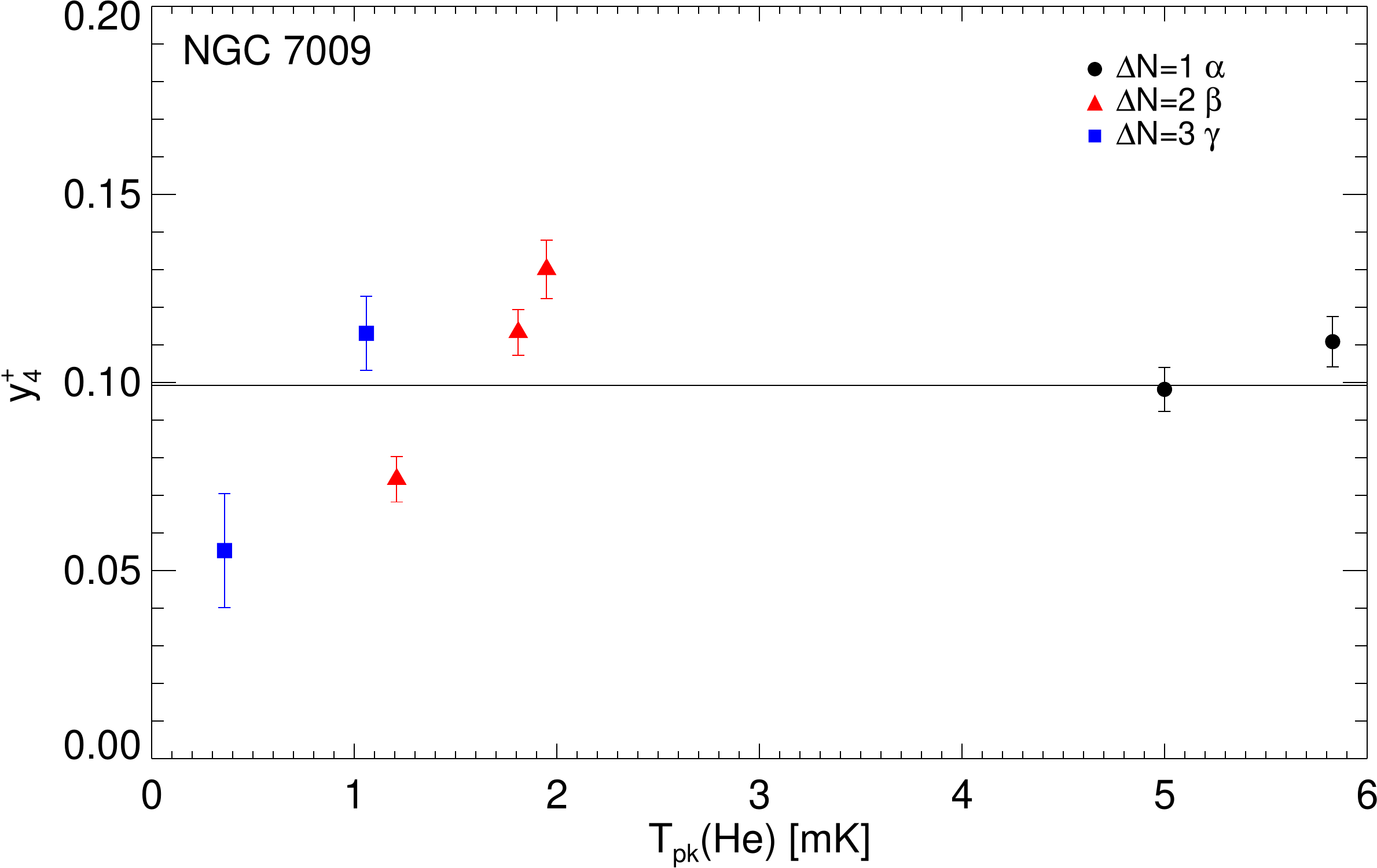} 
\caption{
  The \hepr4\ ratio, \yp4, derived for our sample PNe plotted as a
  function of the He RRL intensity, ${\rm T_{\rm pk}}$(He).  Horizontal
  lines show the weighted average of the $\alpha$, $\beta$, and $\gamma$
  transitions. Error bars stem from the $\pm~1~\sigma$ errors of
  Gaussian fits to the observed RRLs.
}
\label{fig:yptpk}
\end{figure}

Measurements of singly ionized helium RRLs can also be used to assess
the performance of the GBT/ACS spectrometer. Helium RRLs are typically
$\sim$\,10\% the intensity of hydrogen RRLs from Galactic \hii\ regions
and PNe.  Here we use our He and H RRL measurements to calculate the
\hepr4\ ratio, \yp4. For a more accurate determination of \yp4\ we use
the integrated intensity of the RRL transitions:

\begin{equation}
 \yp4 ~=~ { {T_{pk}{\rm (He)} \times \Delta V {\rm (He)}} \over
   {T_{pk}{\rm (H)} \times \Delta V {\rm (H)}} }
\label{eq:yp4def}
\end{equation}

\noindent
where the line peaks, $T_{\rm pk}$, and FWHM line widths, $\Delta\,V$, 
are the values from the Gaussian fits to the observed spectra that are
listed in Appendix~\ref{appen:A}.  Figure~\ref{fig:yptpk} plots these
\yp4\ ratios for our PN sample as a function of the He RRL peak
intensity, $T_{\rm pk}$(He). Because of the intrinsic weakness of
the He emission, \yp4\ can only be derived from our GBT observations for
$\Delta$N\,$\leq$\,4 transitions. The error bars shown are the
$\pm\,1\sigma$ errors of the Gaussian fits. The large difference in
source integration time between \ngc{3242}/\ngc{6543} and
\ngc{6826}/\ngc{7009} is clearly seen.  The poorer spectral sensitivity
for \ngc{6826}/\ngc{7009} means that He RRL emission cannot be detected
for transitions beyond $\Delta$N=3.

The horizontal lines in Figure~\ref{fig:yptpk} show the error weighted
\yp4\ average of the $\alpha$, $\beta$, and $\gamma$ transitions.  For
our PNe sample targets \yp4\ is nominally $\sim$\,0.10.  This
\yp4\ value gives us another metric with which to assess the performance
of the GBT/ACS spectrometer.  It tells us that {\em every} H RRL ought
to have a corresponding He line with $\sim$\,10\% the intensity of the
hydrogen transition.
This is a very important metric because it allows us to interpret what
is {\em not} plotted on Figure~\ref{fig:yptpk} in the context of
instrumental baseline frequency structure. Specifically, there are
observed spectra that ought to show He RRL detections but do not. That
we cannot see these lines means that they are being masked by BN.

It is clear from Figure~\ref{fig:yptpk} that at the $\sim$\,few
\mk\ level for the He RRL intensity we do not measure the same \yp4\ to
within the formal errors.  For example, our \ngc{3242} observations have
sufficient sensitivity to robustly detect He RRLs beyond the $\Delta$N=2
$\beta$ transitions.  The H RRL intensities predict that, for
\yp4\,$\sim$\,0.1, the He RRLs ought to have intensities of $\sim$0.5
and $\sim$0.2\mk\ for the $\gamma$ and $\delta$ transitions.  These
predictions are not reliably confirmed.  Table~\ref{tab:n3242} indicates
that we detect He $\gamma$ emission at the appropriate
$\sim$0.5\mk\ level for the 130, 131, and 132 transition spectra, but
not for the 130a/130b spectra.  Furthermore, although we have nominal He
RRL detections for the 144b, 144g, and 145 $\delta$ transition spectra,
the measured He intensities differ by nearly a factor of 3 and the
inferred \yp4\ values are $\sim$\,0.13, which is 30\% above the nominal
\yp4\ value.

From all these tests we conclude that our ability to measure accurate
spectral line parameters with the GBT/ACS spectrometer system for the
integration times achieved here becomes compromised by instrumental
baseline frequency structure --- BN --- effects at antenna temperatures
$\lsim$\,1\,--\,2\mk. Our GBT/ACS data show that we can {\em detect}
RRLs at these intensities, but the measured line parameters, $T_{\rm
  pk}$ and $\Delta\,V$, are uncertain by as much as $\sim$\,50\% due to
BN.

\section{Radio Recombination Line Emission from Planetary Nebulae}\label{sec:RRLs}

We can use our measured RRL parameters to assess various global physical
properties of our PNe sample.  Our spectra are extremely sensitive. The
\ngc{3242} and \ngc{6543} data in particular have integration times
spanning $\sim$\,200 to $\sim$\,300 hrs. Below we discuss the \hepr4,
\yp4, abundances, FWHM spectral line widths, and excitation state of the
PNsample as a whole.

\newpage
\begin{figure}[h]
\centering
\includegraphics[angle=-90,scale=0.35]{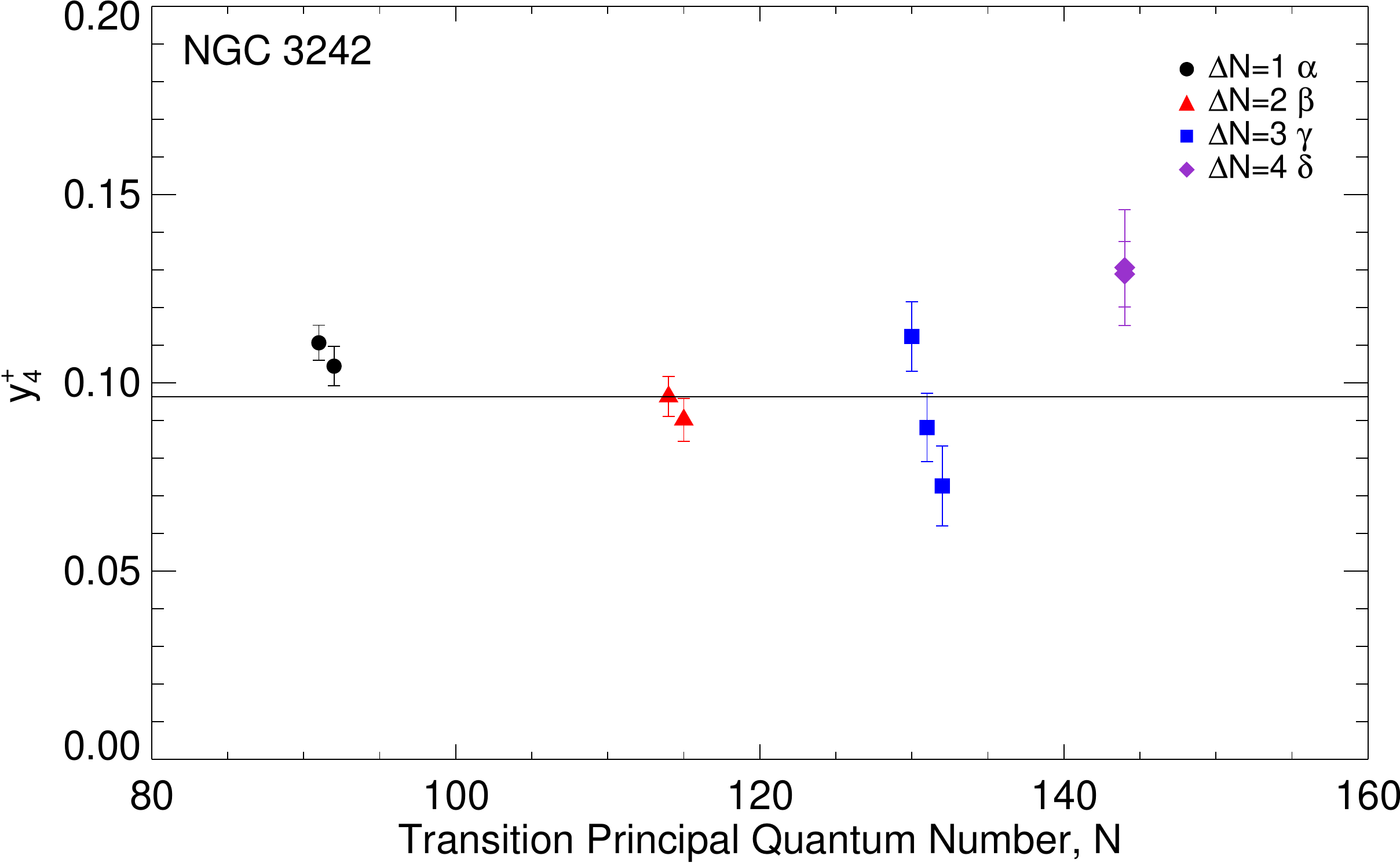}
\includegraphics[angle=-90,scale=0.35]{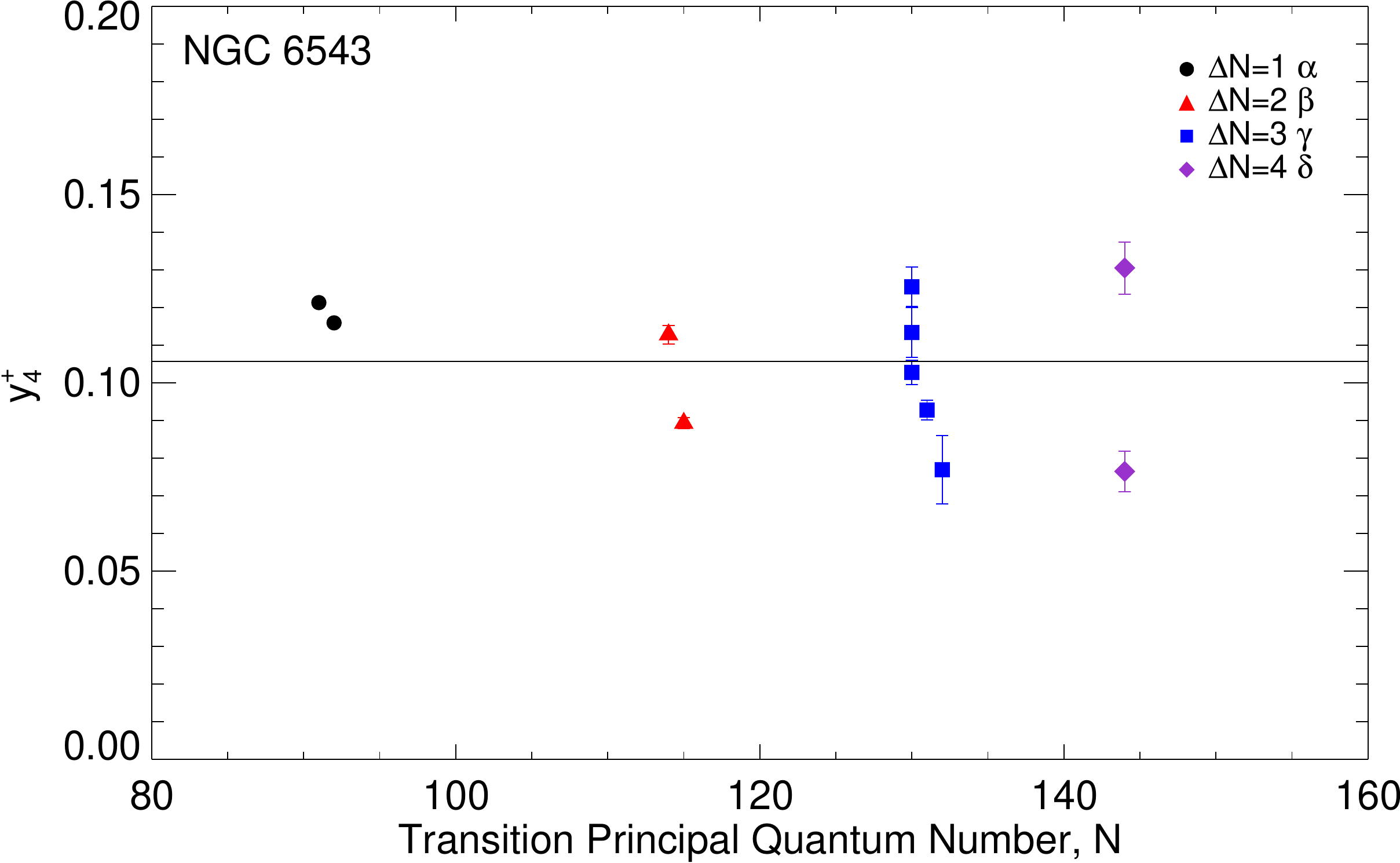}\\
\includegraphics[angle=-90,scale=0.35]{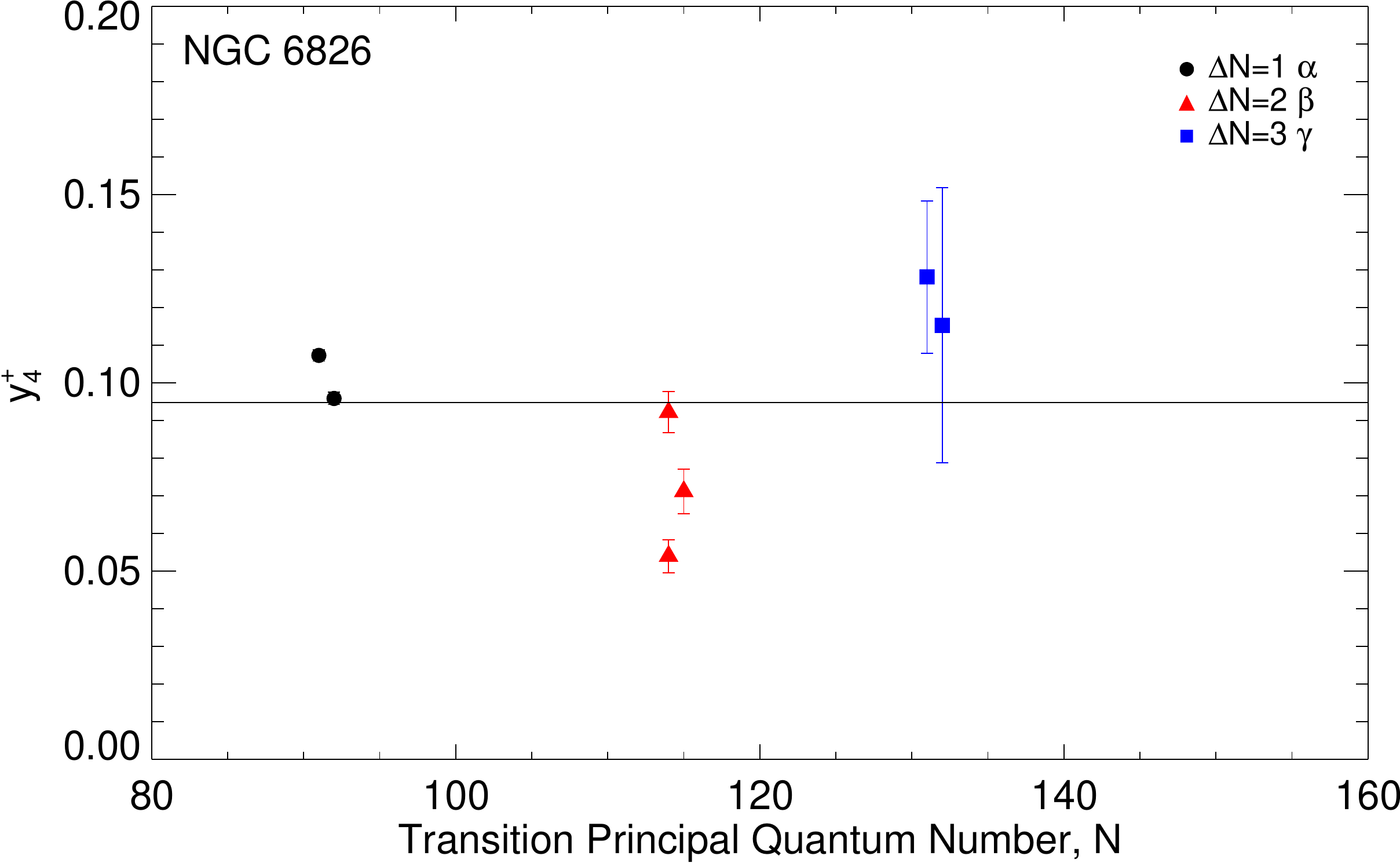}
\includegraphics[angle=-90,scale=0.35]{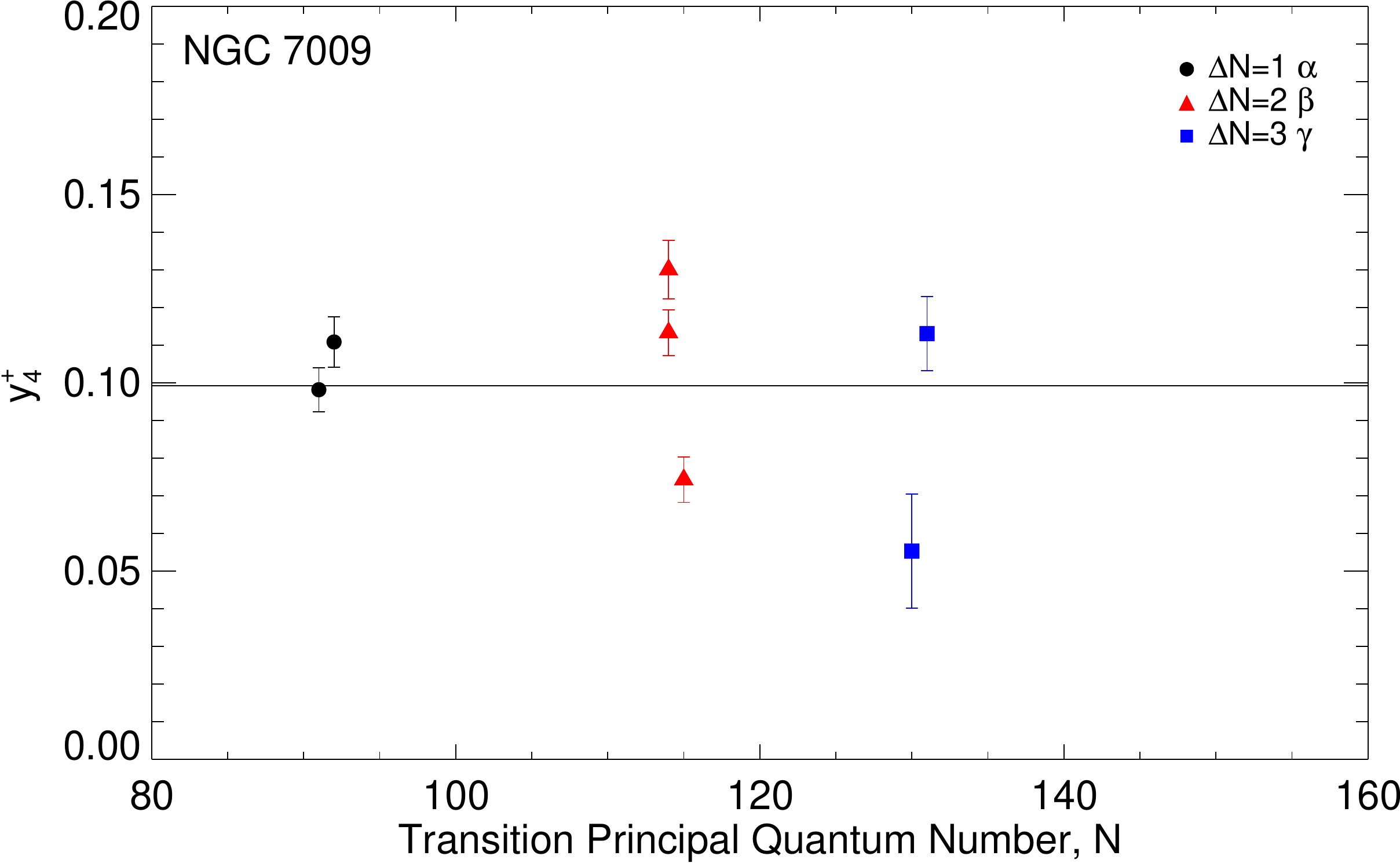} 
\caption{
  Observed RRL \hepr4\ ratio, \yp4, for the planetary nebula sample as a
  function of transition principle quantum number, N.  For sources with
  the largest integration times we can detect He RRL emission from
  transitions with quantum number changes of $\Delta$N = 4. The
  horizontal lines show the average \yp4\ ratio for the $\alpha$,
  $\beta$, and $\gamma$ transitions.  Error bars are $\pm~1~\sigma$ and
  stem from Gaussian fits to the spectra. We find \yp4~$\sim$~0.10 for
  our PNe sample.
  }
\label{fig:yplus}
\end{figure}

\subsection{$^{4}He^{+}$ Abundance}\label{sec:ionization}

We cannot characterize the ionization state of our sample PNe based
solely on the GBT RRL spectra reported here. Our GBT data give no
observational constraints on either the \her4\ or \heppr4\ abundances.
We can, however, use our RRL observations to assess the
\hepr4\ abundance ratio of our PN sample sources.  The central exiting
stars of PNe have very high effective temperatures, $T_{\rm eff} \approx
100,000\K$ \citep{osterbrock}, and these high temperatures are indeed
present in our PNe sample (see Table~\ref{tab:sample}).
One thus expects very high ionization states near these stars.  This
corresponds to the inset regions shown in Figure~\ref{fig:PNeMosaic}.
What ionization state we actually measure is complicated because
approximately 80\% of the Gaussian GBT beam area is sampling only halo
gas, not these inner regions.  This sampling bias is counterbalanced,
however, by the fact that the RRL emission line intensities scale as the
local electron density squared so the RRL emission from these nebulae
overwhelmingly stems from their dense, inner regions.

Thus, given the large GBT beam, our RRL observations provide only a
global characterization of the \hepr4\ abundance of the combination of
the PN cores plus their halos.  Figure~\ref{fig:yplus} shows the \hepr4,
\yp4, abundances for our PNe plotted as a function of transition
principle quantum number, N.  The \yp4\ values are calculated using
Eq.~\ref{eq:yp4def} for each valid He/H RRL measurement pair and the
abundance for each principle quantum number is determined by the average
of all pairs at each N weighted by the spectral integration time.  The
horizontal lines in the figure are the error weighted averages of the
$\alpha$, $\beta$, and $\gamma$ transitions.

The RRL based \yp4\ abundances for our sample PNe are characterized by
\yp4\,$\sim$\,0.10.  Constructing detailed models of PNe requires high
resolution optical data to constrain the complex ionization structure of
these sources. In Section~\ref{sec:PN3He} below we craft numerical
models for \ngc{3242} and \ngc{6543}. There we specify the He ionization
state of these nebulae using \he4\ optical recombination lines.  The
optical and radio \hepr4\ abundances are roughly consistent for both
\ngc{3242} (Table~\ref{tab:n3242model}) and \ngc{6543}
(Table~\ref{tab:n6543model}). The \ngc{3242} model, for example, shows
that its RRL emission arises almost entirely from its inner double shell
--- the inset image region in Figure~\ref{fig:PNeMosaic} --- and that
RRL emission from the halo is negligible.

\begin{figure}[h]
\centering
\includegraphics[angle=-90,scale=0.5]{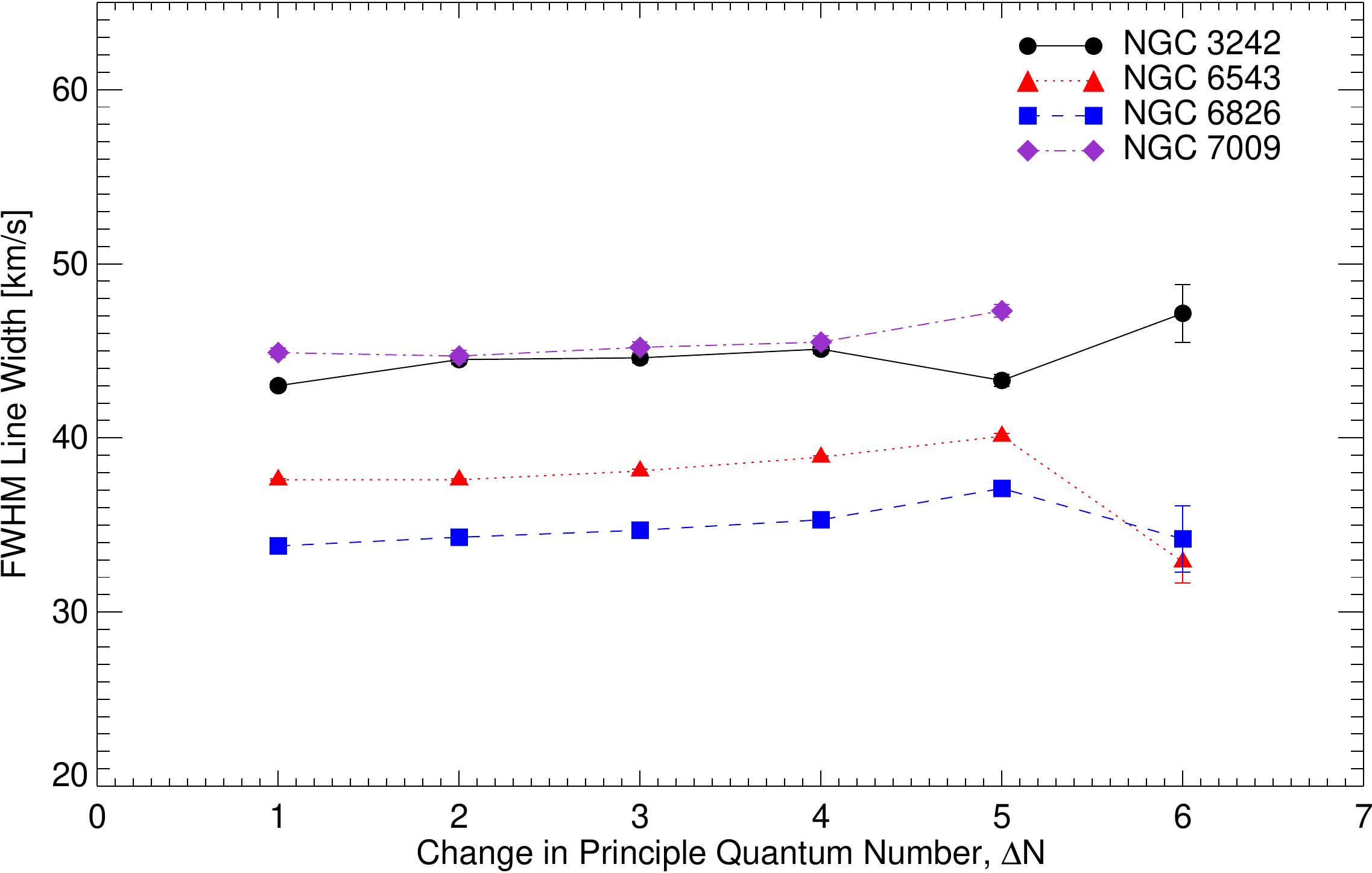} 
\caption{
  Observed RRL FWHM line widths for our sample PNe plotted as a function
  of the transition's change in principle quantum number, $\Delta$N. The
  error bars are $\pm~1.5~\sigma$ and stem from the Gaussian fits to the
  spectra.  The significant expansion velocities of the plasma
  surrounding our PNe are clearly seen in their line widths.
}
\label{fig:fwhm}
\end{figure}

\subsection{Line Widths}\label{sec:fwhm}

The RRL line widths that we measure result from a combination of many
physical processes.  Broadening mechanisms include thermal motions,
non-thermal motions (i.e. turbulence), bulk expansion, other types of
velocity gradients along the line-of-sight, and pressure broadening.  As
discussed above, the GBT's substantial \cm-wavelength beam size,
$\sim\,87\arcsec$, complicates the interpretation of the line widths.
The widths that we measure are a complex combination of the properties
of both the PN inner cores and their extended halos. Our RRL FWHM line
width measurements for our PN sample are summarized in
Figure~\ref{fig:fwhm} which plots the line width as a function of the
transition change in principle quantum number, $\Delta$N.  The line
widths measured for all the RRL transitions with the same $\Delta$N were
averaged weighted by their spectral integration time to give the FWHM
values used in Figure~\ref{fig:fwhm}.

For each PN in our sample, its Gaussian FWHM line width does not
significantly change with $\Delta$N.  We thus appear to see no obvious
evidence for pressure broadening whose effects should appear at high
$\Delta$Ns (or, equivalently, Ns). But \citet{2010linewidth} analyzed
the line shapes of RRL spectra observed for the \hii\ region
SgrB2(N). Their transitions spanned a large range of N and thus ought to
show pressure broadening effects and they do.  For their analysis they
fit Voigt, Gaussian, and Lorentzian line shapes to their spectra.
Their Figure~1 shows that the FWHMs for these different line shapes are
essentially identical.

The FWHM of a RRL is thus not a useful indicator of pressure broadening.
The effects of pressure broadening are only apparent in the extended
line wings or in the peak emission.  Our RRL sensitivity is too poor for
us to be able to make any reliable fits to the spectral line wings but
we do have the sensitivity to detect pressure broadening in the line
peak emission (see sections 5.3 and 6.2).

The expansion of the nebular gas surrounding our sample PNe, however, is
clearly seen in their $\sim$\,35\,--\,45\kms\ line widths.  Both PNe and
\hii\ regions are fully ionized plasmas produced by the ionizing
radiation from their exciting stars. The majority of Galactic
\hii\ regions do not have substantial expansion velocities.  Our PNe
line widths thus are $\sim$\,10\,--\,20\kms\ larger than the average
Galactic \hii\ region FWHM line width of 25\kms \citep{2019shrds}.  Most
of this line broadening must be due to expansion. It certainly is not
due to temperature effects because the plasma electron temperatures of
Galactic \hii\ regions and our sample PNe are almost identical.  Using
the \halpha\ and continuum measurements, we can derive electron
temperatures, \te, for our PN sample's plasmas from these observed
line-to-continuum ratios. We find \te's spanning the range $\sim$\,7,000
to $\sim$\,13,000\K. These temperatures are typical values for Galactic
\hii\ regions \citep[see][for the \te\ derivation and typical
  \hii\ region temperatures]{2019WengerTe}.

\subsection{Intensities and Excitation}\label{sec:excite}

\begin{figure}[h]
\centering
\includegraphics[angle=+90,scale=0.5]{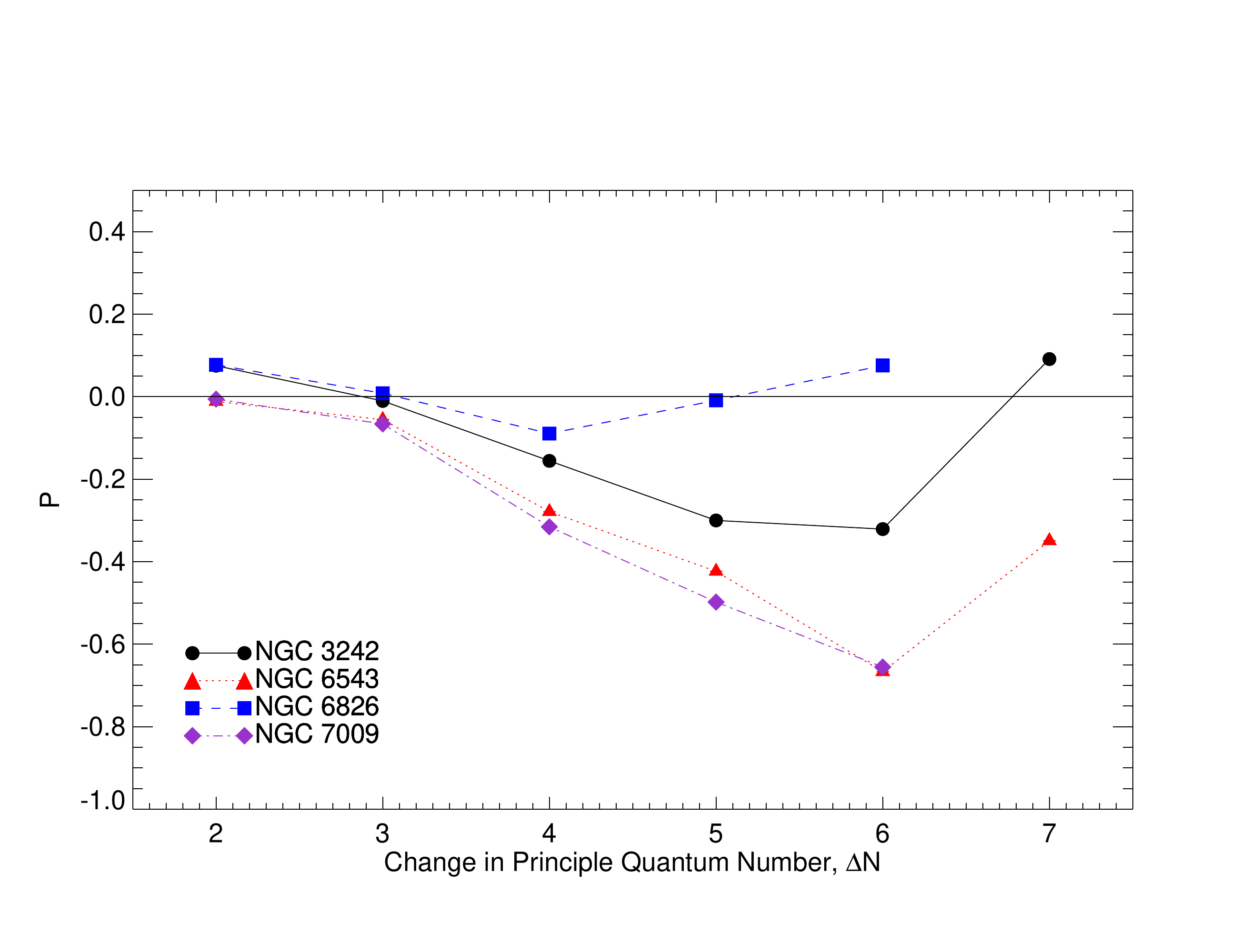} 
\caption{
  Comparison of the fractional difference between the observed line
  ratio and the LTE ratio as a function of the change in the transition
  principle quantum number, $\Delta$N. Plotted here is the normalized
  fractional difference, P = R $-$ 1, where R = Observed Ratio / LTE
  Ratio. Each R ratio is that between the integrated intensity (\mk
  \kms) of the RRL transition and the $<$H$\alpha>$ transition divided
  by the LTE ratio (see text).  The $\pm~1~\sigma$ error bars of these
  data are those of the Gaussian fits to the RRLs. These error bars are
  the size of the plotted symbols.  For most of the PNe in our sample
  apparent departures from LTE are seen with increasing $\Delta$N. The
  RRL intensities at these high $\Delta$N transitions, however, are
  quite weak and their measurement accuracy is compromised by
  instrumental baseline frequency structure (see text
  Section~\ref{sec:baseline}).
}
\label{fig:LTEtest}
\end{figure}

To what extent can our PNe excitation states be described by Local
Thermodynamic Equilibrium (LTE)? To make this assessment we need to know
the expected LTE ratio between a (N, $\Delta$N) transition and a
fiducial (N, $\Delta$N=1) transition. Our observations dictate our
choices. We use \halpha\ as the fiducial transition and calculate the
LTE ratios expected for a set of our observed RRL transitions.
We use the prescription in \citet{1968Menzel} to derive the oscillator
strengths, $f_{NM}$, for recombination transitions between principle
quantum numbers N$\rightarrow$M. Table~\ref{tab:oscillator} lists the
oscillator strengths for some of the radio recombination line
transitions observed here. The Table~\ref{tab:oscillator} transitions
span order changes in principle quantum number, $\Delta$N, between 1 and
24.  Not all these transitions were detected; we list this information
here for completeness. 

The LTE RRL intensity ratio between two optically thin transitions is
given by the ratio of their oscillator strengths, $f_{NM} /
f^{\prime}_{NM}$, and their statistical weights, $g_N / g^{\prime}_N $,
where $g_N~=~2 \times N^2$ :

\begin{equation}
{\rm LTE~Intensity~Ratio}~=~ {{g_N \times f_{NM}} \over {g^{\prime}_N \times f^{\prime}_{NM}}}.
\end{equation}

\noindent
The LTE intensity ratios between some of our observed transitions and
\halpha\ are listed in the final column of Table~\ref{tab:oscillator}.

\begin{deluxetable}{lcccc}
\tablecolumns{5} \tablewidth{0pt}
\tablecaption{RRL Oscillator Strengths and LTE Intensity Ratios\label{tab:oscillator}}
\tablehead{
           \colhead{Transition} & \colhead{N} &
           \colhead{$\Delta$N} & \colhead{$f_{\rm NM}$} &
           \colhead{LTE Ratio} 
}\vspace{-10pt}
\startdata
\phn\halpha   & \phn91 & \phn1 & 0.1764667825E+02   & 1.00000 \\
\phn\hbeta    & 114    & \phn2 & 0.3080855700E+01   & 0.27399 \\  
\phn\hgamma   & 130    & \phn3 & 0.1090205890E+01   & 0.12608 \\
\phn\hdelta   & 144    & \phn4 & 0.5237520000E+00   & 0.07432 \\
\phn\hepsilon & 154    & \phn5 & 0.2926137750E+00   & 0.04749 \\
\phn\hzeta    & 164    & \phn6 & 0.1831153100E+00   & 0.03370 \\
\phn\heta     & 171    & \phn7 & 0.1218369570E+00   & 0.02438 \\
\phn\htheta   & 180    & \phn8 & 0.8682048000E$-$01 & 0.01925 \\
\phn\hiota    & 187    & \phn9 & 0.6395970050E$-$01 & 0.01531 \\ 
\phn\hkappa   & 193    & 10    & 0.4854865600E$-$01 & 0.01238 \\
\phn\hlambda  & 198    & 11    & 0.3773162250E$-$01 & 0.01012 \\
\phn\hmu      & 203    & 12    & 0.3002285000E$-$01 & 0.00847 \\
\phn\hnu      & 208    & 13    & 0.2436411250E$-$01 & 0.00721 \\
\phn\hxi      & 213    & 14    & 0.2010511620E$-$01 & 0.00624 \\
\phn\homicron & 221    & 15    & 0.1704013000E$-$01 & 0.00570 \\
\phn\hpi      & 222    & 16    & 0.1420736100E$-$01 & 0.00479 \\
\phn\hrho     & 227    & 17    & 0.1217512075E$-$01 & 0.00429 \\
\phn\hsigma   & 232    & 18    & 0.1053422930E$-$01 & 0.00388 \\
\phn\htau     & 236    & 19    & 0.9157915750E$-$02 & 0.00349 \\
\phn\hupsilon & 238    & 20    & 0.7964236400E$-$02 & 0.00309 \\
\phn\hphi     & 239    & 21    & 0.6950849150E$-$02 & 0.00272 \\
\phn\hchi     & 247    & 22    & 0.6263460000E$-$02 & 0.00261 \\
\phn\hpsi     & 249    & 23    & 0.5554700550E$-$02 & 0.00236 \\
\phn\homega   & 249    & 24    & 0.4918615500E$-$02 & 0.00209 \\
\enddata
\tablecomments{LTE Ratio~=~Transition / \halpha }
\end{deluxetable}

To make the LTE assessment we use the observed integrated intensity of
each RRL transition, $W_{\rm RRL} = T_{\rm pk} \times \Delta\,V$ (\mk
\kms), where the line peaks, $T_{\rm pk}$, and FWHM line widths,
$\Delta\,V$, are the values from the Gaussian fits to the observed RRL
spectra that are listed in Appendix~\ref{appen:A}. Our observations
include both measurements of adjacent N RRL transitions with the same
$\Delta$N as well as identical (N, $\Delta$N) transitions sampled by
different spectrometer SPWs. For each $\Delta$N we use here a single
integrated intensity, $<W_{\Delta {\rm N}}>$, that is the average value
of these multiple measurements weighted by the integration time of their
spectra. In particular, the fiducial comparison transition we use is the
weighted average of the \halpha\ and \hrrl{92}$\alpha$ transitions,
$<$H$\alpha>$.

All the PNe in our sample show apparent departures from LTE for
$\Delta$N\,$>$\,3.  This can be seen in Figure~\ref{fig:LTEtest} which
compares the fractional difference between the observed line ratio and
the LTE ratio as a function of the principle quantum number change,
$\Delta$N. Plotted is the normalized fractional difference, P = R $-$ 1,
where R = Observed Ratio / LTE Ratio. Each R ratio is that between
$<W_{\Delta {\rm N}}>$, the observed average ratio for a $\Delta$N, and
$<$H$\alpha>$, divided by the LTE ratio listed in
Table~\ref{tab:oscillator}. The $\pm~1~\sigma$ error bars for
Figure~\ref{fig:LTEtest} stem from the Gaussian fits to the RRLs and are
the size of the plotted symbols.

From the analysis in Section~\ref{sec:baseline}, however, we know that
systematic instrumental frequency structure in the spectral baselines
--- the spectral BN --- is compromising the RRL measurements for
$\sim$\,2\mk\ line intensities. Our ability to measure RRL parameters
with the GBT/ACS spectrometer becomes increasingly inaccurate for
$\Delta$N$\,\gsim$\,3 transitions.  The Figure~\ref{fig:LTEtest} error
bars reflect only the error in the Gaussian fits to the RRLs. The
precise magnitude of the BN for any particular RRL spectral region is
unknown so the overall effect on this LTE analysis is uncertain.  Given
this, we deem that numerical modeling of our PN sources (see
Section~\ref{sec:PN3He} below) provides a more robust means to assess
any LTE effects.

Some of the apparent non-LTE behavior seen in Figure~\ref{fig:LTEtest},
however, is probably due to pressure broadening. We observe lower
integrated intensities than expected by LTE for the higher $\Delta$N
transitions.  The effects of pressure broadening are to diminish the
line intensity and broaden the line wings as $\Delta$N increases. But,
as we discuss in Section~\ref{sec:fwhm}, our FWHM line width
measurements are insensitive to any broadening in the line wings.
Therefore the only pressure broadening effect that would be apparent in
our RRL data would be the drop in line intensity as $\Delta$N
increases. Pressure broadening would then manifest as a decrease in R
and P as $\Delta$N increases.  At low $\Delta$Ns, however, P would be
$\sim$ 0.  This is consistent with Figure~\ref{fig:LTEtest}.

\clearpage

\section{Planetary Nebula \he3\ Abundances}\label{sec:PN3He}

The GBT \hep3\ spectra for planetary nebulae reported here are more
sensitive and have less instrumental baseline structure than any
previously taken by the MPIfR 100\m\ and NRAO 140 Foot telescopes.  We
do not confirm the detection of \hep3\ emission from \ngc{3242} based on
the independent observations made by those two telescope spectrometers.
Apparently systematic baseline frequency structure was an even bigger
issue that we thought for these telescopes.  Furthermore, we show in
Section~\ref{sec:baseline} that these effects are still an issue with
the unblocked aperture GBT, albeit at a much lower spectral intensity.

Given the importance of the \he3\ abundance in PNe, we derive the
\her3\ abundances for \ngc{3242} and \ngc{6543} as if the line
parameters listed in Table~\ref{tab:he3} described actual detections.
So here we determine upper limits for the \her3\ abundances in these
PNe.  Because \ngc{6826} and \ngc{7009} have substantially poorer
sensitivity, we choose not to calculate \her3\ abundance upper limits
for these sources.

To do this we must determine the abundance of \he3\ relative to H. This
requires a model since the collisionally excited \hep3\ hyperfine line
intensity depends on the density, whereas both the free-free continuum
and RRL intensities are proportional to the density squared. The
free-free thermal continuum intensity is used to derive the H abundance
which is a critical step in the determination of the \her3\ abundance
ratio.  An ionization correction is also necessary to convert the
\hepr3\ abundance to a \her3\ abundance ratio by number.
We use a numerical code, NEBULA\footnote[5]{See
  http://ascl.net/1809.009.} \citep{NEBULA}, originally developed by
\citet{1995PhDBalser}, to model our sources and derive upper limits for
their \her3\ abundance ratios.  \citet{BB2018} used NEBULA to construct
\hii\ region models and they briefly describe the code but full details
are found in \citet{1995PhDBalser}.

A NEBULA model planetary nebula is composed solely of H and He within a
three-dimensional Cartesian grid with arbitrary density, temperature,
and ionization structure. Each numerical cell specifies values for the
following quantities: electron temperature, $T_{\rm e}$, electron
density, $n_{\rm e}$, \hepr4\ abundance ratio, \heppr4\ abundance ratio,
and the \hepr3\ abundance ratio.  The \hep3\ transition is assumed to be
in LTE, but non-LTE effects and pressure broadening from electron
impacts can be included for the RRL modeling.  All spectra are
broadened by thermal and micro-turbulent motions.
NEBULA calculates synthetic observations for \hep3, RRL, and continuum
emission by performing the radiative transfer through the model grid.
The radiative transfer is executed from the back of the grid to the
front to produce the brightness distribution on the sky.  To simulate an
observation NEBULA calculates model spectra by convolving this
brightness distribution with a Gaussian beam that has the GBT's HPBW at
the \hep3\ frequency.

\subsection{\ngc{3242}}\label{sec:n3242}

We used NEBULA to model \ngc{3242} in the past \citep[see,
  e.g.,][]{1997PNe}.  The NEBULA model described here, however,
supersedes all previous efforts.  Much better PNe information is now
available to constrain the model and the GBT observations reported here
are far superior to those taken previously by on-axis, blocked aperture
telescopes.

The NEBULA model for \ngc{3242} is comprised of a set of nested spheres
with uniform properties.  The model has a multi-shell morphology and a
large halo \citep{2009Phillips}. The properties of the three nested,
spherically symmetric shells are summarized in
Table~\ref{tab:n3242model}. We adopt the Gaia DR2 parallax distance
listed in Table~\ref{tab:sample}.  The RRL emission arises primarily
from the denser inner and outer shells, also called the ``rim'' and
``shell'' \citep{1987Balick}. The angular extent of both these
components is defined by observations and can be seen as the black
circular inset core region of the Figure~\ref{fig:PNeMosaic} \ngc{3242}
image.  A significant portion of the \hep3\ emission comes from the
diffuse halo which entirely fills the GBT beam.

The kinematics of \ngc{3242} are captured by the shape of spectral
lines.  These lines are primarily broadened by three mechanisms:
expansion, thermal motions, and non-thermal motions.  We assume a
symmetric expanding nebula with a velocity that is constrained by
optical collisionally excited lines (CELs) toward the center of the
nebula \citep{1989PNeExp}. The electron temperature (see below)
determines the amount of thermal line broadening.  Any non-thermal
broadening, assumed to be constant throughout the nebula, is constrained 
by the observed line width of the RRLs.

The electron temperature, electron density, and He ionization structure
is determined using both optical and radio data.  The RRL-to-continuum
ratio provides a good estimate for the electron temperature.  Our GBT
observations of the \ngc{3242} \halpha\ and continuum emission yield
$T_{\rm e}$ = 12,700\K\ which is consistent with the optical
integral-field spectroscopy value determined by \citet{2013N3242ion}.
The electron density that is predicted from optical CELs varies
significantly depending on the tracer.  We therefore adopt an average
value based on data from \citet{2013N3242ion} for each component. These
values are adjusted slightly to match the observed RRL peak intensities.

The He ionization structure adopted for our \ngc{3242} model is
summarized in Table~\ref{tab:n3242model}. It is constrained by
\he4\ optical recombination lines \citep{2013N3242ion}. These
observations directly measure the singly and doubly ionized He abundance
ratios. Given the high electron temperature, we assume there is no
neutral helium within \ngc{3242} and that the \he3\ ionization structure
follows that of \he4.  We also adopt a halo abundance, \yp4, of
\hepr4\ = 0.1 and assume that the halo contains no doubly ionized \he4.
The \yp4\ value derived from our RRL data is between the optical
\yp4\ values for the inner and outer shells (0.08 and 0.13,
respectively).  Since the RRL emission comes primarily from the shells
and not the halo, the radio and optical data are roughly
consistent. Using this ionization structure we make an ionization
correction to the NEBULA model input \her3\ abundance : $\hepr3\ =
\left[{\hepr4}~/~{\her4} \right] \times \her3$.

\begin{deluxetable}{lccccccc}
\tablecaption{NEBULA Model Parameters for \ngc{3242} \label{tab:n3242model}}
\tablehead{
\colhead{} & \colhead{$V_{\rm exp}$} & \colhead{Angular Size} &
\colhead{$T_{\rm e}$} & \colhead{$n_{\rm e}$} &
\colhead{} & \colhead{} & \colhead{}
\\
\colhead{Component} & \colhead{(\kms)} & \colhead{(arcsec)} &
\colhead{(K)} & \colhead{(\percc)} &
\colhead{\hepr4} & \colhead{\heppr4} & \colhead{\hepr3}
}
\startdata
Inner Shell & 20 & $10-25$  & 13,000 & 2079 & 0.080 & 0.020 & \nexpo{3.6}{-5} \\
Outer Shell & 20 & $25-44$  & 13,000 &  747 & 0.130 & 0.001 & \nexpo{4.5}{-5} \\
Halo        & 11 & $44-120$ & 13,000 &   10 & 0.100 & 0.000 & \nexpo{4.5}{-5} \\
\enddata
\end{deluxetable}

This NEBULA model does an excellent job in accounting for the continuum
and RRL emission observed for \ngc{3242}. Altogether, 42 hydrogen and
helium RRL transitions were calculated. NEBULA synthetic spectra are
shown in Appendix~\ref{appen:NEBULA} for eight representative
transitions: \halpha, \hep3, \hbeta, and
\hbbeta\ (Figure~\ref{fig:n3242modelA}) and \hgamma, \hggamma, \hdelta,
and \heepsilon\ (Figure~\ref{fig:n3242modelB}). Except for the
144$\delta$ H and He transitions, there is excellent agreement between
the modeled and observed spectra. There is a well-known impedance
mismatch in the GBT's X-band feed horn that produces a ``suck-out'' at
frequencies very near those of the 144$\delta$ H and He transitions that
severely compromises the spectral baselines in that particular SPW.

\subsection{\ngc{6543}}\label{sec:n6543}

As with \ngc{3242} we treat the \hep3\ parameters listed in
Table~\ref{tab:he3} for \ngc{6543} as if this were an actual detection
and use them to derive an upper limit for the \her3\ abundance using a
NEBULA model. Here, however, our model is oversimple. \ngc{6543} is an
extremely complex source. We deem it unnecessary to construct a more
sophisticated model in order to derive what is a notional upper limit to
the \her3\ abundance.

\begin{deluxetable}{lccccccc}
\tablecaption{NEBULA Model Parameters for \ngc{6543} \label{tab:n6543model}}
\tablehead{
\colhead{} & \colhead{$V_{\rm exp}$} & \colhead{Angular Size} &
\colhead{$T_{\rm e}$} & \colhead{$n_{\rm e}$} &
\colhead{} & \colhead{} & \colhead{}
\\
\colhead{Component} & \colhead{(\kms)} & \colhead{(arcsec)} &
\colhead{(K)} & \colhead{(\percc)} &
\colhead{\hepr4} & \colhead{\heppr4} & \colhead{\hepr3}
}
\startdata
Shell & 17 & $9.2-19.4$ &  8,000 & 3750 & 0.12 & 0.0 & \nexpo{7.3}{-5} \\
Halo  & 10 & $19.4-165$ & 14,700 &   35 & 0.12 & 0.0 & \nexpo{7.3}{-5} \\
\enddata
\end{deluxetable}

The NEBULA model for \ngc{6543} has an inner shell surrounded by a large
halo. The properties of these two nested, spherically symmetric shells
are summarized in Table~\ref{tab:n6543model}. We adopt the Gaia DR2
parallax distance listed in Table~\ref{tab:sample}.  The angular extent
of both model components is defined by observations.  The RRL emission
arises primarily from the denser inner shell, the ``rim''
\citep{2004Balick}. In order to better match the RRL spectra we include
non-LTE and pressure broadening effects in the NEBULA model.
%
%
As with \ngc{3242}, a significant portion of the \hep3\ emission comes
from the diffuse halo which entirely fills the GBT beam.
We assume that both components exhibit symmetric expansion velocities
and constrain them using optical CELs.  For the shell we adopt the rim
velocity measured by \citet{2014Schonberner} using [OIII]/[NII]. The
halo expansion velocity as well as all its other physical properties are
adopted from \citet{1989Middlemass}.

The electron temperature and He ionization structure is constrained by
both optical and radio data.  The RRL-to-continuum ratio provides an
estimate for the electron temperature.  Our GBT observations of the
\ngc{6543} \halpha\ and continuum emission yield $T_{\rm e}$ = 7,300\K.
This is consistent with optical measurements \citep{2001Hyung}.  We
adopt the ionization structure derived by \citet{2003Bernard}. This
gives a \yp4\ of $\hepr4 = 0.12$.  The model has no neutral or doubly
ionized helium.

For \ngc{6543} there is clear evidence of pressure broadening when
comparing the NEBULA model RRL intensities with the GBT observations.
To produce a reasonable NEBULA model requires that non-LTE/pressure
broadening effects together with a higher local density specified with a
volume filling factor of 0.2 \citep{1999He3abundance} be included.  This
\ngc{6543} NEBULA model also accounts for the observed continuum
emission.  The synthetic RRL spectra are shown in
Appendix~\ref{appen:NEBULA} for \halpha, \hep3, \hbeta, and
\hbbeta\ (Figure~\ref{fig:n6543modelA}) and \hgamma, \hggamma, \hdelta,
and \heepsilon\ (Figure~\ref{fig:n6543modelB}). The model does a
reasonable job in reproducing the RRL observations for low order
transitions. Its ability to reproduce the RRL spectra, however, clearly
becomes problematic at the higher $\Delta$N transitions.

\subsection{Upper Limits to the \he3\ Abundance in Planetary Nebulae}

\begin{figure}[h]
\centering
\includegraphics[angle=0,scale=0.40]{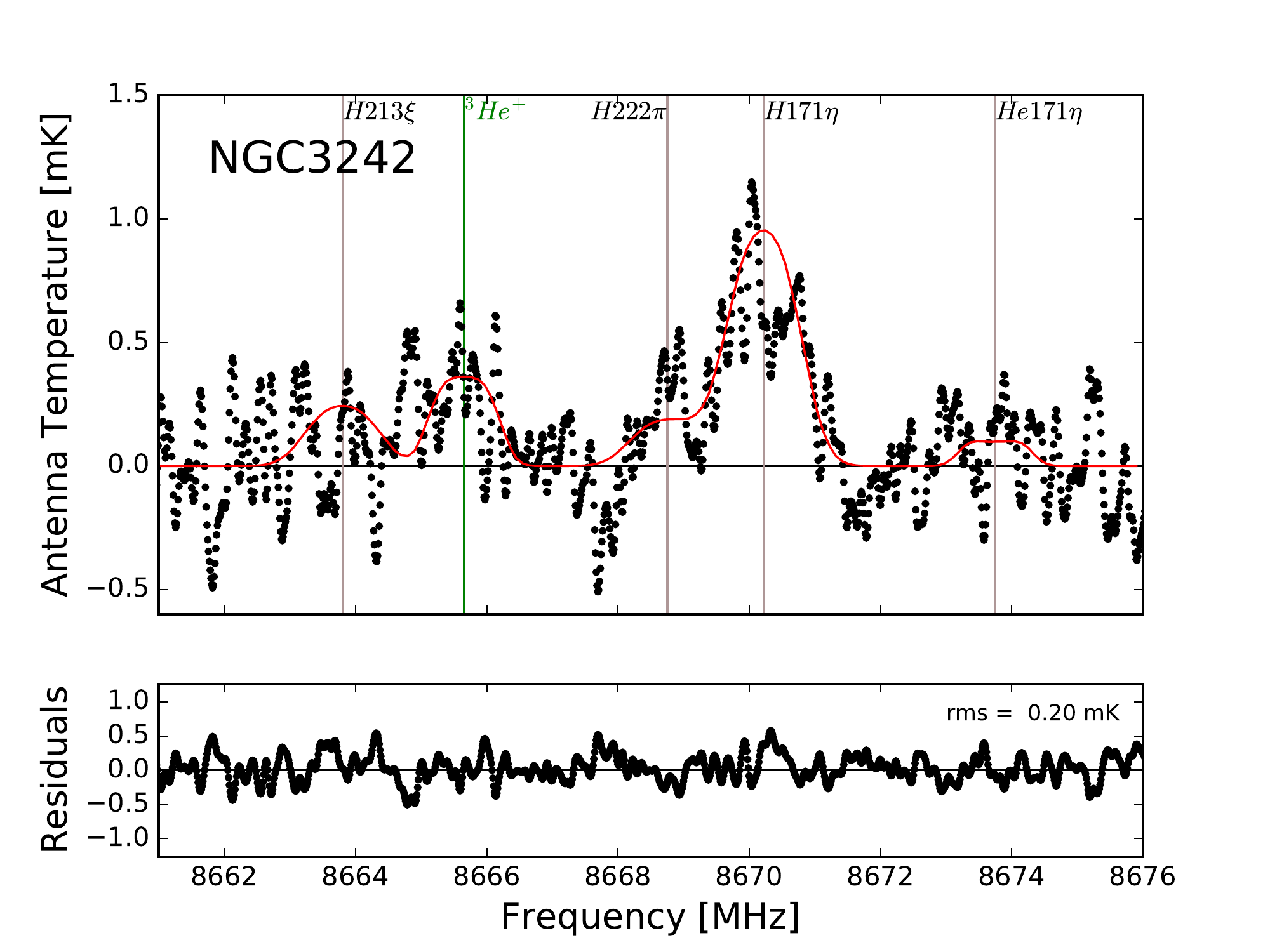}
\includegraphics[angle=0,scale=0.40]{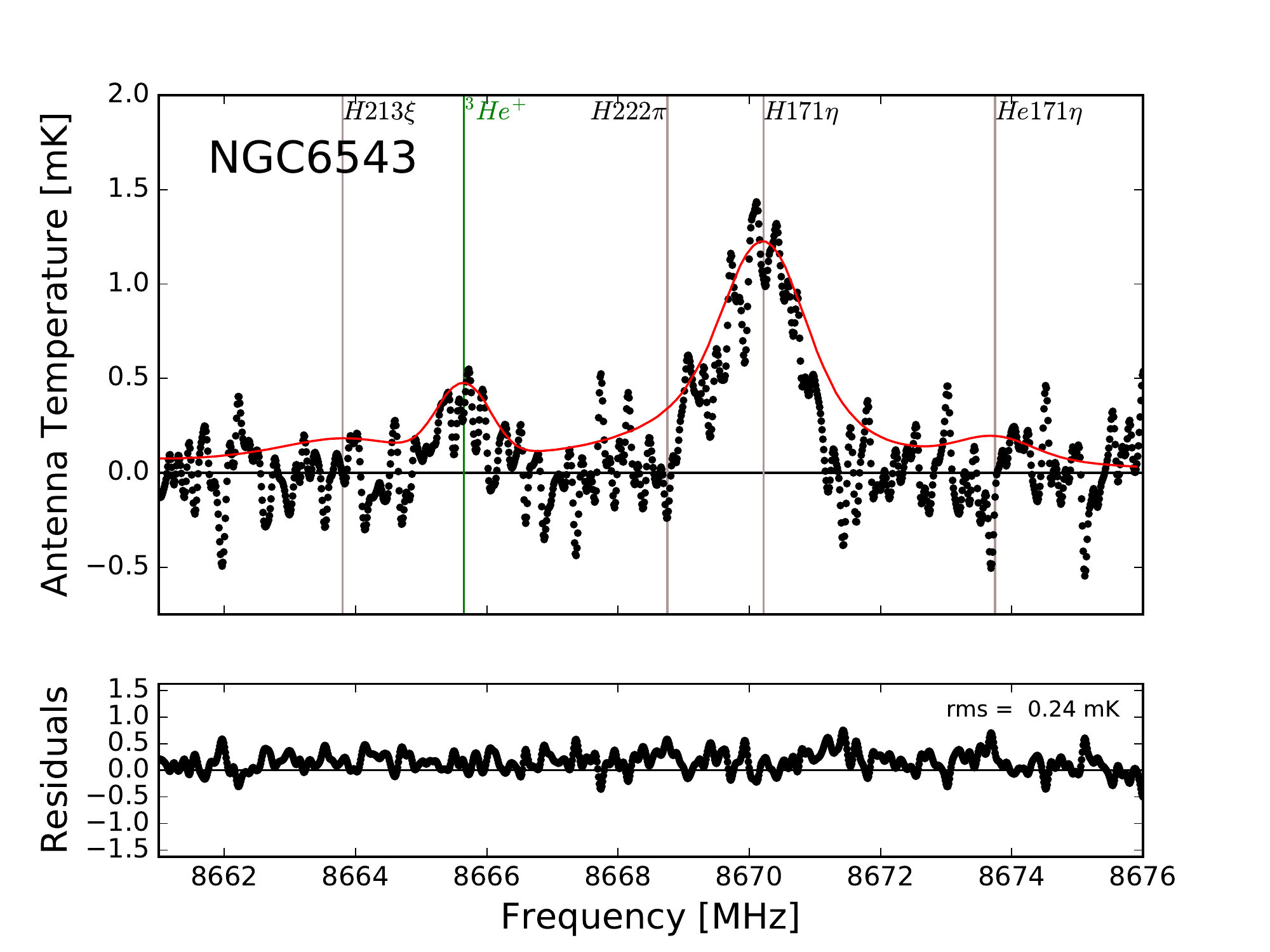}\\

\caption{
  \ngc{3242} and \ngc{6543} NEBULA models for line emission in the
  \hep3\ spectral band (red) compared with observations (black).  The
  residuals between the observed spectra and the models are shown in the
  plots below the spectra.  Vertical lines flag the recombination
  transitions present in each spectral band. {\em These models set the
    upper limits for the \her3\ abundances quoted here.}
}
\label{fig:he3models}
\end{figure}

We have not detected \hep3\ emission from any of our sample PNe so we
can only set upper limits on the \her3\ abundance ratio from these
observations using our NEBULA models.  To derive a \he3\ abundance, a
NEBULA model \her3\ ratio is set so that the synthetic \hep3\ hyperfine
transition intensity is consistent with the observed spectrum.  The
\hep3\ NEBULA model spectra for \ngc{3242} and \ngc{6543} are shown in
Figure~\ref{fig:he3models}.  We find that a \he3\ abundance ratio of
\her3\ = \nexpo{4.5}{-5} for \ngc{3242} and \nexpo{7.3}{-5} for
\ngc{6543} by number gives a reasonable approximation to the observed
spectrum.

Here we did not attempt to fit the NEBULA \he3\ abundance to the
\ngc{3242} and \ngc{6543} spectra in any formal numerical way because
each model takes many CPU hours to run. We are, after all, modeling an
upper limit for these abundances.  Furthermore, the NEBULA model for
\ngc{3242} is quite good and does an excellent job in accounting for the
observed RRL and continuum emission. On the other hand, the \ngc{6543}
NEBULA model is really just a notional one.  Our model is clearly too
simple, even with the inclusion of non-LTE effects, pressure broadening,
and clumping.  We therefore deem that our \ngc{6543} \her3\ upper limit
is not very reliable.

\clearpage

\section{Discussion}\label{sec:discussion}

Our measurements of \he3\ emission from Galactic \hii\ regions and
planetary nebulae are at the sensitivity limit of the GBT/ACS
spectrometer.  The RRL spectra show that this sensitivity is sufficient
to robustly detect and to measure the properties of spectral lines that
have $\gsim$\,1\mk\ intensities and line widths of $\sim$\,25\kms.
\citet{BB2018} analyzed the \hii\ region \he3\ sample. They found that
the \her3\ abundances of \hii\ regions show a shallow negative gradient
when plotted as a function of Galactic radius that is roughly consistent
with the \citet{2012Lagarde} stellar evolution models that include
thermohaline mixing.

Our \her3\ upper limits, especially that for \ngc{3242}, support the
extra mixing solution to the ``\he3\ Problem''.  Unfortunately these
results are still insufficient to prove that thermohaline mixing in
particular occurs in {\em all} low mass stars. Should \he3\ be
definitively detected in even one planetary nebula its \her3\ abundance
would place significant constraints on extra mixing processes in
low-mass stars and, in turn, models for stellar evolution and Galactic
chemical evolution.  A high \her3\ abundance ratio derived for even a
single PN would indicate that some mechanism must be at play to inhibit
the extra mixing in this object. Extra mixing should otherwise occur in
all low-mass stars.

\subsection{Has $^{\it 3}He^{+}$ emission been detected in any planetary nebula?}\label{sec:any}

We have not made any reliable detections of \hep3\ emission from our PN
sample objects with our GBT/ACS observations.  In particular, we can
only set an upper limit on the \her3\ abundance in \ngc{3242}, thus
failing to confirm previously reported detections of \hep3\ emission
based on observations made with the NRAO 140 Foot and MPIfR
100\m\ telescopes.  It is therefore reasonable to ask whether
\hep3\ emission has been detected from {\em any} planetary nebula. Two
other PNe have published \hep3\ detections: \ic{418} and J\,320.

\subsubsection{\ic{418}}\label{sec:ic418}

Using the NASA Deep Space Station 63 (DSS-63) telescope,
\citet{2016IC418} report the detection of \hep3\ emission from the PN
\ic{418}.  The lack of any serious tests of the DSS-63 spectral
baselines, however, together with large discrepancies in the observed
RRL line parameters make this claimed detection highly suspect.  From
their Table 1 the measured $\alpha$ and $\beta$ RRL transition
intensities differ by 7\% and 14\%, respectively.  Furthermore, the
\hepr4\ ratios reported are 0.112, 0.037, 0.258, 0.095, and 0.117 for
the 91$\alpha$, 92$\alpha$, 114$\beta$, 115$\beta$, and 116$\beta$
transitions, respectively. The intensities and line ratios should be the
same for these transitions within the errors. Also compare the variation
in their \yp4\ values with those reported here for our PN sample.  These
are systematic errors in the \citet{2016IC418} measurements hidden in
plain sight.  From this we judge that the level of instrumentally caused
systematic uncertainty in the DSS-63 data is at least $\sim$\,5\mk,
making their reported $\sim$10\mk\ \hep3 detection a $\sim$2\,$\sigma$
signal.

\subsubsection{J\,320}\label{sec:j320}

The advantage of interferometers like the NRAO Very Large Array (VLA) is
that much, but not all, of the instrumental baseline structure is
correlated out.  We therefore deem that the \citet{2006J320}
\hep3\ detection for J320 with the VLA is more robust than that reported
for any other PN.  The signal to noise ratio (SNR) is only 4, but when
averaging over a halo region the SNR increases to 9.  Nevertheless, the
J320 VLA observations suffer from three problems: (1) a 3.3\mhz\ ripple
caused by reflections within the waveguide (common to all antennas); (2)
a limited number of channels and bandwidth which together provide only a
very small spectral region with which to determine the instrumental
baseline; and (3) only one RRL transition to assess the accuracy of the
measurements. The latter two deficiencies are due to the capabilities of
the VLA correlator.

The next step in the study of \he3\ production in PNe is to confirm the
J320 detection. We plan to use the enhanced VLA, the Jansky Very Large
Array (JVLA), to do this.  The JVLA overcomes all three problems with
the VLA observations \citep{2011jvla}.  Optical fiber has replaced the
old waveguides and the 3.3\mhz\ ripple is no longer present.  The new
Wideband Interferometric Digital ARchitecture (WIDAR) correlator
provides an ample number of channels across a large bandwidth to measure
the spectral baseline accurately.  Finally, the flexibility of WIDAR
allows one to tune to many RRLs simultaneously. This will provide the
data needed to carefully assess the quality of the spectral baselines.

\begin{figure}
\centering
\includegraphics[angle=-90,scale=0.4]{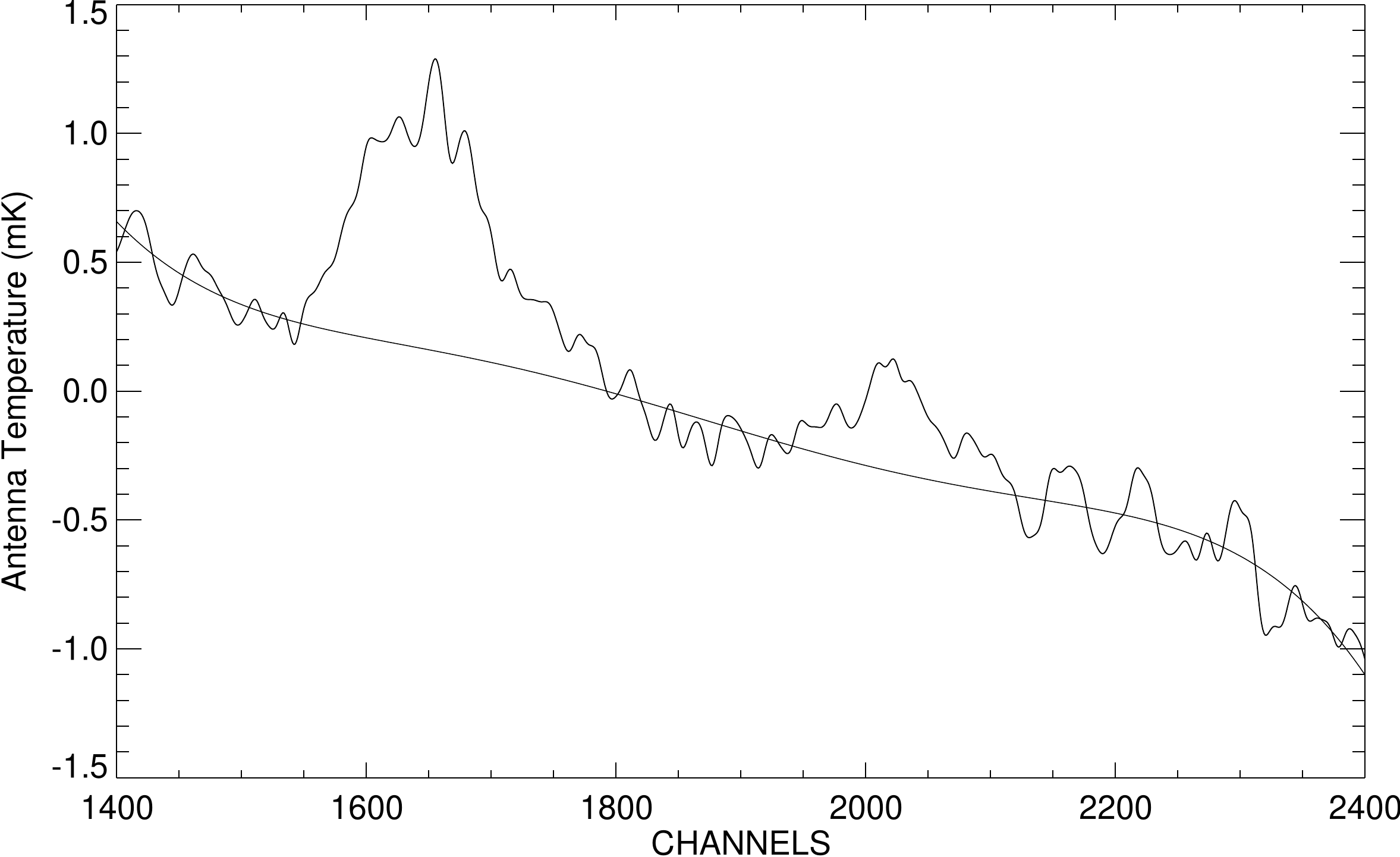}
\includegraphics[angle=-90,scale=0.4]{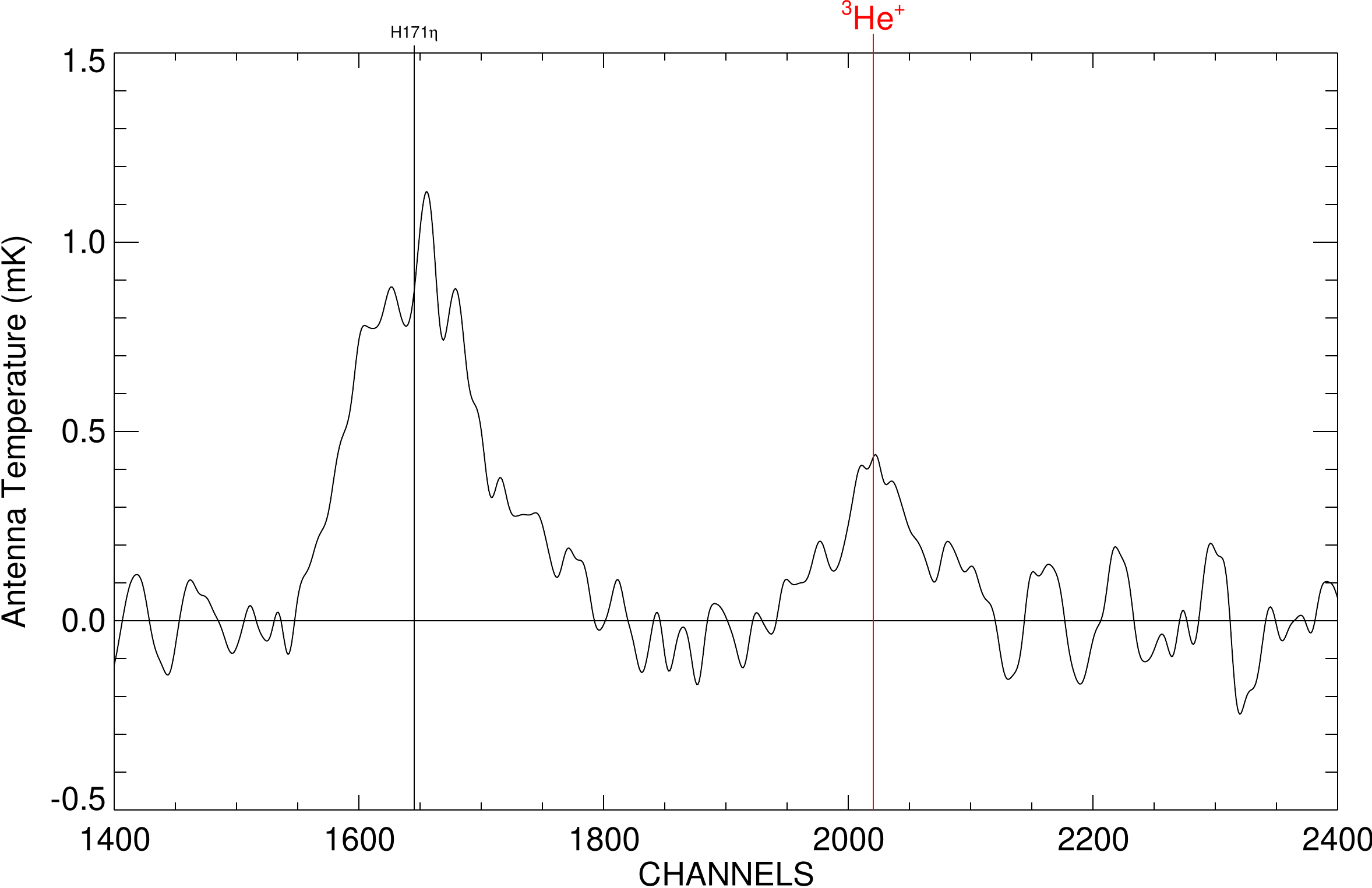}\\

\caption{Composite average \hep3\ spectrum for \ngc{3242} and
  \ngc{6543}.  The spectrum was smoothed to 5\kms\ resolution. After
  smoothing this 557.7 hr integration has 98 \microK\ RMS noise.
  {\em Top:} The 5-th order instrumental baseline fit is shown
  superimposed atop the data. 
  {\em Bottom:} Spectrum that results when this baseline model is subtracted.  
Vertical lines flag the locations of the \heta\ and \hep3\ transitions.
}
\label{fig:composite}
\end{figure}

\subsection{PN Composite Spectrum}

A composite \hep3\ spectrum that is the average of our \ngc{3242} and
\ngc{6543} observations is shown in Figure~\ref{fig:composite}.  It was
processed in an identical manner as that for individual PN spectra.
This spectrum nearly doubles the integration time of the \hep3\ spectral
band. After smoothing to 5\kms\ resolution, this 557.7 hr integration
has 98 \microK\ RMS noise and a quality factor, Q, of 1.60. 
Figure~\ref{fig:composite} shows the spectral baseline and its model
before subtraction.  Even with existing spectrometer systems such as the
GBT/ACS, this composite \hep3\ spectrum suggests that given sufficiently
large integration times robust detections of \hep3\ emission from
planetary nebulae can be made.

We cannot construct a NEBULA model for this composite spectrum which
also means that a \her3\ abundance cannot be derived. The
Figure~\ref{fig:composite} spectrum does give an estimate for the
sensitivity required to detect \hep3\ emission from PNe.  Nonetheless,
the \her3\ abundances derived for such extremely weak \hep3\ emission
from individual PN will imply that there is little if any net production
of \he3\ in low-mass stars.


\subsection{Mixing Processes in Low-Mass Stars}

Can our upper limits for the \her3\ abundance in PNe still provide any
insight about mixing mechanisms in low-mass stars?  \citet{2011Lagarde}
make specific predictions for \he3\ production by such stars from
various mixing processes.  They calculate a grid of stellar models and
report the \he3\ produced during main sequence stellar evolution as a
mass fraction yield, ${\rm Y_{\rm 3}}$.
\citet{2011Lagarde} find that the \he3\ yields depend on stellar mass
and metallicity. We deem our most robust \her3\ upper limit abundance is
that for \ngc{3242}. \citet{1997Galli} cite a mass of 1.2$\,\pm\,$0.2 \msun
for the \ngc{3242} progenitor star which has a mass fraction metallicity
Z = \zsun/2 = 0.0061 \citep{1985Barker} where the Sun's metallicity,
\zsun, is 0.0122 \citep{2006aAsplund, 2006bAsplund}. This \ngc{3242}
metallicity is closest to the \citet{2011Lagarde} models listed in their
Table 3 where Z = 0.004. Their \he3\ model yields for various mixing
mechanisms are summarized in Table~\ref{tab:he3yields}.

\begin{deluxetable}{ccc}[h]
\tablecolumns{3} \tablewidth{0pt}
\tablecaption{Stellar Model \he3\ Yields\tablenotemark{{\rm a}} \label{tab:he3yields}}
\tablehead{
  \colhead{Stellar Mass} & \colhead{Mixing Process} & \colhead{${\rm Y_{\rm 3}}$} \\
  \colhead{\msun} & \colhead{} & \colhead{$\times$\expo{-4}} 
}
\startdata
1.00  & standard                        & 6.39    \\
      & thermohaline                    & 1.35    \\
      & thermohaline + rotation         & 1.50    \\
1.25  & standard\tablenotemark{{\rm b}} & \dots   \\
      & thermohaline                    & 1.96    \\
      & thermohaline + rotation         & 2.04    \\
\enddata
\tablenotetext{{\rm a}}{ Mass fraction \he3\ yields, ${\rm Y_{\rm 3}}$,
  from \citet{2011Lagarde} stellar evolution models.} 
\tablenotetext{{\rm b}}{ No standard mixing model calculated.}
\end{deluxetable}

\begin{figure}
\centering
\includegraphics[angle=0,scale=0.8]{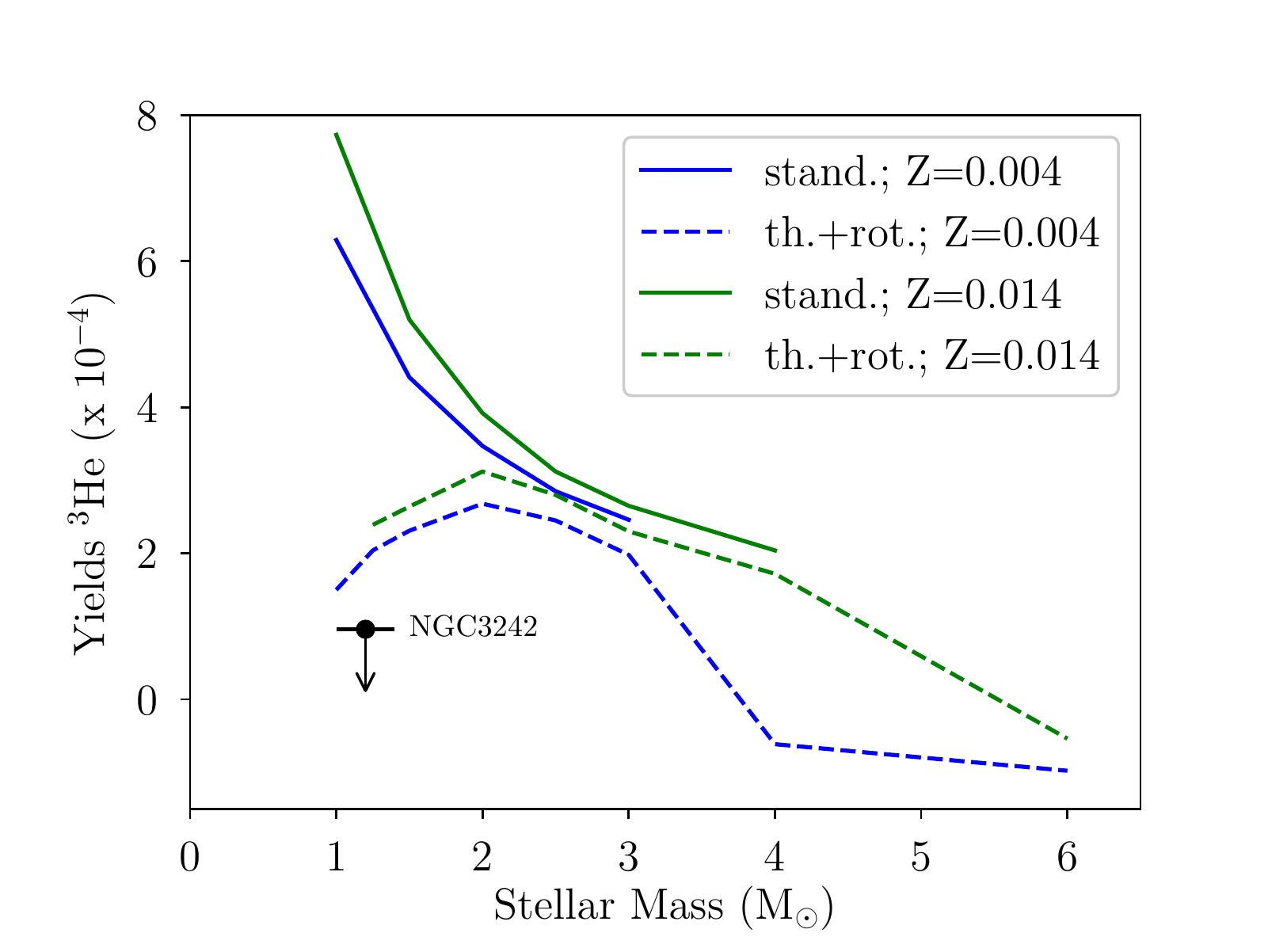}
\caption{
Comparison of our \he3\ mass fraction upper limit for \ngc{3242} with
the \citet{2011Lagarde} stellar evolution models.  The model ${\rm
  Y_{\rm 3}}$ yields are adopted from their Figure~9. The models shown
have a range of metallicities, Z, and employ either standard mixing or a
combination of thermohaline and rotational mixing.
}
\label{fig:Y3yields}
\end{figure}

Our GBT upper limit for \ngc{3242} is \her3\ $<$ \nexpo{4.5}{-5} by
number. Since the stellar yields are given as mass fractions 
this abundance ratio must be converted to mass fraction, ${\rm Y_{\rm 3}}$: 

\begin{equation}
{\rm Y_{\rm 3}} = \frac{3\,{\rm y_{\rm 3}}\,(1 - {\rm Z})}{1 + 4{\rm y}} < \nexpo{9.6}{-5},
\label{eq:3he}
\end{equation}
where ${\rm y_{\rm 3}}$ is the \he3\ abundance ratio upper limit, \her3,
y is the \he4\ abundance ratio, \her4\,=\,0.1 \citep{2013N3242ion}, and
Z is the heavy element mass fraction.
(Details of this conversion of number density to mass fraction are given
  in Appendix~\ref{appen:C}.) 

Our \he3\ mass fraction upper limit for \ngc{3242} is compared in
Figure~\ref{fig:Y3yields} with the \citet{2011Lagarde} stellar evolution
models.  The model ${\rm Y_{\rm 3}}$ yields are adopted from their
Figure~9. The models shown have a range of metallicities, Z, and employ
either standard mixing or a combination of thermohaline and rotational
mixing.  Since we derive here only a notional upper limit for the \ngc{6543}
\her3\ abundance, its ${\rm Y_{\rm 3}}$ limit is unreliable to include
in Figure~\ref{fig:Y3yields}.

It is clear that this \ngc{3242} \her3\ upper limit is inconsistent with
standard stellar production of \he3.  The limit thus requires that some
type of extra mixing process operates in low-mass stars.  This is now in
conflict with \citet{2002Palla} who derive a \cratio\ abundance ratio
limit for \ngc{3242} that is consistent with standard stellar mixing
yields.  Their result is based on {\em Hubble Space Telescope}
observations of the C[III] multiplet near 1908\,\AA\ wavelength.

Thermohaline mixing is currently the best model that is broadly
consistent with our \he3\ observations, but open questions remain.  Our
upper limit on ${\rm Y_{\rm 3}}$ for \ngc{3242} is in fact {\em lower}
than all the \citet{2011Lagarde} model yields summarized in
Table~\ref{tab:he3yields} including thermohaline mixing.  One issue with
thermohaline mixing models is in determining a value for the coefficient
C, which is related to the aspect ratio of the mixing fingers
(length/width).  This is effectively a free parameter that must be
estimated. \citet{2010thermohaline} use C = 1000 which is adopted from
\citet{2007aCharZahn}.  This produces abundance ratios that are
consistent with observations in stars (e.g., their \cratio\ abundances).
But models using lower values of C will not solve the \he3\ problem
\citep[see][]{2010Cantiello}.  Hydrodynamic models are not consistent
with C\,$\sim$\,1000, but there are possible solutions
\citep[see][]{2011Denissenkov}.  Finally, the interplay between
thermohaline and rotation-induced mixing may be important
\citep[see][]{2013Maeder}.
\citet{2018Sengupta}, for example, studied the effects of rotation on
the thermohaline instability and found that rotation could in some cases
significantly increase its mixing efficiency.

\section{Summary}\label{sec:summary}

The \he3\ abundance in \hii\ regions and planetary nebulae provides
important constraints to Big Bang nucleosynthesis, stellar evolution,
and Galactic evolution.  Standard stellar evolution models that predict
the production of copious amounts of \he3\ in low-mass stars, consistent
with the high \her3\ abundance ratios previously reported for a few
planetary nebulae, are at odds with the approximately primordial values
determined for Galactic \hii\ regions.  This inconsistency is called the
``\he3\ Problem''.  Models that include mixing from the thermohaline
instability and rotation provide mechanisms that reduce the enhanced
\her3\ abundances dredged up during the RGB stage.  These models give
\he3\ yields that predict modest \he3\ production by stars over the
lifetime of the Milky Way and can reconcile the ``\he3\ Problem''.

Attempting to detect \hep3\ in PNe challenges the capabilities of all
existing radio spectrometer systems.  Systematic instrumental baseline
frequency structure compromises the ability of these spectrometer
systems to measure the wide, weak \hep3\ emission lines from PNe
accurately.  We previously reported detections of \hep3\ emission from
\ngc{3242} made independently with the MPIfR 100\m\ and NRAO 140 Foot
telescopes.  The GBT observations discussed here do not confirm these
results.  Furthermore, our analysis calls into question all previously
reported detections of \hep3\ emission from planetary nebulae, save
perhaps for J320.

We derive {\em upper limits} for the \her3\ abundance by number of
\nexpo{<4.5}{-5} and \nexpo{<7.3}{-5} for the planetary nebulae
\ngc{3242} and \ngc{6543}, respectively. These are well below previously
reported determinations and also well below the abundances predicted by
standard stellar evolution theory.  In fact, they approach the
abundances found for the ``The 3-Helium Plateau'' defined by
\hii\ region \her3\ abundances.

\acknowledgments

We dedicate this paper to our late colleague Bob Rood whose seminal 1976
paper with Gary Steigman and Beatrice Tinsley led to the formation of
our \he3\ research team. We thank the GBT telescope operators whose
diligence and expertise in executing our observing scripts during
Semester 11A were exceptional.  Bruce Balick and Nazar Budaiev helped
prepare some figures.  This research was partially supported by NSF
award AST-1714688 to TMB.

\vspace{5mm}
\facilities{GBT}
\software{TMBIDL \citep{TMBIDL}, NEBULA \citep{NEBULA} }

\bibliography{bania}

\begin{thebibliography}{}
\expandafter\ifx\csname natexlab\endcsname\relax\def\natexlab#1{#1}\fi

\bibitem[{{Asplund} {et~al.}(2006{\natexlab{a}}){Asplund}, {Grevesse}, \&
  {Jacques Sauval}}]{2006aAsplund}
{Asplund}, M., {Grevesse}, N., \& {Jacques Sauval}, A. 2006{\natexlab{a}},
  \nphysa, 777, 1

\bibitem[{{Asplund} {et~al.}(2006{\natexlab{b}}){Asplund}, {Grevesse}, \&
  {Sauval}}]{2006bAsplund}
{Asplund}, M., {Grevesse}, N., \& {Sauval}, A.~J. 2006{\natexlab{b}},
  Communications in Asteroseismology, 147, 76

\bibitem[{{Balick}(1987)}]{1987Balick}
{Balick}, B. 1987, \aj, 94, 671

\bibitem[{{Balick} \& {Hajian}(2004)}]{2004Balick}
{Balick}, B., \& {Hajian}, A.~R. 2004, \aj, 127, 2269

\bibitem[{{Balser}(1995)}]{1995PhDBalser}
{Balser}, D.~S. 1995, PhD thesis, Boston University

\bibitem[{{Balser} \& {Bania}(2018{\natexlab{a}})}]{BB2018}
{Balser}, D.~S., \& {Bania}, T.~M. 2018{\natexlab{a}}, \aj, 156, 280

\bibitem[{{Balser} \& {Bania}(2018{\natexlab{b}})}]{NEBULA}
---. 2018{\natexlab{b}}, {NEBULA: Radiative transfer code of ionized nebulae at
  radio wavelengths}, , , ascl:1809.009

\bibitem[{{Balser} {et~al.}(1994){Balser}, {Bania}, {Brockway}, {Rood}, \&
  {Wilson}}]{1994Balser}
{Balser}, D.~S., {Bania}, T.~M., {Brockway}, C.~J., {Rood}, R.~T., \& {Wilson},
  T.~L. 1994, \apj, 430, 667

\bibitem[{{Balser} {et~al.}(1997){Balser}, {Bania}, {Rood}, \&
  {Wilson}}]{1997PNe}
{Balser}, D.~S., {Bania}, T.~M., {Rood}, R.~T., \& {Wilson}, T.~L. 1997, \apj,
  483, 320

\bibitem[{{Balser} {et~al.}(1999{\natexlab{a}}){Balser}, {Bania}, {Rood}, \&
  {Wilson}}]{1999He3abundance}
---. 1999{\natexlab{a}}, \apj, 510, 759

\bibitem[{{Balser} {et~al.}(2006){Balser}, {Goss}, {Bania}, \&
  {Rood}}]{2006J320}
{Balser}, D.~S., {Goss}, W.~M., {Bania}, T.~M., \& {Rood}, R.~T. 2006, \apj,
  640, 360

\bibitem[{{Balser} {et~al.}(1999{\natexlab{b}}){Balser}, {Rood}, \&
  {Bania}}]{1999n3242}
{Balser}, D.~S., {Rood}, R.~T., \& {Bania}, T.~M. 1999{\natexlab{b}}, \apjl,
  522, L73

\bibitem[{{Bania} {et~al.}(2016){Bania}, {Wenger}, {Balser}, \&
  {Anderson}}]{TMBIDL}
{Bania}, T., {Wenger}, T., {Balser}, D., \& {Anderson}, L. 2016, {TMBIDL:
  Single dish radio astronomy data reduction package}, , , ascl:1605.005

\bibitem[{{Bania} {et~al.}(2002){Bania}, {Rood}, \& {Balser}}]{2002Nature}
{Bania}, T.~M., {Rood}, R.~T., \& {Balser}, D.~S. 2002, \nat, 415, 54

\bibitem[{{Bania} {et~al.}(2007){Bania}, {Rood}, \& {Balser}}]{2007bozo}
---. 2007, \ssr, 130, 53

\bibitem[{{Barker}(1985)}]{1985Barker}
{Barker}, T. 1985, \apj, 294, 193

\bibitem[{{Bernard-Salas} {et~al.}(2003){Bernard-Salas}, {Pottasch},
  {Wesselius}, \& {Feibelman}}]{2003Bernard}
{Bernard-Salas}, J., {Pottasch}, S.~R., {Wesselius}, P.~R., \& {Feibelman},
  W.~A. 2003, \aap, 406, 165

\bibitem[{{Cantiello} \& {Langer}(2010)}]{2010Cantiello}
{Cantiello}, M., \& {Langer}, N. 2010, \aap, 521, A9

\bibitem[{{Charbonnel}(1995)}]{1995Charbonnel}
{Charbonnel}, C. 1995, \apjl, 453, L41

\bibitem[{{Charbonnel} \& {Lagarde}(2010)}]{2010thermohaline}
{Charbonnel}, C., \& {Lagarde}, N. 2010, \aap, 522, A10

\bibitem[{{Charbonnel} \& {Zahn}(2007{\natexlab{a}})}]{2007bCharZahn}
{Charbonnel}, C., \& {Zahn}, J.~P. 2007{\natexlab{a}}, \aap, 476, L29

\bibitem[{{Charbonnel} \& {Zahn}(2007{\natexlab{b}})}]{2007aCharZahn}
---. 2007{\natexlab{b}}, \aap, 467, L15

\bibitem[{{Chiappini} {et~al.}(2002){Chiappini}, {Renda}, \&
  {Matteucci}}]{2002Chiappini}
{Chiappini}, C., {Renda}, A., \& {Matteucci}, F. 2002, \aap, 395, 789

\bibitem[{{Cyburt}(2004)}]{2004bbns}
{Cyburt}, R.~H. 2004, \prd, 70, 023505

\bibitem[{{Denissenkov} \& {Merryfield}(2011)}]{2011Denissenkov}
{Denissenkov}, P.~A., \& {Merryfield}, W.~J. 2011, \apjl, 727, L8

\bibitem[{{Faulkner}(1970)}]{1970PNeEject}
{Faulkner}, D.~J. 1970, \apj, 162, 513

\bibitem[{{Fisher} {et~al.}(2003){Fisher}, {Norrod}, \& {Balser}}]{Baselines}
{Fisher}, J.~R., {Norrod}, R.~D., \& {Balser}, D.~S. 2003, {NRAO Electronics
  Division Internal Rep. No. 312}, ,

\bibitem[{{Galli} {et~al.}(1995){Galli}, {Palla}, {Ferrini}, \&
  {Penco}}]{1995Galli}
{Galli}, D., {Palla}, F., {Ferrini}, F., \& {Penco}, U. 1995, \apj, 443, 536

\bibitem[{{Galli} {et~al.}(1997){Galli}, {Stanghellini}, {Tosi}, \&
  {Palla}}]{1997Galli}
{Galli}, D., {Stanghellini}, L., {Tosi}, M., \& {Palla}, F. 1997, \apj, 477,
  218

\bibitem[{{Geiss}(1993)}]{1993Geiss}
{Geiss}, J. 1993, in Origin and Evolution of the Elements, ed. N.~{Prantzos},
  E.~{Vangioni-Flam}, \& M.~{Casse}, 89--106

\bibitem[{{Gigho} {et~al.}(2001){Gigho}, {Maddalena}, {Balser}, \&
  {Langston}}]{2001Gigho}
{Gigho}, F., {Maddalena}, R., {Balser}, D.~S., \& {Langston}, G. 2001, {GBT
  Commissioning Memo 10}, ,

\bibitem[{{Gloeckler} \& {Geiss}(1996)}]{1996LISM}
{Gloeckler}, G., \& {Geiss}, J. 1996, \nat, 381, 210

\bibitem[{{Greisen} {et~al.}(2006){Greisen}, {Calabretta}, {Valdes}, \&
  {Allen}}]{2006Vopt}
{Greisen}, E.~W., {Calabretta}, M.~R., {Valdes}, F.~G., \& {Allen}, S.~L. 2006,
  \aap, 446, 747

\bibitem[{{Guzman-Ramirez} {et~al.}(2016){Guzman-Ramirez}, {Rizzo}, {Zijlstra},
  {Garc{\'\i}a-Mir{\'o}}, {Morisset}, \& {Gray}}]{2016IC418}
{Guzman-Ramirez}, L., {Rizzo}, J.~R., {Zijlstra}, A.~A., {et~al.} 2016, \mnras,
  460, L35

\bibitem[{{Hyung} {et~al.}(2001){Hyung}, {Mellema}, {Lee}, \&
  {Kim}}]{2001Hyung}
{Hyung}, S., {Mellema}, G., {Lee}, S.~J., \& {Kim}, H. 2001, \aap, 378, 587

\bibitem[{{Iben}(1967{\natexlab{a}})}]{1967aIben}
{Iben}, Icko, J. 1967{\natexlab{a}}, \apj, 147, 650

\bibitem[{{Iben}(1967{\natexlab{b}})}]{1967bIben}
---. 1967{\natexlab{b}}, \apj, 147, 624

\bibitem[{{Lagarde} {et~al.}(2011){Lagarde}, {Charbonnel}, {Decressin}, \&
  {Hagelberg}}]{2011Lagarde}
{Lagarde}, N., {Charbonnel}, C., {Decressin}, T., \& {Hagelberg}, J. 2011,
  \aap, 536, A28

\bibitem[{{Lagarde} {et~al.}(2012){Lagarde}, {Romano}, {Charbonnel}, {Tosi},
  {Chiappini}, \& {Matteucci}}]{2012Lagarde}
{Lagarde}, N., {Romano}, D., {Charbonnel}, C., {et~al.} 2012, \aap, 542, A62

\bibitem[{{Maeder} {et~al.}(2013){Maeder}, {Meynet}, {Lagarde}, \&
  {Charbonnel}}]{2013Maeder}
{Maeder}, A., {Meynet}, G., {Lagarde}, N., \& {Charbonnel}, C. 2013, \aap, 553,
  A1

\bibitem[{{Markwardt}(2009)}]{2009Markwardt}
{Markwardt}, C.~B. 2009, in Astronomical Society of the Pacific Conference
  Series, Vol. 411, Astronomical Data Analysis Software and Systems XVIII, ed.
  D.~A. {Bohlender}, D.~{Durand}, \& P.~{Dowler}, 251

\bibitem[{{Menzel}(1968)}]{1968Menzel}
{Menzel}, D.~H. 1968, \nat, 218, 756

\bibitem[{{Middlemass} {et~al.}(1989){Middlemass}, {Clegg}, \&
  {Walsh}}]{1989Middlemass}
{Middlemass}, D., {Clegg}, R.~E.~S., \& {Walsh}, J.~R. 1989, \mnras, 239, 1

\bibitem[{{Monteiro} {et~al.}(2013){Monteiro}, {Gon{\c{c}}alves},
  {Leal-Ferreira}, \& {Corradi}}]{2013N3242ion}
{Monteiro}, H., {Gon{\c{c}}alves}, D.~R., {Leal-Ferreira}, M.~L., \& {Corradi},
  R.~L.~M. 2013, \aap, 560, A102

\bibitem[{{Osterbrock} \& {Ferland}(2006)}]{osterbrock}
{Osterbrock}, D.~E., \& {Ferland}, G.~J. 2006, {Astrophysics of Gaseous Nebulae
  and Active Galactic Nuclei}, 2nd edn. (Mill Valley, California: University
  Science Books)

\bibitem[{{Palla} {et~al.}(2000){Palla}, {Bachiller}, {Stanghellini}, {Tosi},
  \& {Galli}}]{2000Palla}
{Palla}, F., {Bachiller}, R., {Stanghellini}, L., {Tosi}, M., \& {Galli}, D.
  2000, \aap, 355, 69

\bibitem[{{Palla} {et~al.}(2002){Palla}, {Galli}, {Marconi}, {Stanghellini}, \&
  {Tosi}}]{2002Palla}
{Palla}, F., {Galli}, D., {Marconi}, A., {Stanghellini}, L., \& {Tosi}, M.
  2002, \apjl, 568, L57

\bibitem[{{Peng} {et~al.}(2000){Peng}, {Kraus}, {Krichbaum}, \&
  {Witzel}}]{2000Peng}
{Peng}, B., {Kraus}, A., {Krichbaum}, T.~P., \& {Witzel}, A. 2000, \aaps, 145,
  1

\bibitem[{{Perley} {et~al.}(2011){Perley}, {Chandler}, {Butler}, \&
  {Wrobel}}]{2011jvla}
{Perley}, R.~A., {Chandler}, C.~J., {Butler}, B.~J., \& {Wrobel}, J.~M. 2011,
  \apjl, 739, L1

\bibitem[{{Phillips} {et~al.}(2009){Phillips}, {Ramos-Larios}, {Schr{\"o}der},
  \& {Contreras}}]{2009Phillips}
{Phillips}, J.~P., {Ramos-Larios}, G., {Schr{\"o}der}, K.~P., \& {Contreras},
  J.~L.~V. 2009, \mnras, 399, 1126

\bibitem[{{Ramos-Larios} {et~al.}(2016){Ramos-Larios}, {Santamar{\'\i}a},
  {Guerrero}, {Marquez-Lugo}, {Sabin}, \& {Toal{\'a}}}]{2016PNrings}
{Ramos-Larios}, G., {Santamar{\'\i}a}, E., {Guerrero}, M.~A., {et~al.} 2016,
  \mnras, 462, 610

\bibitem[{{Romano} {et~al.}(2003){Romano}, {Tosi}, {Matteucci}, \&
  {Chiappini}}]{2003RTMC}
{Romano}, D., {Tosi}, M., {Matteucci}, F., \& {Chiappini}, C. 2003, \mnras,
  346, 295

\bibitem[{{Rood}(1972)}]{1972Rood}
{Rood}, R.~T. 1972, \apj, 177, 681

\bibitem[{{Rood} {et~al.}(1984){Rood}, {Bania}, \& {Wilson}}]{1984RBW}
{Rood}, R.~T., {Bania}, T.~M., \& {Wilson}, T.~L. 1984, \apj, 280, 629

\bibitem[{{Rood} {et~al.}(1992){Rood}, {Bania}, \& {Wilson}}]{1992Nature}
---. 1992, \nat, 355, 618

\bibitem[{{Rood} {et~al.}(1976){Rood}, {Steigman}, \& {Tinsley}}]{RST1976}
{Rood}, R.~T., {Steigman}, G., \& {Tinsley}, B.~M. 1976, \apjl, 207, L57

\bibitem[{{Rood} {et~al.}(1979){Rood}, {Wilson}, \& {Steigman}}]{1979RWS}
{Rood}, R.~T., {Wilson}, T.~L., \& {Steigman}, G. 1979, \apjl, 227, L97

\bibitem[{{Sch{\"o}nberner} {et~al.}(2018){Sch{\"o}nberner}, {Balick}, \&
  {Jacob}}]{2018SBJ}
{Sch{\"o}nberner}, D., {Balick}, B., \& {Jacob}, R. 2018, \aap, 609, A126

\bibitem[{{Sch{\"o}nberner} {et~al.}(2014){Sch{\"o}nberner}, {Jacob},
  {Lehmann}, {Hildebrand t}, {Steffen}, {Zwanzig}, {Sandin}, \&
  {Corradi}}]{2014Schonberner}
{Sch{\"o}nberner}, D., {Jacob}, R., {Lehmann}, H., {et~al.} 2014, Astronomische
  Nachrichten, 335, 378

\bibitem[{{Sch{\"o}nberner} \& {Steffen}(2019)}]{2019PNeDist}
{Sch{\"o}nberner}, D., \& {Steffen}, M. 2019, \aap, 625, A137

\bibitem[{{Sengupta} \& {Garaud}(2018)}]{2018Sengupta}
{Sengupta}, S., \& {Garaud}, P. 2018, \apj, 862, 136

\bibitem[{{Tosi}(1998)}]{1998Tosi}
{Tosi}, M. 1998, \ssr, 84, 207

\bibitem[{{von Proch{\'a}zka} {et~al.}(2010){von Proch{\'a}zka}, {Remijan},
  {Balser}, {Ryans}, {Marshall}, {Schwab}, {Hollis}, {Jewell}, \&
  {Lovas}}]{2010linewidth}
{von Proch{\'a}zka}, A.~A., {Remijan}, A.~J., {Balser}, D.~S., {et~al.} 2010,
  \pasp, 122, 354

\bibitem[{{Weinberger}(1989)}]{1989PNeExp}
{Weinberger}, R. 1989, \aaps, 78, 301

\bibitem[{{Wenger} {et~al.}(2019{\natexlab{a}}){Wenger}, {Balser}, {Anderson},
  \& {Bania}}]{2019WengerTe}
{Wenger}, T.~V., {Balser}, D.~S., {Anderson}, L.~D., \& {Bania}, T.~M.
  2019{\natexlab{a}}, \apj, 887, 114

\bibitem[{{Wenger} {et~al.}(2019{\natexlab{b}}){Wenger}, {Dickey}, {Jordan},
  {Balser}, {Armentrout}, {Anderson}, {Bania}, {Dawson}, {McClure-Griffiths},
  \& {Shea}}]{2019shrds}
{Wenger}, T.~V., {Dickey}, J.~M., {Jordan}, C.~H., {et~al.} 2019{\natexlab{b}},
  \apjs, 240, 24

\bibitem[{{Wilson} \& {Rood}(1994)}]{1994WR}
{Wilson}, T.~L., \& {Rood}, R. 1994, \araa, 32, 191

\end{thebibliography}

\clearpage
\appendix 

\section{Planetary Nebula Radio Recombination Line Parameters}\label{appen:A}

The radio recombination line parameters for the sample of Galactic
planetary nebulae measured using Gaussian fits to observations made with
the Green Bank Telescope near 8665\ghz. Compiled here for each nebula is
the RRL transition, the change in principle quantum number, $\Delta$N,
the line intensity, $T_{\rm L}$ in \mk, and FWHM line width, $\Delta$V
in \kms, together with the RMS noise in \mk\ and integration time,
$t_{\rm intg}$ in hr, of the spectral band containing the transition.
Also listed are the 1$\sigma$ errors of the Gaussian fits to the
intensity, $\sigma T_{\rm L}$, and line width, $\sigma \Delta V$. Some
transitions (H114a/H114b, H130a/H130b and H144b/H144g) appeared in
multiple spectral bands. These transitions were fit independently and
are listed here for completeness.

\begin{deluxetable}{lcrccccr}
\tablecaption{\ngc{3242} Recombination Line Parameters \label{tab:n3242}}
\tablewidth{0pt}
\tablehead{
  \colhead{Transition} & \colhead{$\Delta$N} &
  \colhead{$T_{\rm L}$} & \colhead{$\sigma\,T_{\rm L}$} & 
  \colhead{$\Delta{V}$} & \colhead{$\sigma\,\Delta{V}$} & 
  \colhead{rms} &  \colhead{$t_{\rm intg}$} \\ 
  \colhead{} & \colhead{} & \colhead{(mK)} & \colhead{(mK)} & 
  \colhead{(\kms)} & \colhead{(\kms)} &
  \colhead{(mK)} & \colhead{(hr)}
} 
\startdata 
H91    & 1 &  37.84 &  0.11 & 43.0 &  0.15 & 0.117 & 306.7 \\
He91   & 1 &   4.49 &  0.12 & 40.1 &  1.30 & 0.117 & 306.7 \\
H92    & 1 &  37.05 &  0.12 & 44.5 &  0.17 & 0.313 &  93.2 \\
He92   & 1 &   4.12 &  0.13 & 41.8 &  1.60 & 0.313 &  93.2 \\
H114   & 2 &  10.86 &  0.04 & 44.6 &  0.19 & 0.150 & 308.0 \\
He114  & 2 &   1.13 &  0.04 & 41.3 &  1.73 & 0.150 & 308.0 \\
H115   & 2 &  10.58 &  0.04 & 45.1 &  0.18 & 0.156 & 305.0 \\
He115  & 2 &   1.02 &  0.04 & 42.2 &  2.09 & 0.156 & 305.0 \\
H130   & 3 &   4.80 &  0.02 & 43.3 &  0.24 & 0.116 & 308.0 \\
He130  & 3 &   0.43 &  0.02 & 54.3 &  3.63 & 0.116 & 308.0 \\
H130a  & 3 &   5.15 &  0.04 & 45.4 &  0.42 & 0.238 &  91.7 \\
H130b  & 3 &   5.24 &  0.03 & 45.0 &  0.32 & 0.228 &  94.4 \\
H131   & 3 &   4.47 &  0.02 & 42.0 &  0.26 & 0.159 & 286.1 \\
He131  & 3 &   0.42 &  0.03 & 39.4 &  2.91 & 0.159 & 286.1 \\
H132   & 3 &   4.10 &  0.03 & 43.6 &  0.43 & 0.251 &  91.8 \\
He132  & 3 &   0.43 &  0.04 & 30.2 &  3.38 & 0.251 &  91.8 \\
H144b  & 4 &   2.30 &  0.02 & 42.4 &  0.38 & 0.116 & 305.0 \\
He144b & 4 &   0.31 &  0.02 & 41.1 &  4.01 & 0.116 & 305.0 \\
H144g  & 4 &   2.33 &  0.02 & 43.8 &  0.34 & 0.107 & 286.1 \\
He144g & 4 &   0.46 &  0.02 & 28.6 &  1.44 & 0.107 & 286.1 \\
H145   & 4 &   2.55 &  0.04 & 47.4 &  0.87 & 0.264 &  91.8 \\
He145  & 4 &   0.80 &  0.04 & 51.4 &  3.18 & 0.264 &  91.8 \\
H152   & 5 &   1.50 &  0.01 & 50.9 &  0.60 & 0.117 & 208.7 \\
H154   & 5 &   0.84 &  0.16 & 22.9 &  4.90 & 0.117 & 306.7 \\
H155   & 5 &   1.74 &  0.01 & 45.3 &  0.45 & 0.175 & 305.0 \\
H156   & 5 &   1.75 &  0.03 & 47.0 &  1.20 & 0.263 &  91.8 \\
H164   & 6 &   0.77 &  0.01 & 45.9 &  1.14 & 0.144 & 286.1 \\
H165   & 6 &   0.86 &  0.04 & 51.0 &  2.81 & 0.231 &  93.2 \\
H171   & 7 &   0.75 &  0.02 & 57.9 &  1.52 & 0.145 & 301.7 \\
\enddata 
\end{deluxetable}

\newpage
\begin{deluxetable}{lcrccccr}
\tablecaption{\ngc{6543} Recombination Line Parameters \label{tab:n6543}}
\tablewidth{0pt}
\tablehead{
  \colhead{Transition} & \colhead{$\Delta$N} &
  \colhead{$T_{\rm L}$} & \colhead{$\sigma\,T_{\rm L}$} & 
  \colhead{$\Delta{V}$} & \colhead{$\sigma\,\Delta{V}$} & 
  \colhead{rms} &  \colhead{$t_{\rm intg}$} \\ 
  \colhead{} & \colhead{} & \colhead{(mK)} & \colhead{(mK)} & 
  \colhead{(\kms)} & \colhead{(\kms)} &
  \colhead{(mK)} & \colhead{(hr)}
} 
\startdata 
H91    & 1 &  92.08 &  0.06 & 37.6 &  0.03 & 0.146 & 258.1 \\
He91   & 1 &  11.90 &  0.06 & 35.3 &  0.20 & 0.146 & 258.1 \\
H92    & 1 &  90.22 &  0.06 & 37.6 &  0.03 & 0.326 &  65.7 \\
He92   & 1 &  11.27 &  0.06 & 34.9 &  0.22 & 0.326 &  65.7 \\
H114   & 2 &  24.72 &  0.04 & 38.1 &  0.07 & 0.151 & 263.5 \\
He114  & 2 &   3.01 &  0.04 & 35.3 &  0.59 & 0.151 & 263.5 \\
H115   & 2 &  23.77 &  0.02 & 38.9 &  0.05 & 0.173 & 257.3 \\
He115  & 2 &   2.63 &  0.03 & 31.4 &  0.38 & 0.173 & 257.3 \\
H130   & 3 &  10.47 &  0.02 & 40.1 &  0.10 & 0.168 & 263.9 \\
He130  & 3 &   1.37 &  0.03 & 31.5 &  0.68 & 0.168 & 263.9 \\
H130a  & 3 &   9.90 &  0.03 & 37.4 &  0.14 & 0.278 &  65.9 \\
He130a & 3 &   1.48 &  0.04 & 31.4 &  0.99 & 0.219 &  65.9 \\
H130b  & 3 &   9.99 &  0.04 & 38.3 &  0.18 & 0.311 &  65.1 \\
He130b & 3 &   1.36 &  0.05 & 31.9 &  1.41 & 0.311 &  65.1 \\
H131   & 3 &  10.17 &  0.02 & 41.7 &  0.08 & 0.139 & 236.2 \\
He131  & 3 &   1.09 &  0.02 & 36.1 &  0.73 & 0.139 & 236.2 \\
H132   & 3 &   9.91 &  0.06 & 39.4 &  0.26 & 0.386 &  53.4 \\
He132  & 3 &   0.96 &  0.07 & 31.3 &  2.86 & 0.386 &  53.4 \\
H144b  & 4 &   4.86 &  0.02 & 38.3 &  0.18 & 0.166 & 257.3 \\
He144b & 4 &   0.60 &  0.02 & 40.5 &  1.65 & 0.166 & 257.3 \\
H144g  & 4 &   4.98 &  0.02 & 40.8 &  0.15 & 0.166 & 236.2 \\
He144g & 4 &   0.42 &  0.02 & 37.0 &  1.89 & 0.166 & 236.2 \\
H145   & 4 &   3.61 &  0.08 & 30.2 &  0.82 & 0.403 &  53.4 \\
H152   & 5 &   3.08 &  0.02 & 47.4 &  0.37 & 0.158 & 191.6 \\
H154   & 5 &   2.07 &  0.06 & 30.4 &  1.05 & 0.146 & 258.1 \\
H155   & 5 &   2.51 &  0.02 & 41.4 &  0.35 & 0.192 & 257.3 \\
H156   & 5 &   1.76 &  0.06 & 31.9 &  1.27 & 0.307 &  53.4 \\
H164   & 6 &   1.13 &  0.03 & 31.4 &  0.91 & 0.188 & 236.2 \\
H165   & 6 &   1.35 &  0.05 & 38.2 &  1.66 & 0.315 &  65.7 \\
H171   & 7 &   1.13 &  0.01 & 48.4 &  0.78 & 0.112 & 256.1 \\
\enddata 
\end{deluxetable}

\newpage
\begin{deluxetable}{lcrccccr}
\tablecaption{\ngc{6826} Recombination Line Parameters \label{tab:n6826}}
\tablewidth{0pt}
\tablehead{
  \colhead{Transition} & \colhead{$\Delta$N} &
  \colhead{$T_{\rm L}$} & \colhead{$\sigma\,T_{\rm L}$} & 
  \colhead{$\Delta{V}$} & \colhead{$\sigma\,\Delta{V}$} & 
  \colhead{rms} &  \colhead{$t_{\rm intg}$} \\ 
  \colhead{} & \colhead{} & \colhead{(mK)} & \colhead{(mK)} & 
  \colhead{(\kms)} & \colhead{(\kms)} &
  \colhead{(mK)} & \colhead{(hr)}
} 
\startdata 
H91    & 1 &  37.18 &  0.04 & 33.8 &  0.04 & 0.255 &  51.6 \\
He91   & 1 &   4.48 &  0.04 & 30.1 &  0.33 & 0.255 &  51.6 \\
H92    & 1 &  36.53 &  0.04 & 34.3 &  0.04 & 0.394 &  51.1 \\
He92   & 1 &   4.40 &  0.05 & 27.3 &  0.34 & 0.394 &  51.1 \\
H114a  & 2 &  10.48 &  0.04 & 34.7 &  0.15 & 0.345 &  51.5 \\
He114a & 2 &   1.28 &  0.05 & 26.2 &  1.16 & 0.345 &  51.5 \\
H114b  & 2 &  10.58 &  0.04 & 35.3 &  0.16 & 0.331 &  51.4 \\
He114b & 2 &   1.09 &  0.06 & 18.5 &  1.10 & 0.331 &  51.4 \\
H115   & 2 &  10.06 &  0.04 & 37.1 &  0.19 & 0.296 &  51.4 \\
He115  & 2 &   1.15 &  0.06 & 23.1 &  1.47 & 0.296 &  51.4 \\
H130a  & 3 &   4.57 &  0.04 & 35.9 &  0.37 & 0.296 &  51.5 \\
H130b  & 3 &   4.72 &  0.04 & 36.6 &  0.34 & 0.292 &  51.4 \\
H131   & 3 &   3.81 &  0.04 & 38.8 &  0.47 & 0.350 &  48.1 \\
He131  & 3 &   0.37 &  0.04 & 51.2 &  5.83 & 0.350 &  48.1 \\
H132   & 3 &   4.05 &  0.05 & 37.0 &  0.56 & 0.342 &  43.1 \\
He132  & 3 &   0.28 &  0.07 & 61.7 & 11.95 & 0.342 &  43.1 \\
H144b  & 4 &   2.25 &  0.04 & 38.6 &  0.70 & 0.336 &  51.4 \\
H144g  & 4 &   2.24 &  0.04 & 37.0 &  0.72 & 0.351 &  48.1 \\
H154   & 5 &   1.41 &  0.04 & 34.5 &  1.25 & 0.255 &  51.6 \\
H155   & 5 &   1.36 &  0.04 & 43.3 &  1.35 & 0.304 &  51.4 \\
H156   & 5 &   1.78 &  0.05 & 40.4 &  1.61 & 0.315 &  43.1 \\
H164   & 6 &   1.33 &  0.04 & 34.2 &  1.27 & 0.366 &  48.1 \\
\enddata 
\end{deluxetable}

\newpage
\begin{deluxetable}{lcrccccr}
\tablecaption{\ngc{7009} Recombination Line Parameters \label{tab:n7009}}
\tablewidth{0pt}
\tablehead{
  \colhead{Transition} & \colhead{$\Delta$N} &
  \colhead{$T_{\rm L}$} & \colhead{$\sigma\,T_{\rm L}$} & 
  \colhead{$\Delta{V}$} & \colhead{$\sigma\,\Delta{V}$} & 
  \colhead{rms} &  \colhead{$t_{\rm intg}$} \\ 
  \colhead{} & \colhead{} & \colhead{(mK)} & \colhead{(mK)} & 
  \colhead{(\kms)} & \colhead{(\kms)} &
  \colhead{(mK)} & \colhead{(hr)}
} 
\startdata 
H91    & 1 &  50.24 &  0.18 & 44.9 &  0.18 & 0.232 &  47.2 \\
He91   & 1 &   5.00 &  0.18 & 44.3 &  2.11 & 0.232 &  47.2 \\
H92    & 1 &  50.70 &  0.22 & 44.7 &  0.22 & 0.398 &  46.8 \\
He92   & 1 &   5.83 &  0.22 & 43.1 &  1.98 & 0.398 &  46.8 \\
H114a  & 2 &  13.46 &  0.05 & 45.2 &  0.20 & 0.318 &  44.1 \\
He114a & 2 &   1.81 &  0.06 & 38.1 &  1.61 & 0.318 &  44.1 \\
H114b  & 2 &  13.54 &  0.06 & 45.5 &  0.24 & 0.312 &  43.9 \\
He114b & 2 &   1.95 &  0.07 & 41.1 &  1.96 & 0.312 &  43.9 \\
H115   & 2 &  13.15 &  0.06 & 47.3 &  0.24 & 0.336 &  47.2 \\
He115  & 2 &   1.21 &  0.06 & 38.2 &  2.46 & 0.336 &  47.2 \\
H130a  & 3 &   5.96 &  0.05 & 46.4 &  0.47 & 0.390 &  44.1 \\
He130a & 3 &   0.36 &  0.06 & 42.5 &  9.19 & 0.390 &  44.1 \\
H130b  & 3 &   6.09 &  0.04 & 45.6 &  0.34 & 0.358 &  43.9 \\
H131   & 3 &   5.96 &  0.05 & 44.5 &  0.41 & 0.411 &  45.4 \\
He131  & 3 &   1.06 &  0.06 & 28.3 &  1.84 & 0.411 &  45.4 \\
H132   & 3 &   5.09 &  0.04 & 47.5 &  0.48 & 0.447 &  39.7 \\
H144b  & 4 &   2.48 &  0.05 & 45.1 &  1.02 & 0.348 &  47.2 \\
H144g  & 4 &   2.30 &  0.05 & 41.1 &  1.03 & 0.402 &  45.4 \\
H145   & 4 &   2.51 &  0.05 & 57.3 &  1.54 & 0.789 &  39.7 \\
H154   & 5 &   0.78 &  0.21 & 30.7 &  9.68 & 0.232 &  47.2 \\
H155   & 5 &   1.54 &  0.04 & 47.1 &  1.39 & 0.402 &  47.2 \\
H156   & 5 &   1.67 &  0.07 & 46.4 &  2.66 & 0.352 &  39.7 \\
H164   & 6 &   1.21 &  0.08 & 21.7 &  1.56 & 0.304 &  45.4 \\
\enddata 
\end{deluxetable}

\clearpage

\section{{\em NEBULA} Model Spectra for Radio Recombination Lines}\label{appen:NEBULA}

Representative model radio recombination line spectra for the planetary
nebulae \ngc{3242} and \ngc{6543} compared with the GBT
observations. These NEBULA models are described in
Section~\ref{sec:PN3He} and their properties summarized in
Tables~\ref{tab:n3242model} and \ref{tab:n6543model}, respectively. The
black circles are the observed spectra whereas the red lines are the
NEBULA model spectra.  The residuals between the data and the models are
shown in the plots below the spectra.  The recombination transitions
present in each spectral band are flagged with vertical lines.

\begin{figure}[!h]
\centering 
\includegraphics[angle=0,scale=0.40]{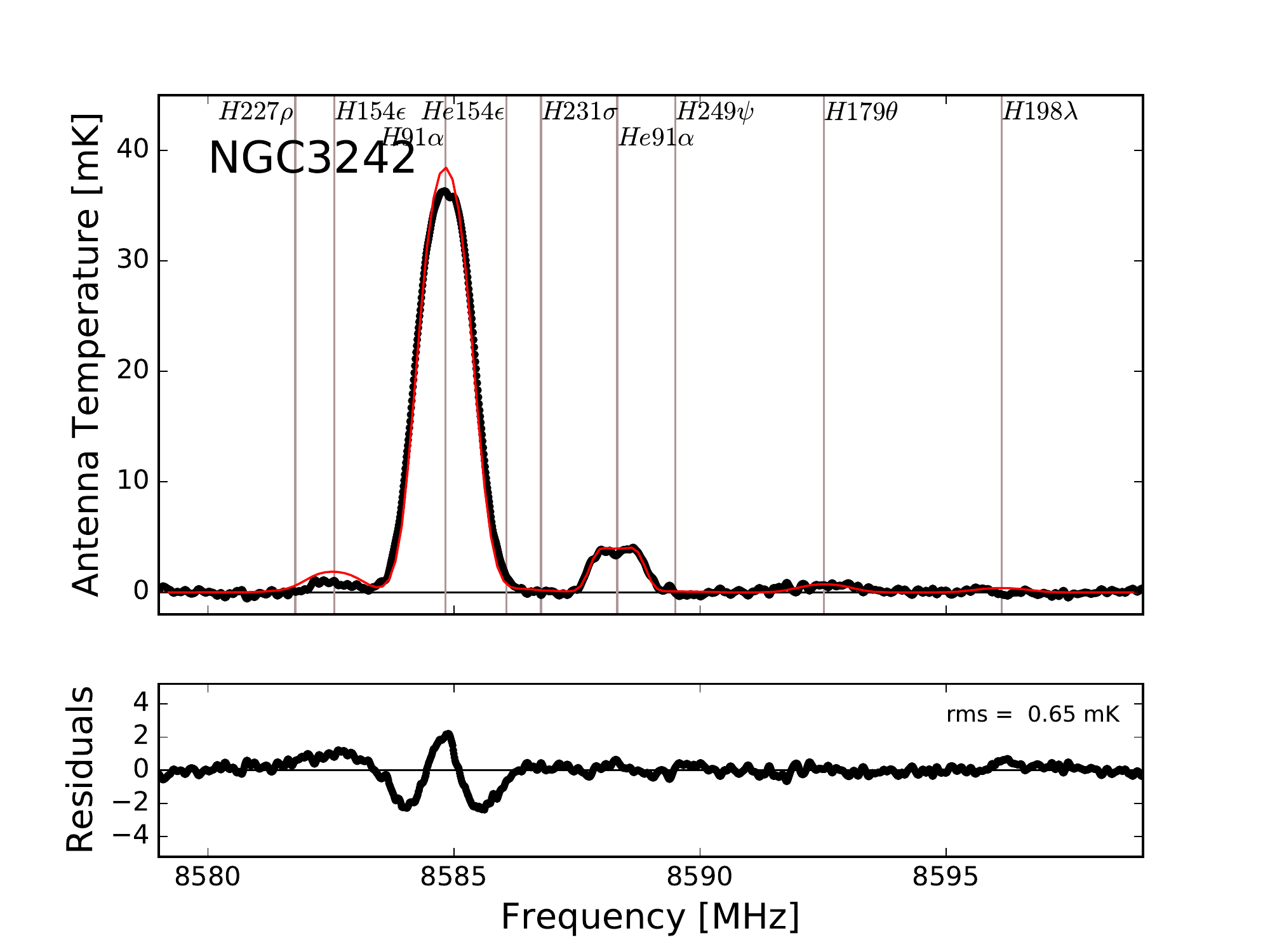}
\includegraphics[angle=0,scale=0.40]{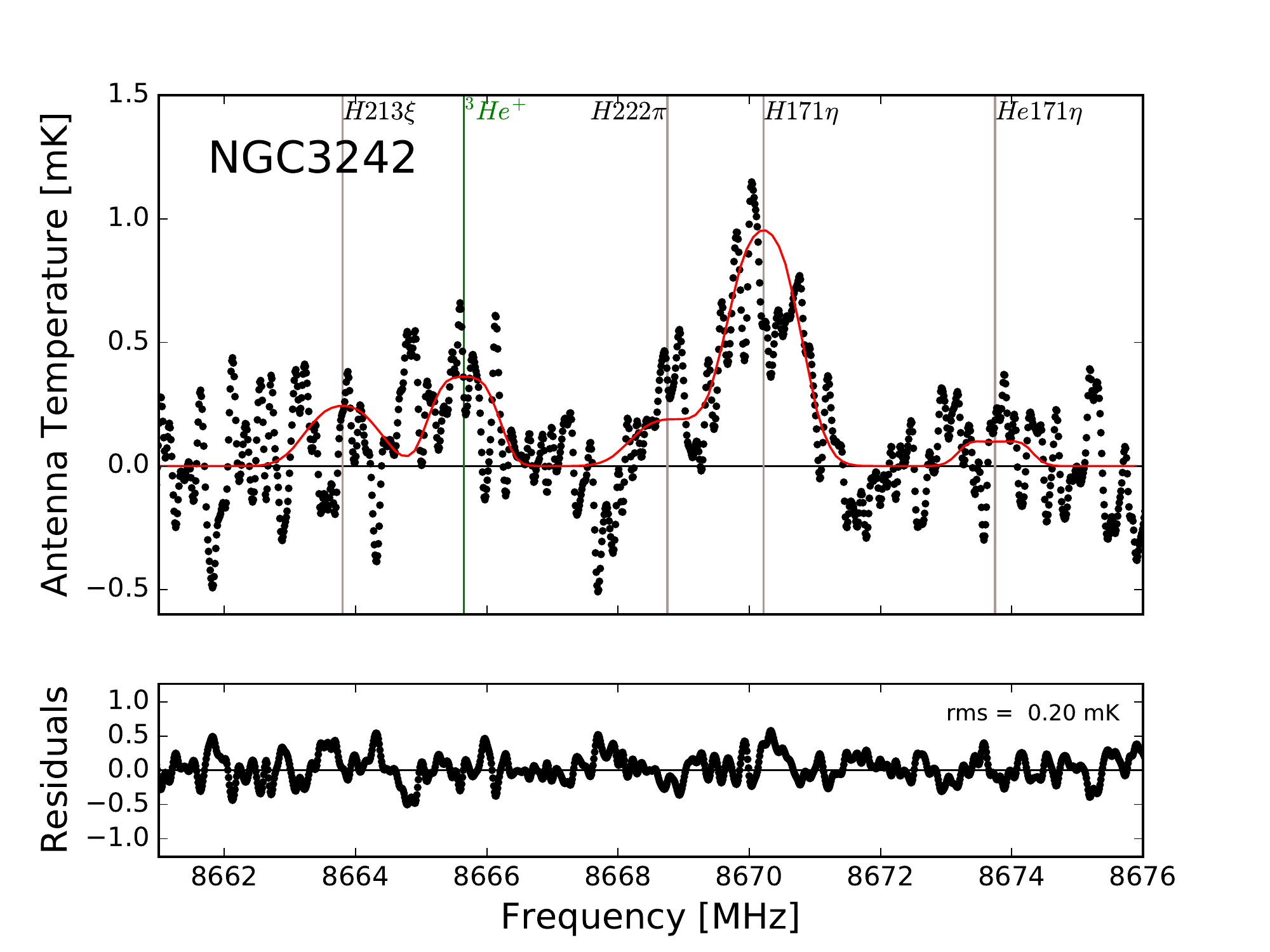}\\


\includegraphics[angle=0,scale=0.40]{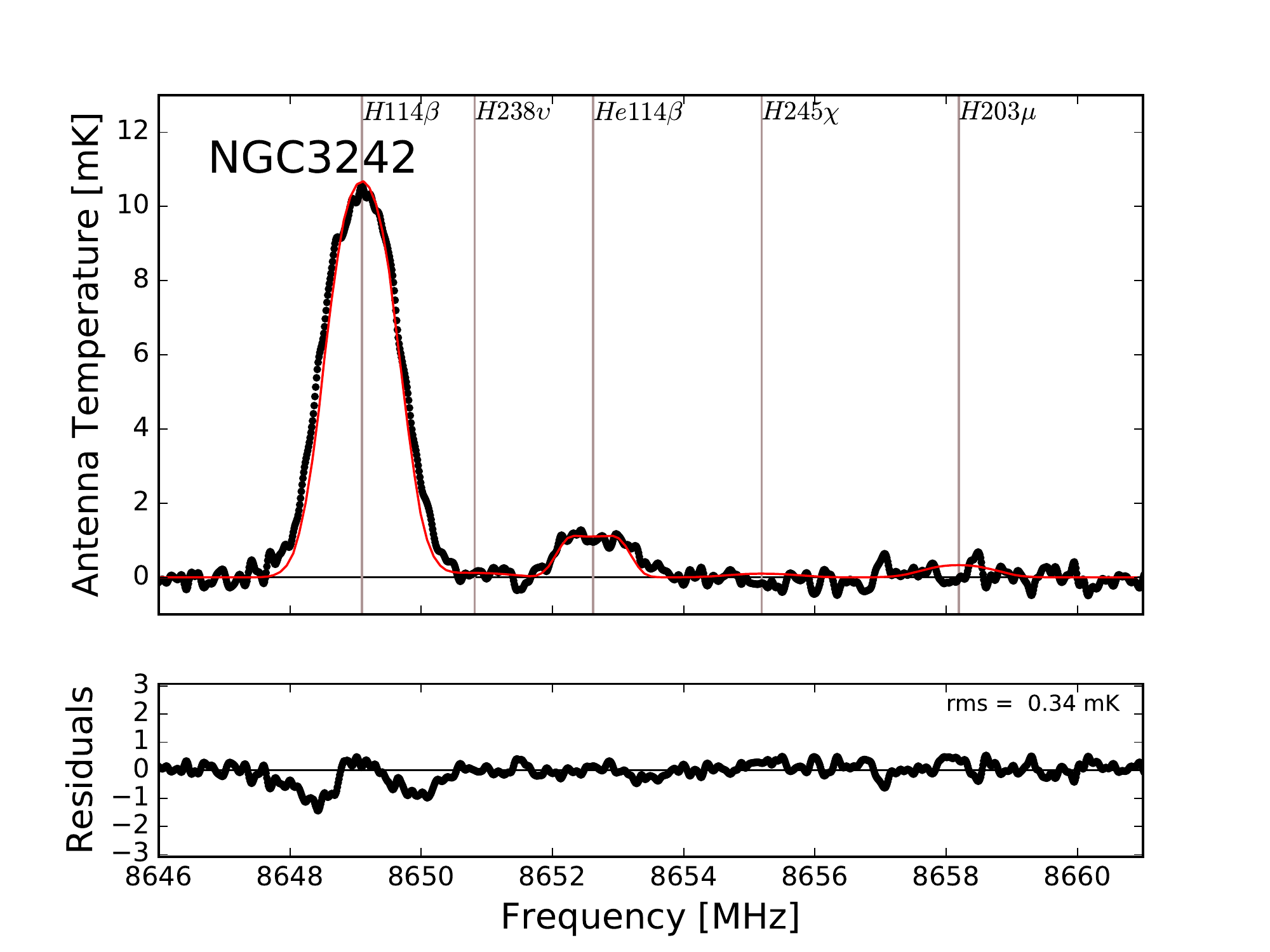}
\includegraphics[angle=0,scale=0.40]{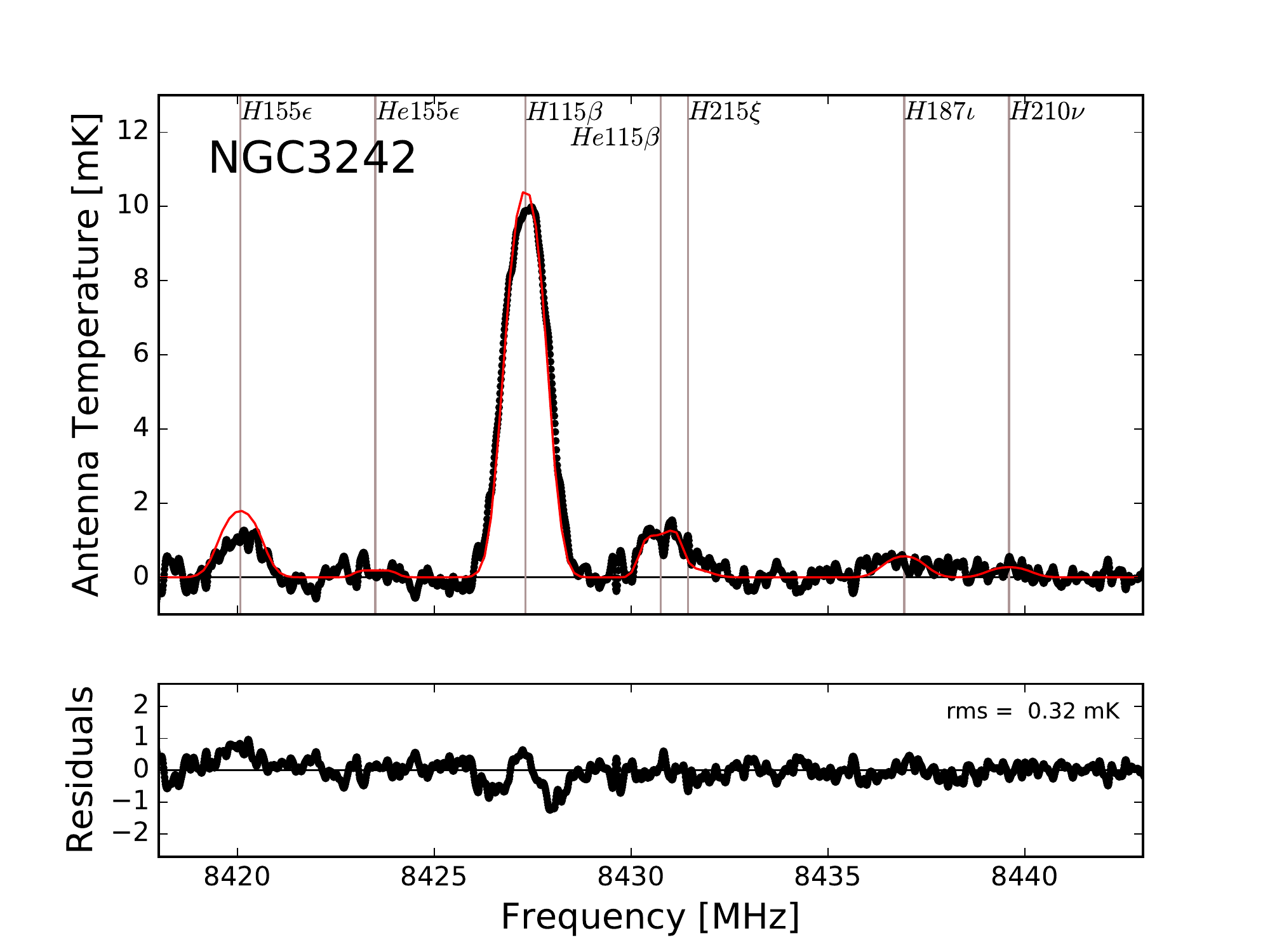}\\


\caption{\ngc{3242} NEBULA models. Starting at the top left the
  following spectral bands were modeled: \halpha, \hep3, \hbeta, 
  and \hbbeta.
}
\label{fig:n3242modelA}
\end{figure}

\begin{figure}[!h]
\centering \vspace{-1.0cm}
\includegraphics[angle=0,scale=0.40]{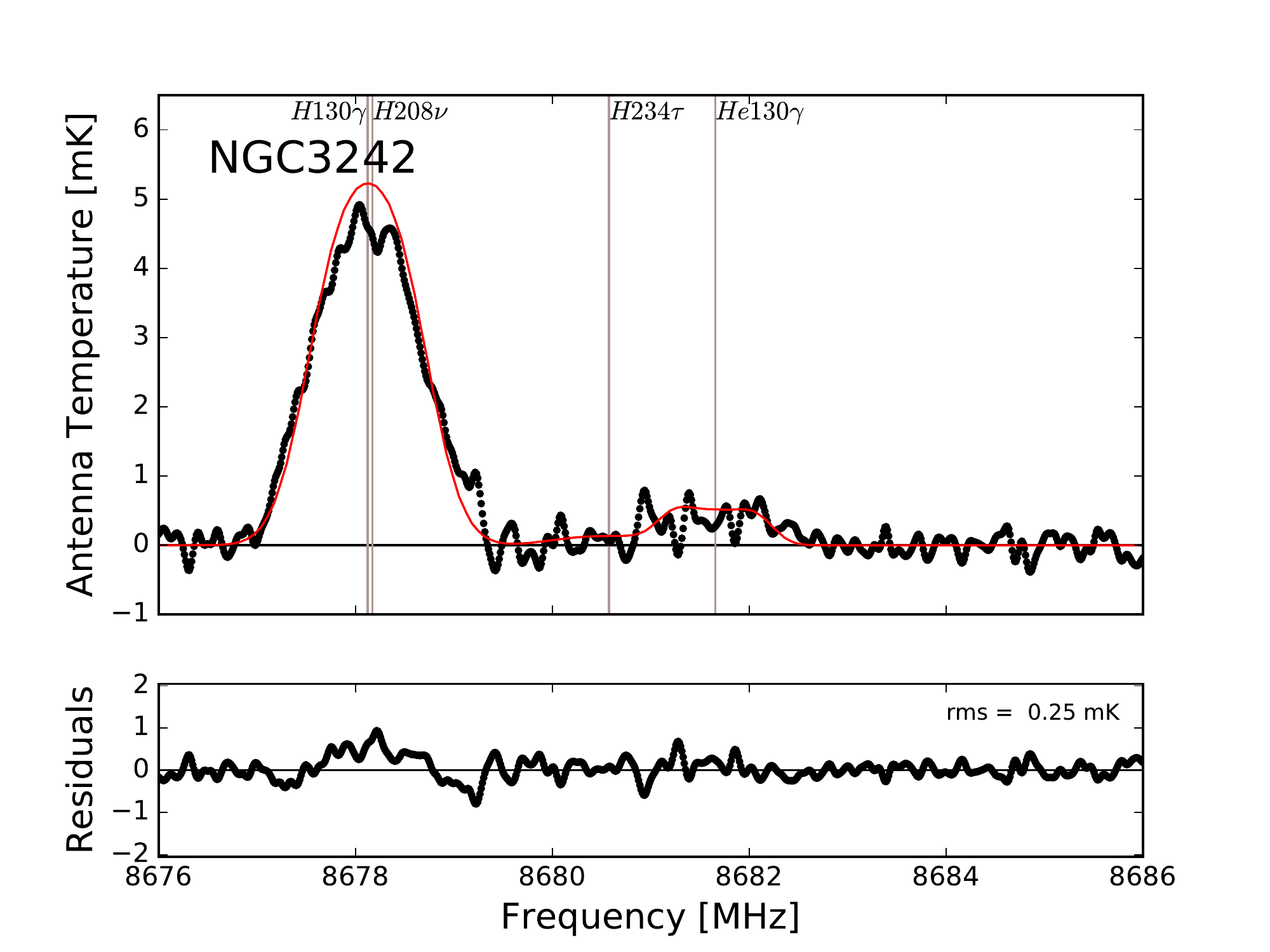}
\includegraphics[angle=0,scale=0.40]{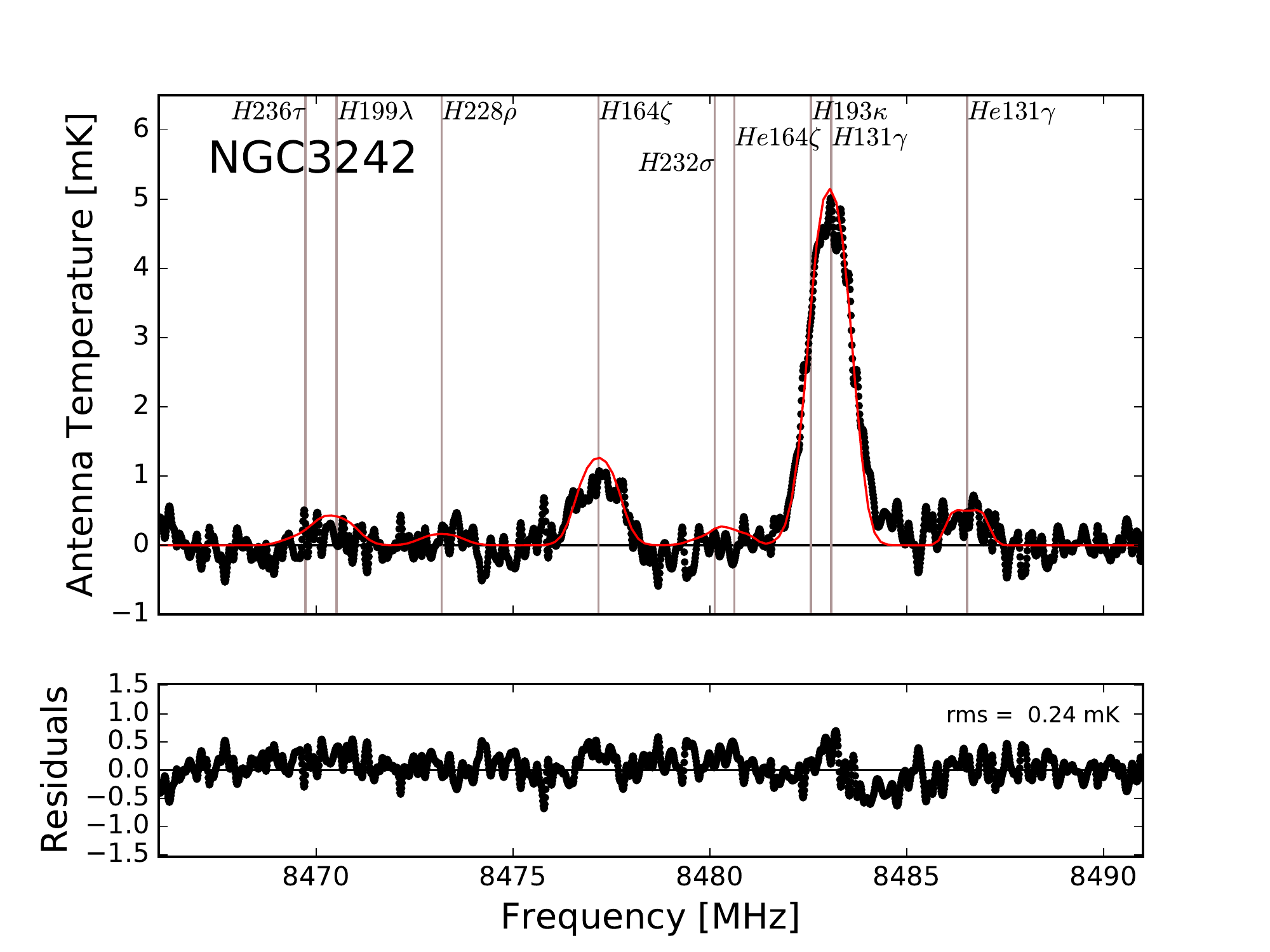}\\

\vspace{-.15cm}

\includegraphics[angle=0,scale=0.40]{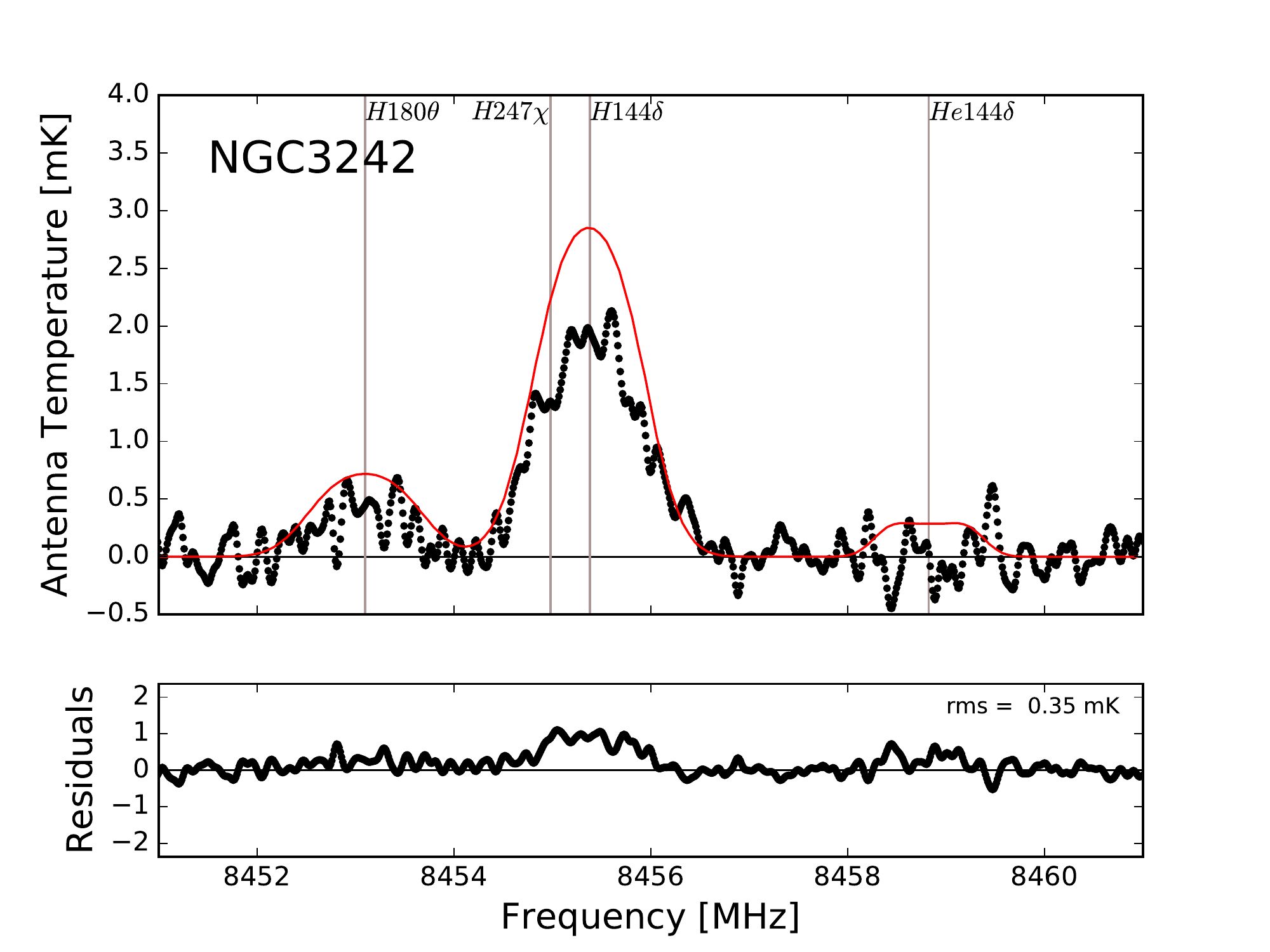}
\includegraphics[angle=0,scale=0.40]{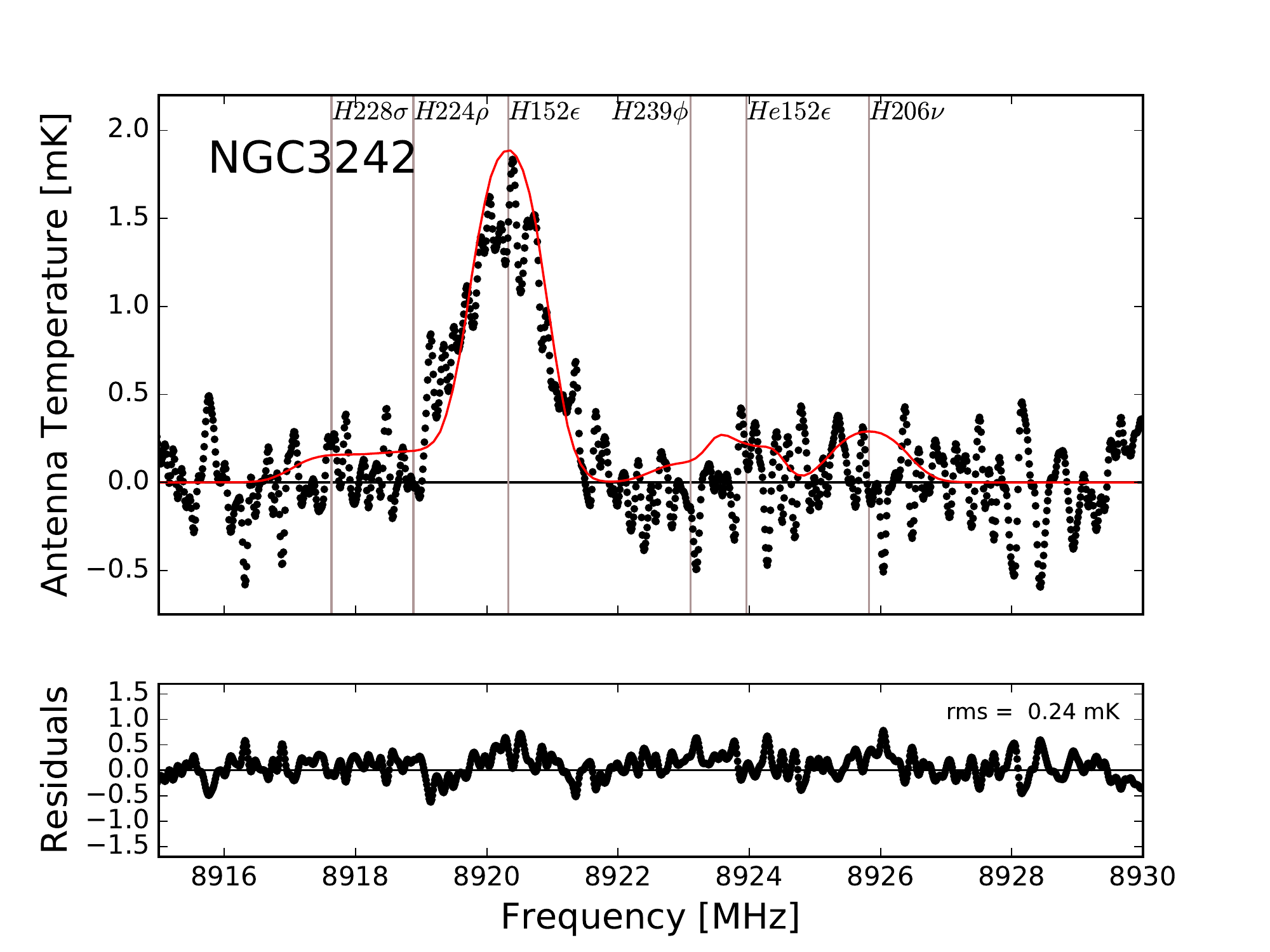}\\

\caption{\ngc{3242} NEBULA models. Starting at the top left the
  following spectral bands were modeled: \hgamma, \hggamma, \hdelta,
  and \heepsilon.
}
\label{fig:n3242modelB}
\end{figure}

\begin{figure}[!h]
\centering 
\includegraphics[angle=0,scale=0.40]{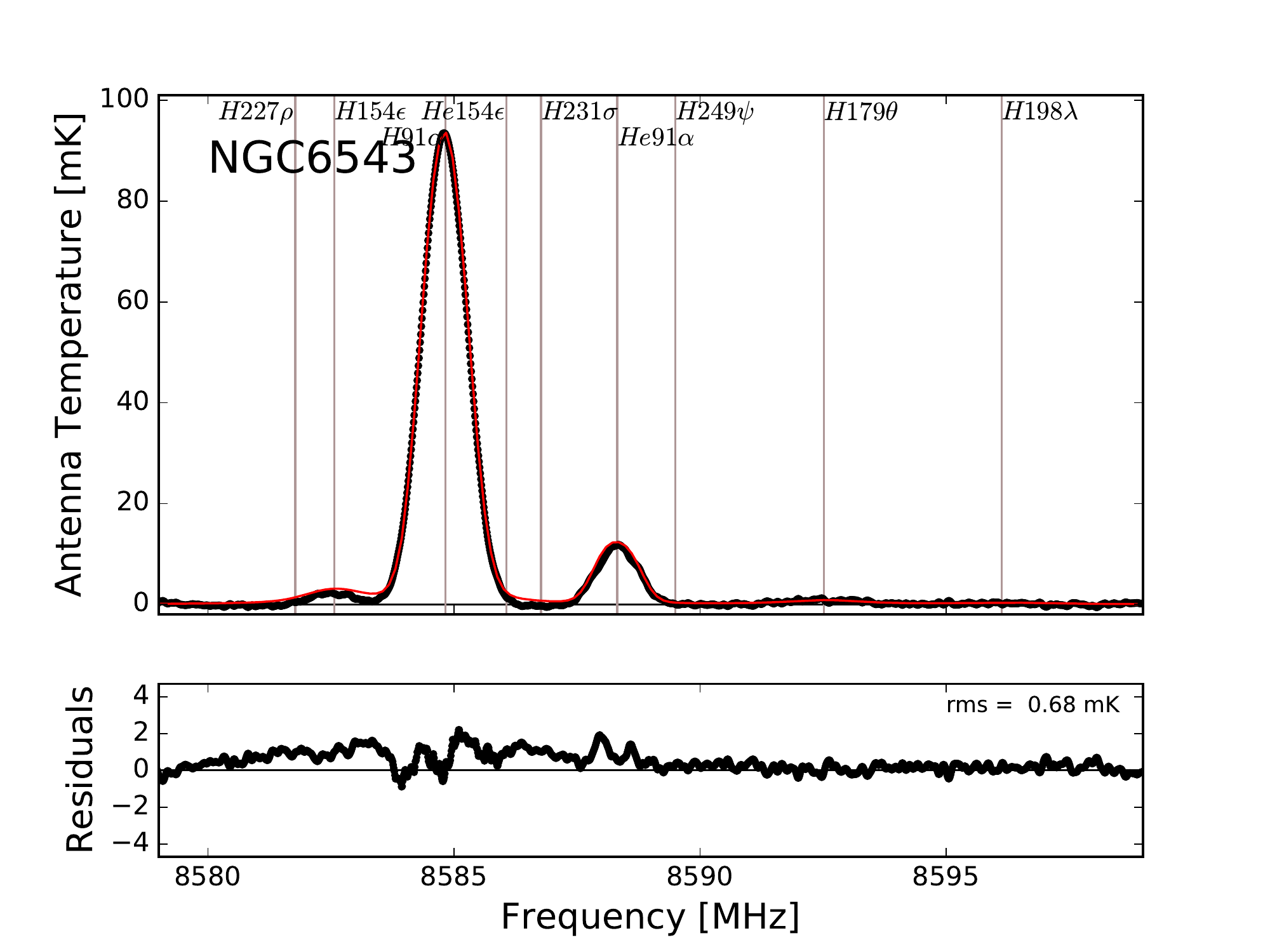}
\includegraphics[angle=0,scale=0.40]{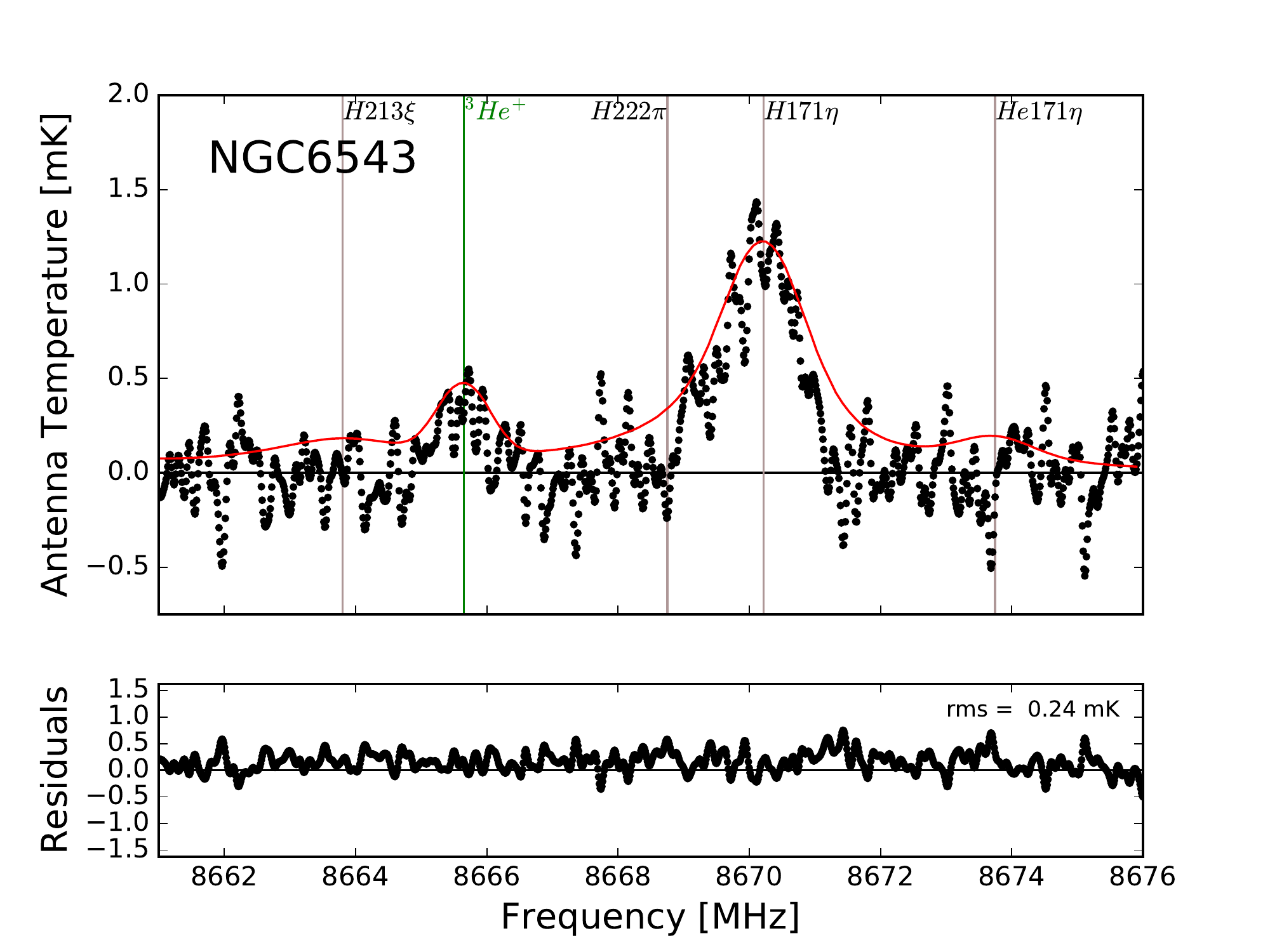}\\


\includegraphics[angle=0,scale=0.40]{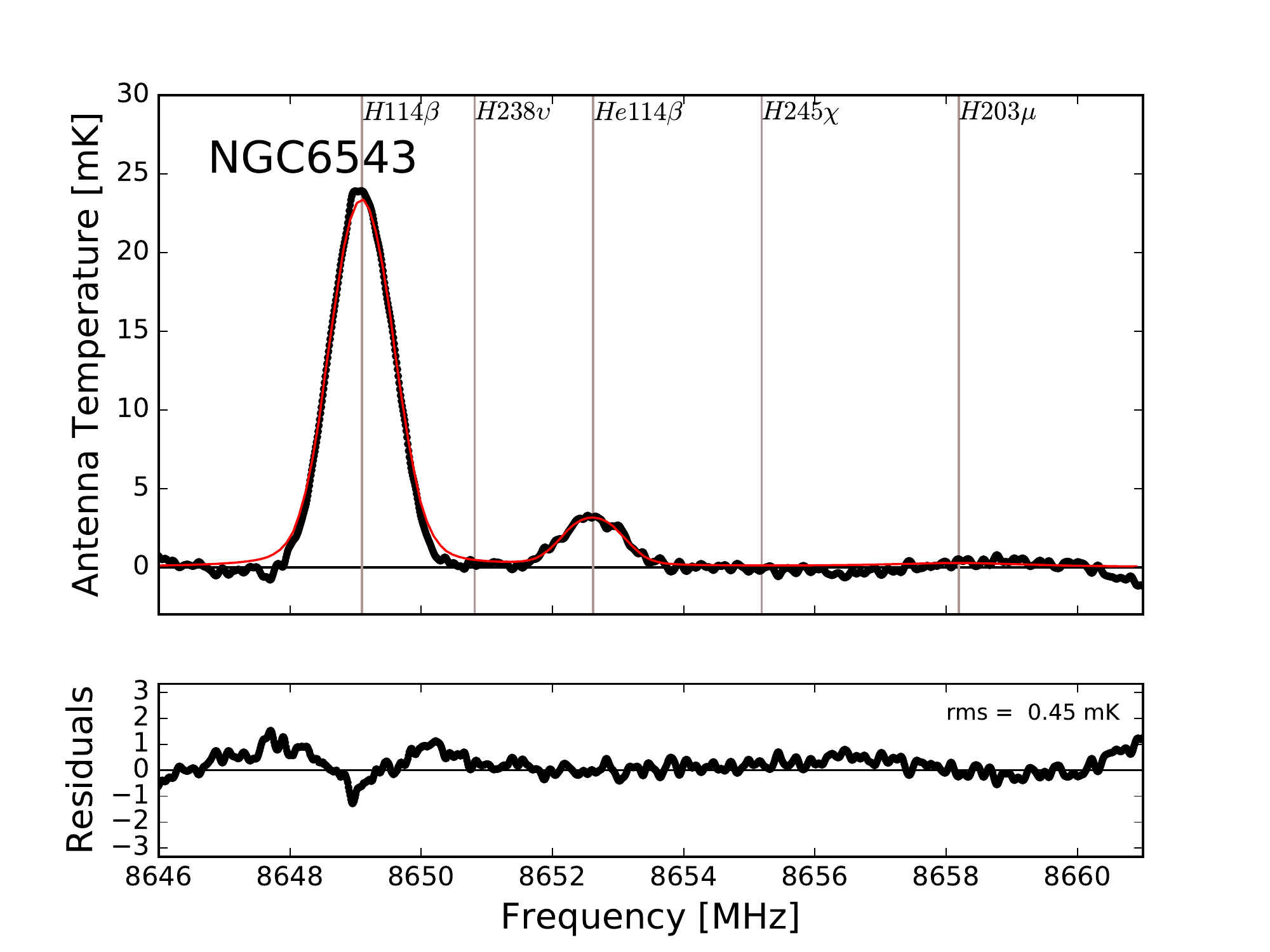}
\includegraphics[angle=0,scale=0.40]{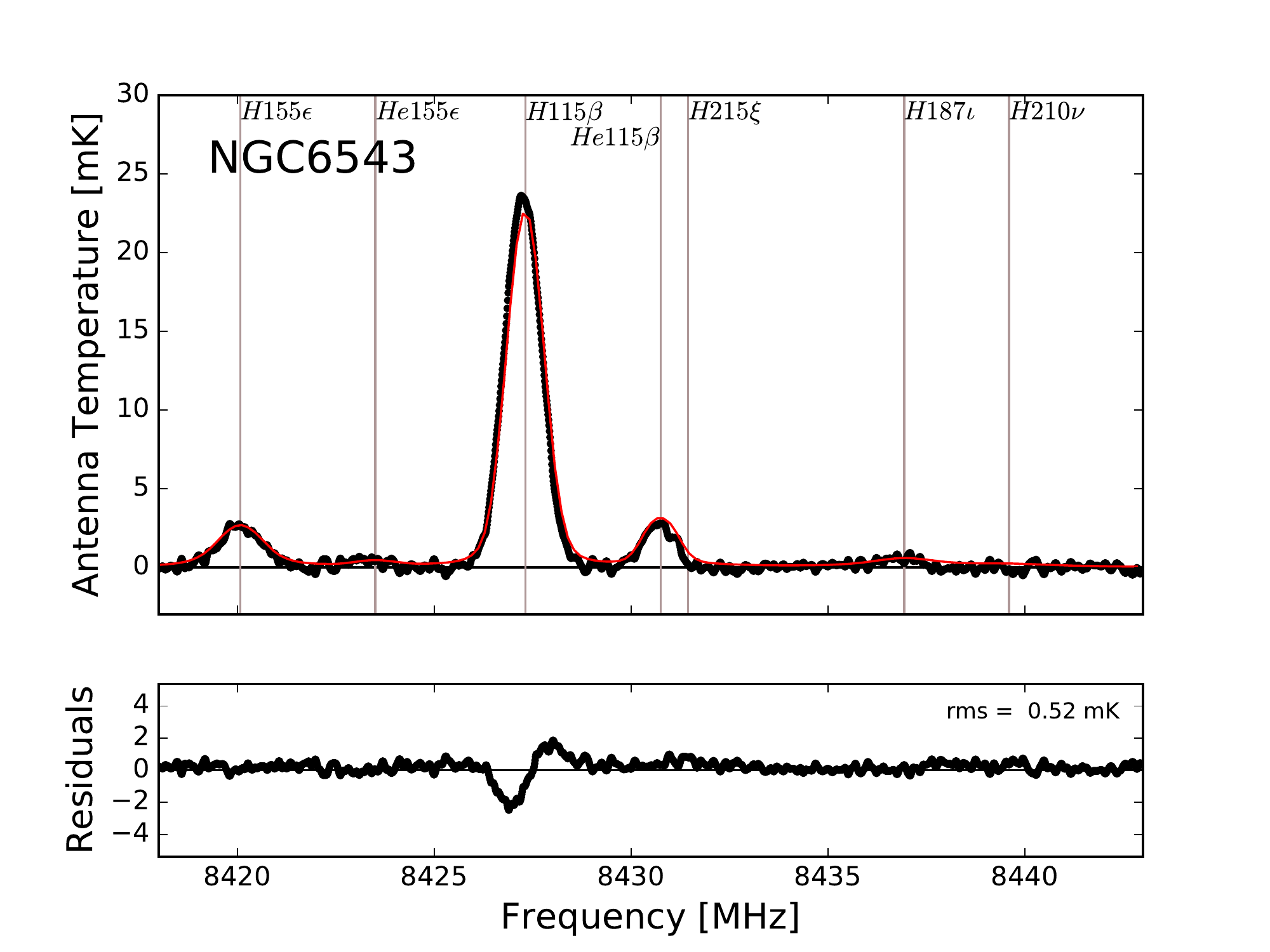}\\


\caption{\ngc{6543} NEBULA models. Starting at the top left the
  following spectral bands were modeled: \halpha, \hep3, \hbeta, 
  and \hbbeta.
}
\label{fig:n6543modelA}
\end{figure}

\begin{figure}[!h]
\centering \vspace{-1.0cm}
\includegraphics[angle=0,scale=0.40]{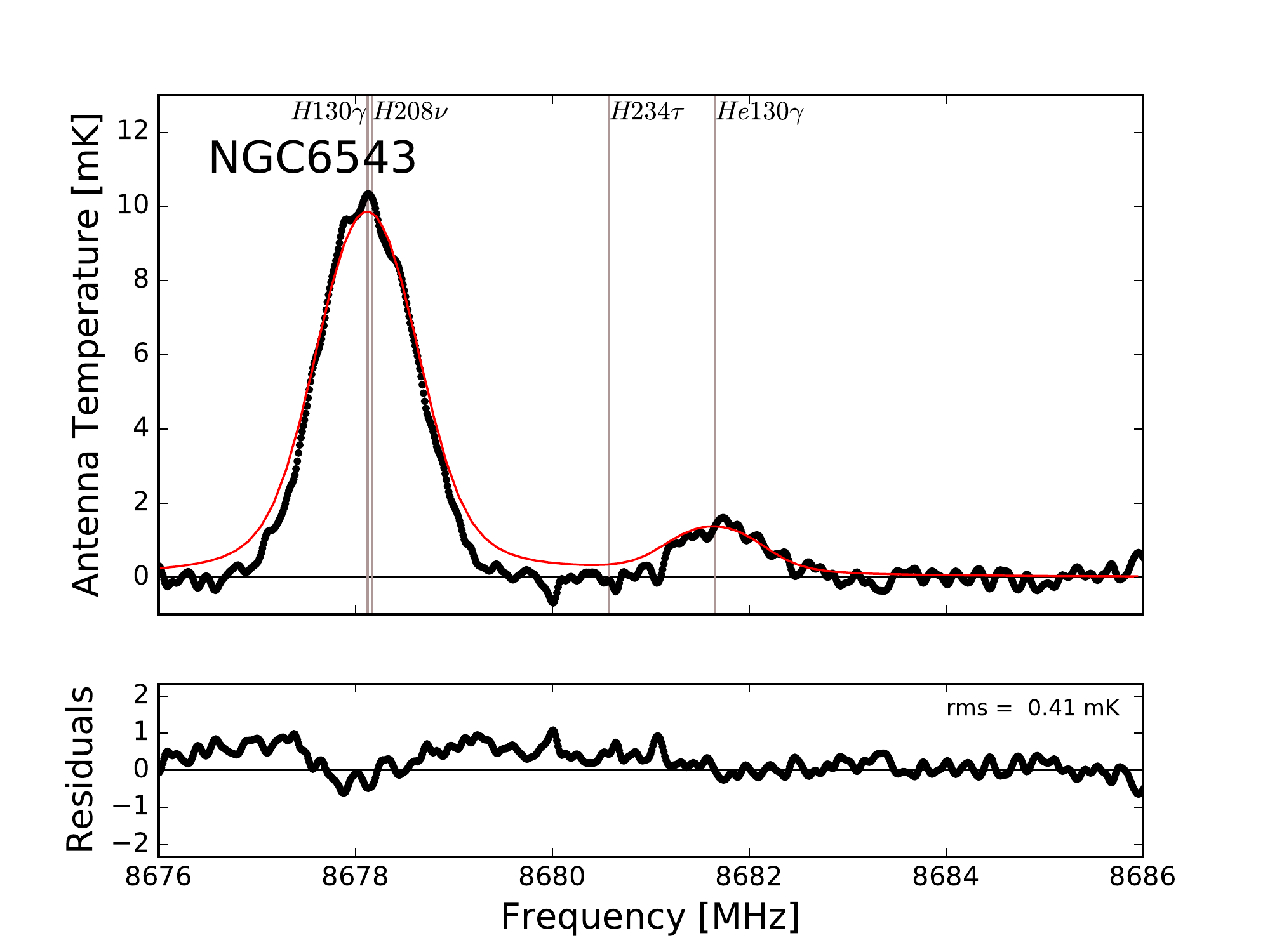}
\includegraphics[angle=0,scale=0.40]{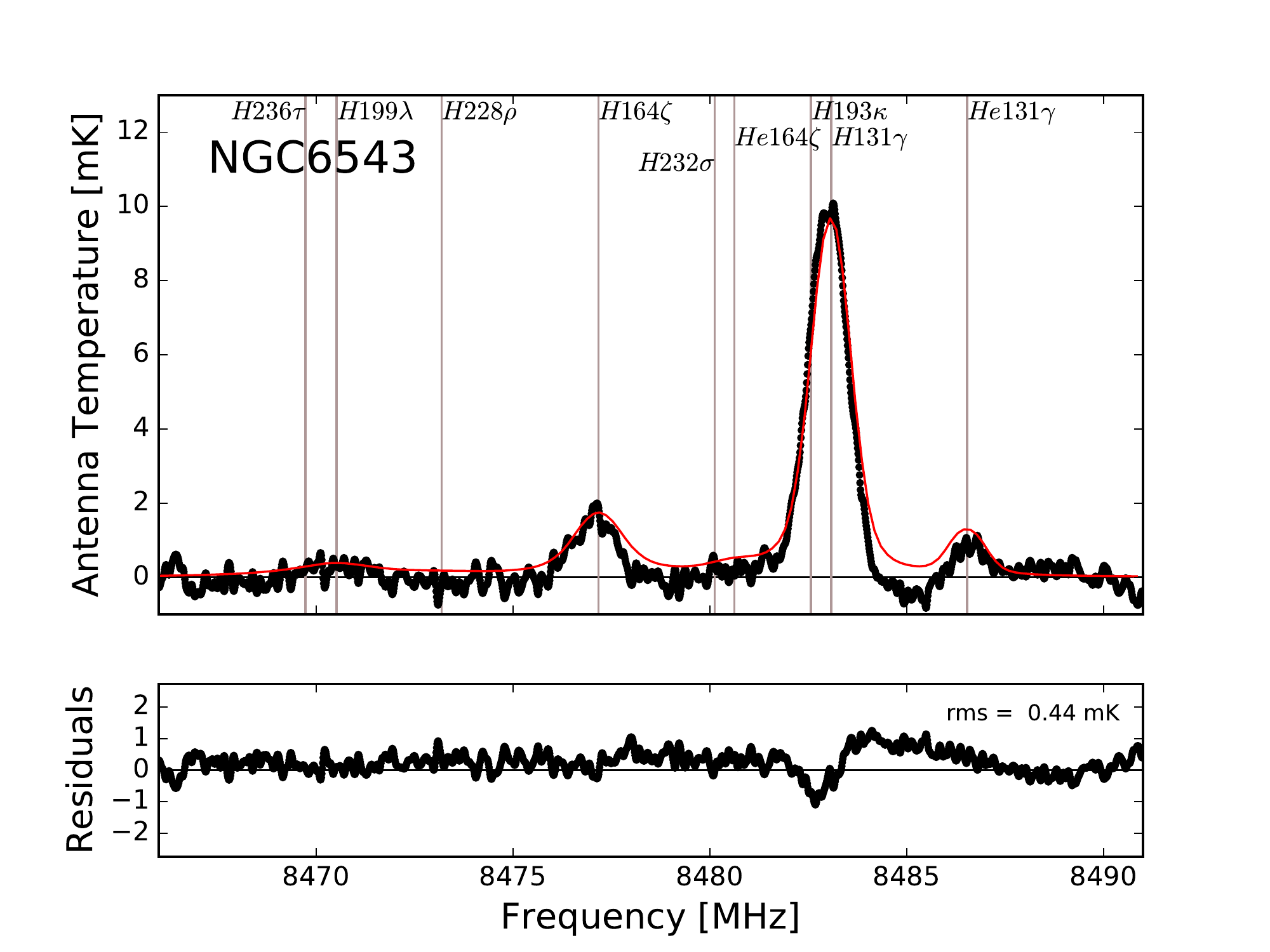}\\

\vspace{-.15cm}

\includegraphics[angle=0,scale=0.40]{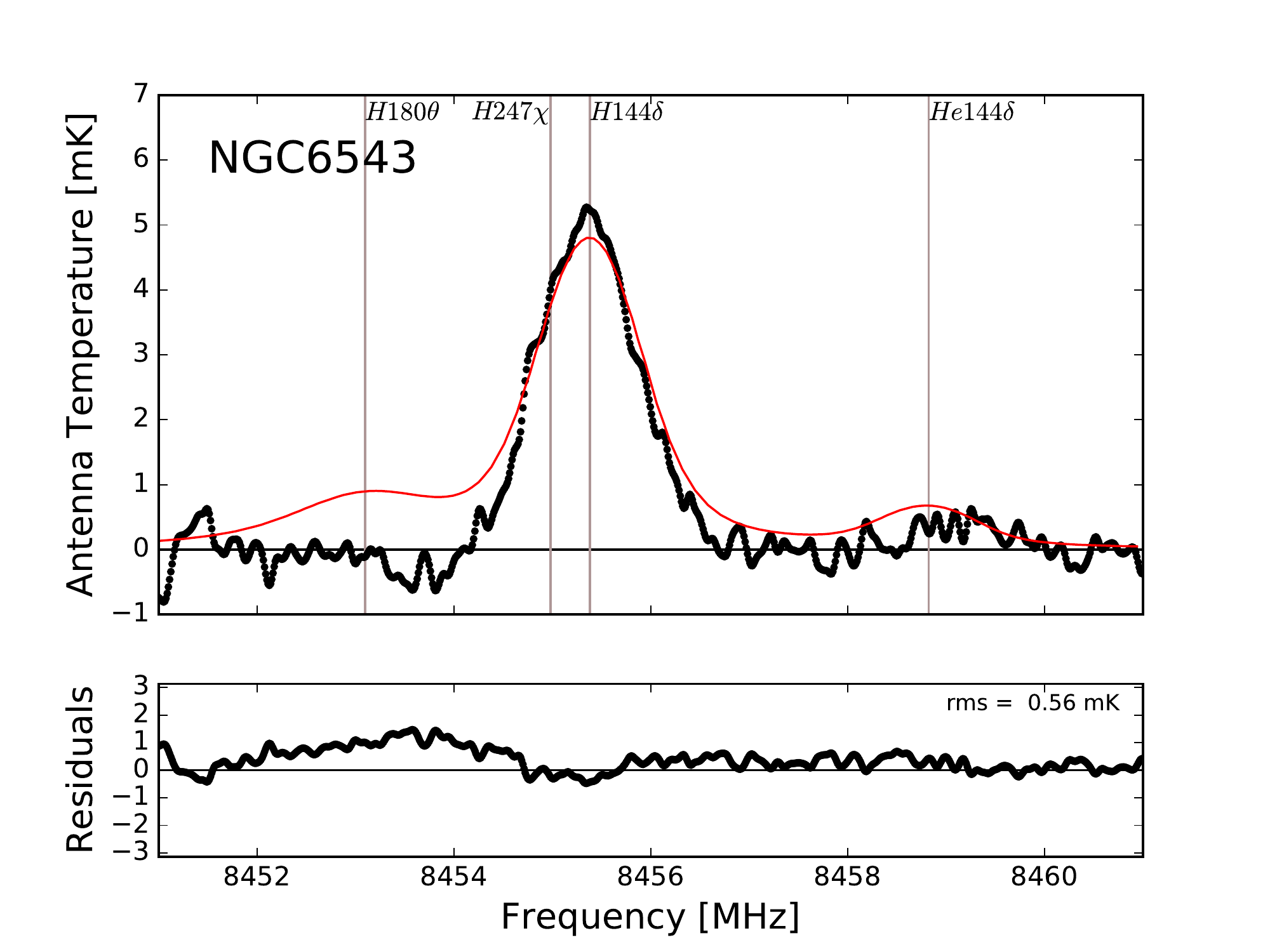}
\includegraphics[angle=0,scale=0.40]{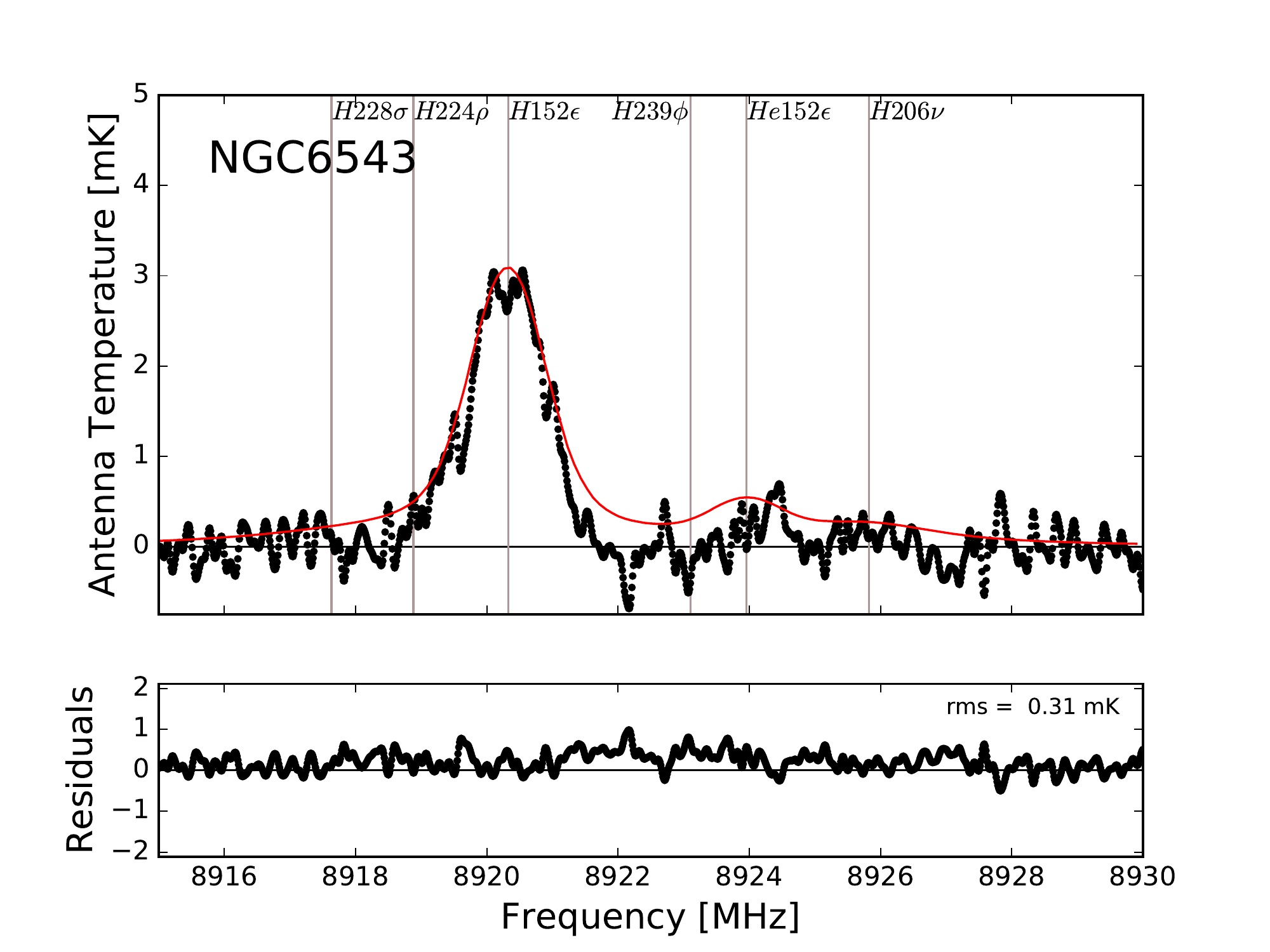}\\

\caption{\ngc{6543} NEBULA models. Starting at the top left the
  following spectral bands were modeled: \hgamma, \hggamma, \hdelta,
  and \heepsilon.
}
\label{fig:n6543modelB}
\end{figure}

\clearpage

\section{Conversion of Number Density to Mass Fraction}\label{appen:C}

Abundance ratios are typically expressed either by number or by mass.
The number density of species i is given by
\begin{equation}
  n_{\rm i} = \frac{X_{\rm i}\,\rho}{m_{\rm i}}
\label{eq:mass}
\end{equation}
where $\rho$ is the total density, $X_{\rm i}$ is the mass fraction of
species i, and $m_{\rm i}$ is the mass of species i.  We define $X$,
$Y$, and $Z$ to be the mass fraction of hydrogen, helium, and metals,
respectively.  Therefore, $X + Y + Z = 1$.

\subsection{\he4\ Abundances}

From Equation~\ref{eq:mass}
\begin{equation}
  n(\he4) = \frac{X(\he4)\,\rho}{m(\he4)} = \frac{Y\,\rho}{m(\he4)}
\end{equation}
and
\begin{equation}
  n({\rm H}) = \frac{X({\rm H})\,\rho}{m({\rm H})} = \frac{X\,\rho}{m({\rm H})}.
\end{equation}
Therefore the \her4\ abundance ratio by number is given by
\begin{equation}
  n(\he4)/n({\rm H}) \equiv y = \frac{Y\,m({\rm H})}{X\,m(\he4)}.
\end{equation}
But $X = 1 - Y - Z$ and $m(\he4) \sim 4\,m({\rm H})$, so
\begin{equation}
  y = \frac{Y}{4\,(1 - Y - Z)}.
\end{equation}
Solving for $Y$ yields
\begin{equation}
  Y = \frac{4\,y\,(1 - Z)}{(1 + 4y)}.
\label{eq:4he}
\end{equation}
Here we ignore nuclear binding energy and neglect the contribution of
the less abundant isotopic abundances.

\subsection{\he3\ Abundances}

From Equation~\ref{eq:mass}
\begin{equation}
  n(\he3) = \frac{X(\he3)\,\rho}{m(\he3)} = \frac{Y_{3}\,\rho}{m(\he3)}
\end{equation}
where we have defined the \he3\ mass fraction as $Y_{3}$.  Therefore
the \her3\ abundnace ratio by number is given by
\begin{equation}
  n(\he3)/n({\rm H}) \equiv y_{3} = \frac{Y_{3}\,m({\rm H})}{X\,m(\he3)}.
\end{equation}
But $X = 1 - Y - Z$ and $m(\he3) \sim 3\,m({\rm H})$, so
\begin{equation}
  y_{3} = \frac{Y_{3}}{3\,(1 - Y - Z)}.
\end{equation}
Solving for $Y_{3}$ yields
\begin{equation}
  Y_{3} = 3\,y_{3}\,(1 - Y - Z).
\end{equation}
Substituting $Y$ from Equation~\ref{eq:4he} and rearranging terms yields
\begin{equation}
Y_{3} = \frac{3\,y_{3}\,(1 - Z)}{1 + 4y}.
\label{eq:3he}
\end{equation}

\end{document}